\documentclass[12pt,a4paper]{article}
\usepackage[utf8]{inputenc}
\usepackage[T1]{fontenc}
\usepackage{amsmath,amsfonts,amsthm,mathabx}
\usepackage{amssymb}
\usepackage{graphicx}
\usepackage{lineno}
\usepackage{geometry}
\usepackage{xcolor}
\usepackage{dsfont}
\usepackage{mathtools}
\usepackage{tgtermes}
\numberwithin{equation}{section}
\usepackage{authblk}
\usepackage{hyperref} 
\usepackage{cite}

\numberwithin{equation}{section}
\newcommand{\be}{\begin{equation}}
\newcommand{\ee}{\end{equation}}
\newcommand{\bal}{\begin{aligned}}
\newcommand{\eal}{\end{aligned}}

\def\bJ{\boldsymbol{J}}
\def\bP{\boldsymbol{P}}
\def\bZ{\boldsymbol{Z}}
\def\bk{\boldsymbol{k}}
\def\bsigma{\boldsymbol{\sigma}}
\def\bG{\boldsymbol{G}}
\def\bH{\boldsymbol{H}}
\def\bM{\boldsymbol{M}}
\def\bS{\boldsymbol{S}}
\def\bL{\boldsymbol{L}}
\def\bT{\boldsymbol{T}}
\def\bmK{\boldsymbol{\mathcalT}}
\def\bSigma{\boldsymbol{\Sigma}}
\def\bSigmat{\tilde{\boldsymbol{\Sigma}}}
\def\ed{\mathrm{d}}
\def\bmG{\boldsymbol{\mathcal G}}
\def\bmH{\boldsymbol{\mathcal H}}
\def\bmM{\boldsymbol{\mathcal M}}
\def\bmS{\boldsymbol{\mathcal S}}
\def\bmL{\boldsymbol{\mathcal L}}
\def\bmP{\boldsymbol{\mathcal P}}
\def\bmK{\boldsymbol{\mathcal K}}
\def\bmJ{\boldsymbol{\mathcal J}}
\def\bmA{\boldsymbol{\mathcal A}}
\def\bmB{\boldsymbol{\mathcal B}}
\def\bmC{\boldsymbol{\mathcal C}}
\def\bmD{\boldsymbol{\mathcal D}}
\def\bmZ{\boldsymbol{\mathcal Z}}
\def\bmT{\boldsymbol{\mathcal T}}
\def\bmY{\boldsymbol{\mathcal Y}}
\def\bmKt{\tilde{\boldsymbol{\mathcal K}}}
\def\sp{\mathrm{Span}}

\def\bF{\boldsymbol{F}}
\def\bA{\boldsymbol{A}}
\def\nw{\mathfrak{nw}}

\def\nw{\mathfrak{nw}}

\def\hsnwi{\mathfrak{hs_3 nw1}}
\def\hsnwii{\mathfrak{hs_3 nw2}}
\def\hsnwiii{\mathfrak{hs_3 nw1,2}}

\def\hsbi{\mathfrak{hs_3 barg1_{2+1}}}
\def\hsbii{\mathfrak{hs_3 barg2_{2+1}}}
\def\hsbiii{\mathfrak{hs_3 barg1,2_{2+1}}}

\def\hsmi{\mathfrak{hs_3 nr\textnormal{-}max1_{2+1}}}
\def\hsmii{\mathfrak{hs_3 nr\textnormal{-}max2_{2+1}}}
\def\hsmiii{\mathfrak{hs_3 nr\textnormal{-}max1,2_{2+1}}}

\def\hsni{\mathfrak{hs_3 nh1_{2+1}}}
\def\hsnii{\mathfrak{hs_3 nh2_{2+1}}}
\def\hsniii{\mathfrak{hs_3 nh1,2_{2+1}}}

\def\hsali{\mathfrak{hs_3 nr\textnormal{-}adslor1_{2+1}}}
\def\hsalii{\mathfrak{hs_3 nr\textnormal{-}adslor2_{2+1}}}
\def\hsaliii{\mathfrak{hs_3 nr\textnormal{-}adslor1,2_{2+1}}}

\def\hsrelp{\mathfrak{hs_3 poin_{2+1}}}
\def\hsrela{\mathfrak{hs_3 ads_{2+1}}}
\def\hsrelm{\mathfrak{hs_3 max_{2+1}}}
\def\hsrelal{\mathfrak{hs_3 adslor_{2+1}}}

\def\reli{\mathfrak{sl}(3,\mathbb R)\oplus \mathfrak{so}(3)}
\def\relii{\mathfrak{sl}(3,\mathbb R)\oplus \mathfrak{sl}(2,\mathbb R)\oplus \mathfrak{so}(2)}

\def\uo{\mathfrak{so}(2)}
\def\sltw{\mathfrak{sl}(2,\mathbb R)}
\def\sotr{\mathfrak{so}(3)}
\def\sltr{\mathfrak{sl}(3,\mathbb R)}
\def\sltwu{\mathfrak{sl}(2,\mathbb R)\oplus \mathfrak{so}(2)}

\def\wed{\,}

\def\r{{\rm rel}}

\def\bX{\boldsymbol{X}}
\def\bY{\boldsymbol{Y}}

\vspace{-10pt}
\date{}

\begin{document}
\title{\textbf{Non-relativistic spin-$\mathbf 3$ symmetries in $\mathbf{2+1}$ dimensions from expanded/extended Nappi-Witten algebras}}

\author[$a\,\ast$]{Ricardo Caroca }
\author[$b\,\star$]{Diego M. Pe\~nafiel}
\author[$c\,\dagger$]{Patricio Salgado-Rebolledo}
\vspace{10pt}
\affil[$a$]{\it {\normalsize Departamento de Matem\'atica y F\'isica Aplicadas\break Universidad Cat\'olica de la Sant\'isima Concepci\'on, Alonso de Ribera 2850, Concepci\'on, Chile}
\vspace{1em}}

\affil[$b$]{\it {\normalsize 
Instituto de Ciencias Exactas y Naturales, Facultad de Ciencias\break Universidad Arturo Prat, Avenida Arturo Prat Chac\'on 2120, 1110939, Iquique, Chile} \vspace{1em}}

\affil[$c$]{\it {\normalsize  Institute of Theoretical Physics\break Wroc\l{}aw University of Science and Technology, 50-370 Wroc\l{}aw, Poland}}

\maketitle

\begin{abstract}
\noindent We show that infinite families of Galilean spin-$3$ symmetries in $2+1$  dimensions, which include higher-spin extensions of the Bargmann, Newton-Hooke, non-relativistic Maxwell, and non-relativistic AdS-Lorentz algebras, can be obtained as Lie algebra expansions of two different spin-$3$ extensions of the Nappi-Witten symmetry. These higher-spin Nappi-Witten algebras, in turn, are obtained by means of In\"on\"u-Wigner contractions applied to suitable direct product extensions of $\sltr$. Conversely, we show that the same result can be obtained by considering contractions of expanded $\sltr$ algebras. The method can be used to define non-relativistic higher-spin Chern-Simon gravity theories in $2+1$  dimensions in a systematic way. 
\end{abstract}

\vfill
\noindent {\small{\it
 \noindent\rule{2cm}{0.4pt} \\
$\ast${\,\tt rcaroca@ucsc.cl}; \\
$\star${\,\tt dimolina@unap.cl}}; \\
$\dagger${\,\tt patricio.salgado-rebolledo@pwr.edu.pl}
}

\newpage
\tableofcontents
\newpage

\section{Introduction}

\noindent The study of higher-spin fields in $2+1$ dimensions has received considerable attention in recent years due to the possibility to overcome some of the difficulties appearing in higher-dimensional cases, where in order to surpass a series of old no-go theorems \cite{Bekaert:2010hw} and obtain a consistent theory, the inclusion of a cosmological constant as well as an infinite tower of interacting higher-spin fields is needed \cite{Vasiliev:1990en,Vasiliev:2003ev}. In fact, in the massless sector, a three-dimensional higher-spin gravity theory with negative cosmological constant that includes a finite number of higher-spin fields can be defined by means of a Chern-Simons (CS) action \cite{Blencowe:1988gj,Bergshoeff:1989ns,Vasiliev:1995dn}.  Even though CS theories unavoidably lead to no local propagating degrees of freedom in the bulk, higher-spin CS gravity has been shown to admit black hole solutions with non-trivial properties (see the reviews \cite{Bunster:2014mua,Castro:2016tlm} and references therein). Furthermore, considering suitable boundary conditions gives rise to rich boundary dynamics and asymptotic symmetries given by $\mathcal W$-algebras \cite{Henneaux:2010xg,Campoleoni:2010zq}. Higher-spin CS theories were later generalized to the case of vanishing cosmological constant \cite{Afshar:2013vka,Gonzalez:2013oaa} and non-Lorentzian gravity \cite{Bergshoeff:2016soe,Chernyavsky:2019hyp}.\\

\noindent The recent interest in higher-spin fields in three space-time dimensions goes beyond gravitational physics. A remarkable $(2+1)$-dimensional system where higher-spin symmetries play an important role is the fractional quantum Hall effect, where $\mathcal W$-algebras appear as the quantum realization of the area-preserving diffeormorphism algebra \cite{Cappelli:1992yv,Iso:1992aa,Cappelli:2021kxd}. Moreover, it has been conjectured that, apart from the Magnetoroton \cite{Girvin:1986zz}, a spin-$2$ graviton-like mode, additional collective excitations described by higher-spin fields should be introduced in the known effective field theory description to reproduce the full Girvin-Macdonald-Platzman algebra \cite{Golkar:2016thq,Gromov:2017qeb}. Another exotic scenario where higher-spin fields can be relevant is Fracton physics (see the reviews \cite{Nandkishore:2018sel,Grosvenor:2021hkn} and references therein), which introduces quasiparticles with restricted mobility whose local interactions are described by higher-rank gauge fields. The higher-spin theories capable of describing these types of physical systems are, however, non-relativistic (NR).\\

\noindent NR symmetries and NR geometry have attracted recent interest due to their applicability in extensions of holography \cite{Taylor:2008tg,Balasubramanian:2008dm,Herzog:2008wg,Bagchi:2009my} and condensed matter physics \cite{Duval:2001hu,Son:2008ye,Geracie:2014nka,Gromov:2014vla}. The extension of kinematical symmetries to include higher-spin generators was introduced in \cite{Bergshoeff:2016soe}, where in particular two different spin-$3$ extensions of the Extended Bargmann and Newton-Hooke algebras were given\footnote{Galilean and Carrollian conformal higher-spin symmetries in arbitrary dimensions have been recently constructed in \cite{Campoleoni:2021blr}.}. On the other hand, Lie algebra expansions methods \cite{Hatsuda:2001pp,deAzcarraga:2002xi,Boulanger:2002bt,Izaurieta:2006zz,deAzcarraga:2007et,Penafiel:2016ufo}  have been extensively used in the construction of extended non-Lorentzian symmetries \cite{Bergshoeff:2019ctr,deAzcarraga:2019mdn,Hansen:2019vqf,Penafiel:2019czp,Gomis:2019nih,Kasikci:2020qsj,Concha:2020eam,Kasikci:2021atn,Gomis:2022spp,Ekiz:2022wbi}. Particularly, in \cite{Penafiel:2019czp} it has been shown that applying the expansion method to the Nappi-Witten algebra allows one to define infinite families of NR symmetries in $2+1$ dimensions that generalize the Bargmann and the Newton-Hooke symmetries. These are central extensions of the Generalized Galilean symmetries presented in \cite{Gonzalez:2016xwo} for $D=2+1$. This result can be contrasted with the families of relativistic spin-$3$ symmetries in $2+1$ dimensions found in \cite{Caroca:2017izc} by means of Lie algebra expansions applied to the $\sltr$ algebra. In particular, two families were constructed explicitly, which define generalized spin-$3$ symmetries of AdS type and Poincar\'e type. In this setup, the first \emph{level} of the expansion yields the spin-$3$ extension of the AdS algebra, $\sltr\times\sltr$, and the spin-$3$ extension of the Poincar\'e algebra, respectively. At the second level, spin-$3$ extensions of the $(2+1)$-dimensional version of the Maxwell algebra and the AdS-Lorentz algebra arise, whereas higher levels correspond to spin-$3$ extensions of more complicated examples of the so-called Generalized AdS \cite{Concha:2016kdz} and Generalized Poincar\'e \cite{Izaurieta:2009hz} algebras in three space-time dimensions. Therefore, if higher-spin extensions of the Nappi-Witten symmetry can be defined, expansions of such symmetries can be used to define NR limits of the higher-spin symmetries defined in \cite{Caroca:2017izc}. In $2+1$ dimensions, NR limits usually require to extend relativistic symmetries by introducing extra Abelian generators as a way to obtain a non-degenerate invariant form in the resulting NR algebra (see for example \cite{Bergshoeff:2022eog}). Similarly, to define proper NR contractions of $\sltr$, it will be necessary to consider suitable extensions thereof. This time, however, the necessary extensions to be introduced are non-Abelian. In particular, we find two higher-spin extensions of the $\nw$ symmetry, which we dub $\hsnwi$ and $\hsnwii$, and follow from two inequivalent contractions applied to $\reli$ and $\relii$, respectively.\\
\\
\noindent 
In this article, i) we define two different spin-$3$ extensions of the Nappi-Witten symmetry, which we denominate $\hsnwi$ and $\hsnwii$, by means of In\"on\"u-Wigner contractions \cite{inonu1993contraction} of suitable direct product extensions of the $\sltr$ algebra, ii) we show that expansions of these spin-$3$ Nappi-Witten algebras define NR higher-spin symmetries in $2+1$ dimensions, iii) we show that the same result can be achieved by first expanding the $\sltr$ algebra and subsequently defining a NR contraction of (an extension of) the resulting expanded symmetry, iv) we construct non-degenerate invariant bilinear forms on the resulting NR higher-spin algebras and define the corresponding higher-spin CS gravity theories.\\

\noindent The paper is organized as follows: in Section \ref{HSNW} we show how the $\nw$ algebra can be defined as a NR contraction of the $\sltwu$ algebra and extend this result to define a NR contraction of the $\reli$ algebra and the $\relii$ algebra, which lead to two different higher-spin extensions of the $\nw$ symmetry. In Section \ref{ExpNW} we construct expanded Nappi-Witten algebras by means of Lie algebra expansions based on Abelian semigroups. The method is also used to define invariant bilinear forms on the expanded algebras and CS actions. In Section \ref{Examples}, we show that the spin-$3$ extensions of the Extended Bargmann algebra and the NR Maxwell algebra in $2+1$ dimensions correspond to particular examples of expanded higher-spin Nappi-Witten algebras. By considering a different semigroup, a cosmological constant can be introduced, leading to a spin-$3$ extension of the Newton-Hooke and the NR AdS-Lorentz algebras. The algebraic structure of these symmetries is discussed and the explicit form of the corresponding CS higher-spin gravity theories is given. Finally, we end with a general discussion of our results and elaborate on possible future directions.\\

\noindent Note added: during the preparation of the manuscript, it came to our knowledge the development of Ref. \cite{Concha:2022muu}, where a spin-$3$ generalization of the Lie algebra expansions considered \cite{Bergshoeff:2019ctr,Gomis:2019nih} for $D=2+1$ are worked out, and presents some overlap with the results included in Section \ref{sectionNH} of our article.

\section{Spin-$3$ extensions of the Nappi-Witten algebra}\label{HSNW}

In this section, we construct two different spin-$3$ extensions of the Nappi-Witten algebra $\nw$ by considering suitable In\"o\"u-Wigner contractions of extensions of the $\sltr$ algebra. To make our construction more transparent, we start by reviewing how the $\nw$ algebra itself can be obtained as a contraction of an Abelian extension of $\sltw$\footnote{This type of contraction was introduced by Saletan \cite{saletan1961contraction} and later generalized to higher dimensions in \cite{Olive:1993hk,Sfetsos:1994vz}.}. The starting point is the $\sltw$ algebra
\be
\left[\bL_A,\bL_B \right]=\epsilon_{AB}^{\quad C}\bL_C,\qquad A=0,1,2,
\ee
which is isomorphic to the Lorentz algebra in $2+1$  dimensions, $\mathfrak{so} (2,1)$. Here $\epsilon_{012}=1$ and the relativistic indices are lowered and raised with the Minkowski metric $\eta_{AB}={\rm diag}(-1,1,1)$ and its inverse. A NR limit can be defined by introducing a $\uo$ generator $\bX$ and redefining the Lie algebra generators in the following way
\be\label{slu}
\bal
\bL_{0}  &=  \dfrac{1}{2}\bk+c^{2}\bsigma,\qquad
\bL_a &=  c\,\bk_a,\qquad
\bX & =  \frac{1}{2}\bk-c^{2}\bsigma ,
\eal
\ee
where have introduced the parameter $c$, which stands for the speed of light, and decomposed the relativistic indices as $A=(0,a=1,2)$. The new generators $\left\{\bk, \bk_a, \bsigma\right\}$ are then expressed in terms of $\bL_a$ and $\bX$ through the relations
\be\label{invcontNW}
\bal
\bk & =  \bL_{0}+\bX,\qquad
\bk_a & =  \frac{1}{c}\bL_a,\qquad
\bsigma & =  \frac{1}{2c^2}\left(\bL_{0}-\bX\right),
\eal
\ee
and, in the limit $c\rightarrow\infty$, their commutators reproduce the Nappi-Witten algebra $\nw$
\be\label{nwalg}
\left[\bk,\bk_a\right]=\epsilon_{ab}\,\bk_b
,\qquad
\left[\bk_a,\bk_b\right]=-\epsilon_{ab}\,\bsigma \qquad (\epsilon_{ab}\equiv\epsilon_{0ab}).
\ee
Due to the central term $\bsigma$, the $\nw$ algebra admits a non-degenerate invariant bilinear form (see Section \ref{IBF}). The Nappi-Witten symmetry is the central extension of the Euclidean symmetry in two dimensions\footnote{This corresponds to the Euclidean version of the Nappi-Witten algebra presented in \cite{Nappi:1993ie}.} \cite{Figueroa-OFarrill:1999cmq}. As we have just shown, it can be interpreted as a centrally extended NR limit of the $\sltw$ algebra, pretty much in the same way as the Bargmann algebra is a centrally extended NR limit of the Galilean one. It is thus evident that, in order to generalize the previous construction and define a symmetry that could be understood as a NR contraction of $\sltr$ in the sense of \eqref{slu}, it will be necessary to properly extend the $\sltr$ algebra. Its commutation relations read\footnote{The convention for the symmetrization of indices is 
$x_{(A}y_{B)}=x_A y_B+x_B y_A$, without normalization factor.} \cite{Campoleoni:2010zq}
\be\label{sl3alg}
\left[\bL_A,\bL_B\right] =\epsilon_{AB}^{\quad C}\bL_{C}\,\quad
\left[\bL_A,\bmL_{BC}\right] =\epsilon_{A(B}^{\quad \;D}\bmL_{C)D}\,\quad
\left[\bmL_{AB},\bmL_{CD}\right]  =- \eta_{(A(C}\epsilon_{D)B)}^{\quad \;E} \bL_E,
\ee
where $\bL_A$ defines an $\sltw$ sublagelbra and $\bmL_{AB}$ is traceless symmetric generator, i.e.
\be
\bmL_{AB}=\bmL_{BA},\quad
\eta^{AB}\bmL_{AB}=0\quad\Rightarrow\quad
\bmL_{00}=\delta^{ab}\bmL_{ab}\equiv \bmL_{aa}.
\ee
Splitting the relativistic indices into temporal and spatial components yields
\be\label{sltrcomm}
\bal
&\left[\bL_{0},\bL_a\right]   =\epsilon_{ab}\bL_b,
&
&\left[\bL_a,\bL_b\right]  = -\epsilon_{ab}\bL_{0},
&
&\left[\bL_{0},\bmL_{a0}\right]   =\epsilon_{ab}\bmL_{b0},
\\
&\left[\bL_{0},\bmL_{ab}\right]   =-\epsilon_{c(a}\bmL_{b)c},
&
&\left[\bL_a,\bmL_{bc} \right] =-\epsilon_{a(b}\bmL_{c)0},
&
&\left[\bL_a,\bmL_{b0}\right]   =-\epsilon_{a(c}\bmL_{b)c},
\\
&\left[\bmL_{a0},\bmL_{b0}\right]  =\epsilon_{ab}\bL_{0},
&
&\left[\bmL_{a0},\bmL_{bc}\right]  =\delta_{a(b}\epsilon_{c)d}\bL_{d},
&
&\left[\bmL_{ab},\bmL_{cd}\right]  =\delta_{(a(c}\epsilon_{d)b)}\bL_{0}.
\eal
\ee
Thus, one would like to keep the definitions \eqref{slu} for the $\sltw$ generators and find some analog relations for the spin-$3$ generators $\bmL_{ab}$ and $\bmL_{a0}$ in such a way that the limit  $c\rightarrow\infty$ produces a spin-$3$ extension of $\nw$ with a non-degenerate invariant form. This requires supplementing the Abelian generator $\bX$ with extra generators that, as we will see, do not commute with $\bX$. In the following, we consider two extensions of $\sltr$ that admit a well-defined NR limit. These are $\reli$ and $\relii$. In each case, NR contractions lead to novel extensions of the $\nw$ algebra that include higher-rank generators, which can be interpreted as NR higher-spin symmetries.

\subsection{NR contraction of $\reli$}
\label{NRcontractions}

In order to extend the $\sltr$ commutation relations \eqref{sltrcomm}, we introduce a $\sotr$ algebra spanned by generators $\{\bX,\bX_a\}$ satisfying  
\be\label{so3alg}
\bal
\left[\bX,\bX_a\right]=&\epsilon_{ab}\bX_b
,\qquad
\left[\bX_a, \bX_b\right]=&\epsilon_{ab} \bX,
\eal
\ee
and consider the direct product $\mathfrak{sl}(3,\mathbb{R})\times\mathfrak{so}(3)$ generated by $\bL_A=\{\bL_0,\bL_a\}$, $\bmL_{AB}= \{ \bmL_{a0},\bmL_{ab}\}$, and $\{\bX, \bX_a\}$. The NR limit can be defined by means of the change of basis
\be\label{redefreli}
\bal
\bL_{0}&= \dfrac{1}{2}\bk+c^{2}\bsigma,
&
\bL_a&=c\, \bk_a,
&
\bX &=\dfrac{1}{2}\bk-c^{2}\bsigma,
\\
\bmL_{a0} &=\dfrac{1}{2}\bmK_a+c^{2}\bSigma_a,
&
\bmL_{ab} &= c\,\bmK_{ab},
&
\bX_a&=\dfrac{1}{2}\bmK_a-c^{2}\bSigma_a,
\eal
\ee
with inverse relations given by
\be\label{gencont1}
\bal
\bk & =  \bL_{0}+\bX, &
\bk_a  & =  \frac{1}{c} \bL_a, &
\bsigma & =  \frac{1}{2c^2}\left(\bL_{0}-\bX\right),
\\
\bmK_a&= \bmL_{a0}+\bX_a,
&
\bmK_{ab}&=  \frac{1}{c}\bmL_{ab},
&
\bSigma_a&=  \frac{1}{2c^{2}}\left(\bmL_{a0}-\bX_a\right).
\eal
\ee
In the limit $c\rightarrow\infty$ we find the following non-vanishing commutation relations for the generators \eqref{gencont1}
\be\label{HSNWalgebra1}
\bal
&\left[\bk, \bk_a\right]  =  \epsilon_{ab} \bk_b,&
&\left[\bk_a,\bk_b\right] = -\epsilon_{ab}\bsigma,&
&\left[\bk,\bmK_a\right]  = \epsilon_{ab}\bmK_b,
\\
&\left[\bk,\bmK_{ab}\right] =  -\epsilon_{c(a}\bmK_{b)c},&
&\left[\bk,\bSigma_a\right] = \epsilon_{ab}\bSigma_b,&
&\left[\bk_a,\bmK_b\right] = -\epsilon_{a(c}\bmK_{b)c},
\\
&\left[\bk_a,\bmK_{bc}\right]  =  -\epsilon_{a(b}\bSigma_{c)}, &
&\left[\bsigma,\bmK_a\right]  =  \epsilon_{ab}\bSigma_b,&
&\left[\bmK_a,\bmK_b\right]  =  \epsilon_{ab}\bk,
\\
&\left[\bmK_a, \bSigma_b\right] = \epsilon_{ab}\bsigma,&
&\left[\bmK_a,\bmK_{bc}\right]  =\delta_{a(b}\epsilon_{c)d}\bk_d,&
&\left[\bmK_{ab},\bmK_{cd}\right] = \delta_{(a(c}\epsilon_{d)b)}\bsigma.
\eal
\ee
We will refer to this algebra as $\hsnwi$ since it defines a higher-spin extension of the Nappi-Witten algebra\footnote{Note that $\hsnwi$ is isomorphic to the homogeneous part of spin-$3$ Bargmann algebra in $2+1$ dimensions $\hsbi$ found in \cite{Bergshoeff:2016soe}.} .

\subsection{NR contraction of $\relii$}
As a second possibility to extend $\sltr$, we introduce the $\sltwu$ algebra 
\be\label{sl2xu1v1}
[\bmA,\bmB]=\bmB,\quad [\bmA,\bmC]=-\bmC,\quad [\bmB,\bmC]=2\bmA,\quad [\bmD,{\rm all}]=0,
\ee
where the set $\{\bmA,\bmB,\bmC\}$ spans $\sltw$, whereas $\bmD$ is the Abelian $\uo$ generator. By redefining the generators as
\be
\bal
\bmA&=\frac{1}{4}\left( \bX_{22} -\bX_{11}\right),&
\bmB&=\frac{1}{2}\left( \bX_{12} +\bX\right),\\
\bmC&=\frac{1}{2}\left( \bX_{12} -\bX\right),&
\bmD&=\frac{1}{4}\left( \bX_{22} +\bX_{11}\right),\\
\eal
\ee
the $\sltwu$ algebra \eqref{sl2xu1v1} can be written in a form that is more convenient for our purposes, namely
\be\label{sl2xu1v2}
\bal
\left[\bX,\bX_{ab}\right]=&-\epsilon_{c(a}\bX_{b)c}
,\qquad
\left[\bX_{ab},\bX_{cd}\right]=&\delta_{(a(c}\epsilon_{d)b)} \bX,
\eal
\ee
with $\bX_{ab}=\bX_{ba}$. Thus, we consider the $\relii$ algebra generated by $\bL_A=\{\bL_0,\bL_a\}$, $\bmL_{AB}= \{ \bmL_{a0},\bmL_{ab}\}$, which satisfy the commutation relations \eqref{sltrcomm}, and $\{\bX, \bX_{ab}\}$. The NR contraction follows from the generator redefinition
\be\label{redefrelii}
\bal
\bL_{0}&= \dfrac{1}{2}\bk+c^{2}\bsigma,
&
\bL_a &=c\,\bk_a ,
&
\bX &=\dfrac{1}{2}\bk-c^{2}\bsigma,
\\
\bmL_{ab} &=\dfrac{1}{2}\bmKt_{ab}+c^{2}\bSigmat_{ab},
&
\bmL_{a0} &= c\,\bmKt_a ,
&
\bX_{ab}&=\dfrac{1}{2}\bmKt_{ab}-c^{2}\bSigmat_{ab}.
\eal
\ee
analog to \eqref{redefreli}. Inverting the transformation yields the relations
\be\label{gencont2}
\bal
\bk & =  \bL_{0}+\bX, &
\bk_a & =  \frac{1}{c}\bL_a, &
\bsigma & =  \frac{1}{2c^2}\left(\bL_{0}-\bX\right),
\\
\bmKt_{ab}&= \bmL_{ab}+\bX_{ab},
&
\bmKt_a &=  \frac{1}{c}\bmL_{a0},
&
\bSigmat_{ab}&=  \frac{1}{2c^{2}}\left(\bmL_{ab}-\bX_{ab}\right).
\eal
\ee
which can be used to find the commutation relations in the new basis. In the limit $c\rightarrow\infty$ the non-vanishing commutators read
\be\label{HSNWalgebra2}
\bal
&[\bk,\bk_a ]    =\epsilon_{ab}\bk_b ,&
&[\bk_a ,\bk_b ] = -\epsilon_{ab}\bsigma,&
&[\bk,\bmKt_a ] =\epsilon_{ab}\bmKt_b ,
\\
&[\bk,\bmKt_{ab}] =-\epsilon_{c(a}\bmKt_{b)c},&
&[\bk,\bSigmat_{ab}] =-\epsilon_{c(a}\bSigmat_{b)c},&
&[\bk_a ,\bmKt_b ]=-\epsilon_{a(c}\bSigmat_{b)c},
\\
&[\bk_a ,\bmKt_{bc}]=-\epsilon_{a(b}\bmKt_{c)},&
&[\bsigma,\bmKt_{ab}] =-\epsilon_{c(a}\bSigmat_{b)c},&
&[\bmKt_{ab},\bmKt_{cd}] =\delta_{(a(c}\epsilon_{d)b)} \bk.
\\
&[\bmKt_a ,\bmKt_b ] =\epsilon_{ab}\bsigma,&
&[\bmKt_a ,\bmKt_{bc}] =\delta_{a(b}\epsilon_{c)d}\bk_d,&
&[\bmKt_{ab},\bSigmat_{cd}] =\delta_{(a(c}\epsilon_{d)b)} \bsigma,
\eal
\ee
and define a second higher-spin extension of the Nappi-Witten symmetry\footnote{As in the previous case, the algebra $\hsnwii$ is isomorphic to the homogenous part of the second spin-$3$ extension of the Extended Bargmann algebra in $2+1$ dimensions $\hsbii$ found in \cite{Bergshoeff:2016soe}, which can be shown using the redefinition given in Eq. \eqref{redefSigma}.}, which we dub as $\hsnwii$. Note that it is is possible to define a new generator $\bSigmat_{ab}$
\be\label{redefSigma}
\bSigmat_{ab}\;\longrightarrow\;\bSigmat^{\prime}_{ab}=\bSigmat_{ab} +\delta_{ab} \bSigmat_{cc} ,
\ee
which has the effect of modifying one of the commutation relations in \eqref{HSNWalgebra2} as
\be
[\bk_a ,\bmKt_b ]=-\epsilon_{a(c}\bSigmat_{b)c}
\quad\longrightarrow\quad
[\bk_a ,\bmKt_b ]=-\epsilon_{ac}\bSigmat^{\prime}_{bc},
\ee
while leaving the other commutation relations involving $\bSigmat_{ab}$ invariant. This relation will be useful in Section \ref{Examples}, when connecting some of our results with the ones found in \cite{Bergshoeff:2016soe}.

\section{Three-dimensional higher-spin gravity from spin-$3$ Nappi-Witten algebras}
\label{ExpNW}

In the previous section we obtained two different higher-spin extensions of the $\nw$ algebra, namely $\hsnwi$ and $\hsnwii$, by defining NR contractions of suitable extensions of $\sltr$. This can be summarized as follows:
\be
\bal
&\reli\;&
\underset{c\rightarrow\infty}{\longrightarrow}& \qquad\hsnwi
\\[6pt]
&\relii\;&\underset{c\rightarrow\infty}{\longrightarrow}& \qquad\hsnwii.
\eal
\ee
The next step in our construction is to apply the Lie algebra expansion method based on Abelian semigroups \cite{Izaurieta:2006zz} to these NR symmetries and construct CS actions invariant under them. In this way, we will obtain two different families of NR spin-$3$ algebras.

\subsection{Expansions of higher-spin Nappi-Witten algebras}
In order to define expansions of the spin-$3$ Nappi-Witten algebras \eqref{HSNWalgebra1} and \eqref{HSNWalgebra2}, we introduce an Abelian semigroup $S=\left\{ \lambda _{0},\lambda _{1},\dots
,\lambda _{N}\right\} $, endowed with the associative product \begin{equation}\label{slaw}
\lambda_i \cdot \lambda_j= \lambda_j \cdot \lambda_i = \lambda_{i\diamond j}\,,
\end{equation}
where $\diamond$ defines the semigroup product law. Since the direct product of $S$ with a Lie algebra is again a Lie algebra \cite{Izaurieta:2006zz}, one can define the expanded Nappi-Witten algebra \cite{Penafiel:2019czp}
\be\label{expnw}
S\times \nw=\sp\left\{ S\otimes \{
\bk,\bk_a,\bsigma \} \right\}.
\ee
whose generators are obtained by considering the tensor product of the semigroup elements and the generators of the $\nw$ algebra, i.e.
\be\label{expandedgen} 
\bk^{(i)}_a=\lambda_i \otimes\bk_{a}\,,\quad
\bk^{(i)}=\lambda_i\otimes\bk \,,\quad
\bsigma^{(i)} =\lambda_i \otimes\bsigma\,,
\ee
The commutation relations of the expanded Nappi-Witten algebra are
\be\label{expNW}
[ \bk^{(i)},\bk^{(j)}_a ] = \epsilon_{a}^{\;\;b} \bk^{(i\diamond j)}_b , \quad
[ \bk^{(i)}_a,\bk^{(j)}_b ] = -\epsilon_{ab}\bsigma^{(i\diamond j)} .
\ee
Similarly, one can apply the expansion procedure to the symmetries $\hsnwi$ and $\hsnwii$, which were defined in Eqs. \eqref{HSNWalgebra1} and \eqref{HSNWalgebra2}, as follows: 

\begin{itemize}
\item \textbf{Expansion of $\hsnwi$}: we consider the direct product of $S$ and the generators of \eqref{HSNWalgebra1}, which has the form
\be\label{exp1}
S\times \hsnwi=\sp\left\{ S\otimes \{
\bk,\bk_a,\bsigma ,\bmK _a,\bmK _{ab},\bSigma _a \} \right\},
\ee
with the expanded generators defined in \eqref{expandedgen} plus
\be\label{expandedgenhs1} 
\bmK^{(i)}_{ab}=\lambda_i \otimes\bmK_{ab}\,,\quad
\bmK_a^{(i)}=\lambda_i\otimes\bmK_a \,,\quad
\bSigma_a^{(i)} =\lambda_i \otimes\bSigma_a\,.
\ee
The commutation relations of the $S\times \hsnwi$ algebra are directly obtained from \eqref{HSNWalgebra1} and the semigroup multiplication law \eqref{slaw}. They are given by
\be\label{HSNWalgebra1exp}
\bal
&[\bk^{(i)}, \bk_a^{(j)}]  =  \epsilon_{ab} \bk_b^{(i\diamond j)},&
&[\bk_a^{(i)},\bk_b^{(j)}] = -\epsilon_{ab}\bsigma^{(i\diamond j)},&
&[\bk^{(i)},\bmK_a^{(j)}]  = \epsilon_{ab}\bmK_b^{(i\diamond j)},
\\
&[\bk^{(i)},\bmK_{ab}^{(j)}] =  -\epsilon_{c(a}\bmK_{b)c}^{(i\diamond j)},&
&[\bk^{(i)},\bSigma_a^{(j)}] = \epsilon_{ab}\bSigma_b^{(i\diamond j)},&
&[\bk_a^{(i)},\bmK_b^{(j)}] = -\epsilon_{a(c}\bmK_{b)c}^{(i\diamond j)},
\\
&[\bk_a^{(i)},\bmK_{bc}^{(j)}]  =  -\epsilon_{a(b}\bSigma_{c)}^{(i\diamond j)}, &
&[\bsigma^{(i)},\bmK_a^{(j)}]  =  \epsilon_{ab}\bSigma_b^{(i\diamond j)},&
&[\bmK_a^{(i)},\bmK_b^{(j)}]  =  \epsilon_{ab}\bk^{(i\diamond j)},
\\
&[\bmK_a^{(i)}, \bSigma_b^{(j)}] = \epsilon_{ab}\bsigma^{(i\diamond j)},&
&[\bmK_a^{(i)},\bmK_{bc}^{(j)}]  =\delta_{a(b}\epsilon_{c)d}\bk_d^{(i\diamond j)},&
&[\bmK_{ab}^{(i)},\bmK_{cd}^{(j)}] = \delta_{(a(c}\epsilon_{d)b)}\bsigma^{(i\diamond j)}.
\eal
\ee

\item \textbf{Expansion of $\hsnwii$}: applying the same method to \eqref{HSNWalgebra2} leads to
\be\label{exp2}
S\times \hsnwii=\sp\left\{ S\otimes  \{
\bk,\bk_a,\bsigma ,\bmKt _a, \bmKt _{ab}, \bSigmat _{ab} \} \right\}.
\ee
The expanded generators in this case are given by \eqref{expandedgen} together with
\be\label{expandedgenhs2} 
\bmKt^{(i)}_{ab}=\lambda_i \otimes\bmKt_{ab}\,,\quad
\bmKt_a^{(i)}=\lambda_i\otimes\bmKt_a \,,\quad
\bSigmat_{ab}^{(i)} =\lambda_i \otimes\bSigmat_{ab}\,,
\ee
and the commutation relations of $S\times \hsnwii$ read
\be\label{HSNWalgebra2exp}
\bal
&[\bk^{(i)},\bk_a^{(j)} ]    =\epsilon_{ab}\bk_b^{(i\diamond j)},&
&[\bk_a^{(i)} ,\bk_b^{(j)} ] = -\epsilon_{ab}\bsigma^{(i\diamond j)},&
&[\bk^{(i)},\bmKt_a^{(j)} ] =\epsilon_{ab}\bmKt_b^{(i\diamond j)} ,
\\
&[\bk^{(i)},\bmKt_{ab}^{(j)}] =-\epsilon_{c(a}\bmKt_{b)c}^{(i\diamond j)},&
&[\bk^{(i)},\bSigmat_{ab}^{(j)}] =-\epsilon_{c(a}\bSigmat_{b)c}^{(i\diamond j)},&
&[\bk_a^{(i)} ,\bmKt_b^{(j)} ]=-\epsilon_{a(c}\bSigmat_{b)c}^{(i\diamond j)},
\\
&[\bk_a^{(i)} ,\bmKt_{bc}^{(j)}]=-\epsilon_{a(b}\bmKt_{c)}^{(i\diamond j)},&
&[\bsigma^{(i)},\bmKt_{ab}^{(j)}] =-\epsilon_{c(a}\bSigmat_{b)c}^{(i\diamond j)},&
&[\bmKt_{ab}^{(i)},\bmKt_{cd}^{(j)}] =\delta_{(a(c}\epsilon_{d)b)} \bk^{(i\diamond j)},
\\
&[\bmKt_a^{(i)} ,\bmKt_b^{(j)} ] =\epsilon_{ab}\bsigma^{(i\diamond j)},&
&[\bmKt_a^{(i)} ,\bmKt_{bc}^{(j)}] =\delta_{a(b}\epsilon_{c)d}\bk_d^{(i\diamond j)},&
&[\bmKt_{ab}^{(i)},\bSigmat_{cd}^{(j)}] =\delta_{(a(c}\epsilon_{d)b)} \bsigma^{(i\diamond j)}.
\eal
\ee
We should note that, as done in \eqref{redefSigma} in the case of the $\hsnwii$ algebra, one can redefine the expanded generator
\be\label{redefSigmaexp}
\bSigmat_{ab}^{(i)}\;\longrightarrow\;\bSigmat^{\prime(i)}_{ab}=\bSigmat_{ab}^{(i)} +\delta_{ab} \bSigmat_{cc}^{(i)}, 
\ee
and modify the commutator of the generators $\bk_a^{(i)}$ and $\bmKt_a^{(i)} $ in \eqref{HSNWalgebra2exp} as
\be
[\bk_a^{(i)} ,\bmKt_b^{(j)} ]=-\epsilon_{a(c}\bSigmat_{b)c}^{(i\diamond j}
\quad\longrightarrow\quad
[\bk_a^{(i)} ,\bmKt_b^{(j)} ]=-\epsilon_{ac}\bSigmat^{\prime(i\diamond j)}_{bc},
\ee
while leaving all the remaining commutation relations of $S\times\hsnwii$ invariant.

\end{itemize}
The algebras defined by the commutation relations \eqref{HSNWalgebra1exp} and \eqref{HSNWalgebra2exp} define two different higher-spin extensions of the expanded Nappi-Witten algebra \eqref{expNW}. We know from Ref. \cite{Penafiel:2019czp} that different choices of the semigroup $S$ lead to expanded Nappi-Witten algebras $S\times \nw$ that are isomorphic to NR symmetries in $2+1$ dimensions, such as the Extended Bargmann algebra \cite{Papageorgiou:2009zc}, the Exotic Newton-Hooke algebra \cite{Alvarez:2007fw}, and NR Maxwell \cite{Aviles:2018jzw} algebra. Higher-spin extension of these NR symmetries can be obtained by considering the same semigroups in the expanded algebras \eqref{HSNWalgebra1exp} and \eqref{HSNWalgebra2exp}.

\subsection{Invariant bilinear forms}
\label{IBF}

The In\"on\"u-Wigner contractions introduced in Section \ref{HSNW} can be used to define non-degenerate invariant bilinear forms on the spin-$3$ Nappi-Witten algebras $\hsnwi$ and $\hsnwii$. We begin by showing how this procedure works in the case of the Nappi-Witten algebra \eqref{nwalg}, by looking again at the contraction \eqref{slu} of the $\sltwu$ algebra, whose invariant form is given by
\be\label{itsltwu1}
\left \langle \bL_A , \bL_B \right \rangle =\varepsilon_0\eta _{AB},\qquad
\left \langle \bX, \bX \right \rangle =\varepsilon_1,
\ee
where $\varepsilon_0$ and  $\varepsilon_1$ are non-vanishing constants. Introducing the new constants 
\be\label{redefitsltwu1}
\gamma_0= \frac{1}{c^2}\varepsilon_0,\qquad \gamma_1= \varepsilon_1-\varepsilon_0,
\ee
the generator redefinition \eqref{invcontNW} yields the invariant bilinear form on the $\nw$ algebra \cite{Figueroa-OFarrill:1999cmq} in the
limit $c\rightarrow\infty$, which is non-degenerate for $\gamma_0\neq0$: 
\be\label{itnw}
\left \langle \bk,\bsigma \right \rangle = - \gamma_0 ,\quad \left \langle \bk_a,
\bk_b \right \rangle = \gamma_0\delta_{ab} ,\quad \left \langle \bk,\bk \right
\rangle = \gamma_1.
\ee

To define invariant tensors on $\hsnwi$ and $\hsnwii$, we follow exactly the same procedure. The non-degenerate invariant bilinear form of the $\sltr$ is \cite{Campoleoni:2010zq}
\be\label{itsltr}
\left \langle \bL_A , \bL_B\right \rangle =\varepsilon_0\eta _{AB},\qquad
\left \langle \bmL_{AB} \bmL_{CD}\right \rangle
=\varepsilon_0 \left( \eta _{A(C}\eta _{D)B}-\frac{2}{3}\eta
_{AB}\eta _{CD} \right),
\ee
which, after splitting temporal and spatial components, takes the form
\be\label{invtsl3}
\left\langle \bL_{0}\bL_{0}\right \rangle
=-\varepsilon_0,\quad 
\left\langle \bL_{a}\bL_{b}\right \rangle=- 
\left\langle \bmL_{a0}\bmL_{b0}\right \rangle
=\varepsilon_0 \delta _{ab}
,\quad 
\left \langle
\bmL_{ab}\bmL_{cd}\right \rangle =\varepsilon_0 \left( \delta _{a(c}\delta _{d)b}-\frac{2}{3}\delta _{ab}\delta _{cd} \right). 
\ee
The extensions we are interested in require the introduction of invariant tensors on $\sotr$ and $\sltwu$ before taking the NR limit.
\begin{itemize}
\item The invariant bilinear form on $\sotr$ can be written as
\begin{equation}
\left \langle \bX,\bX\right \rangle =\varepsilon_1, \qquad \left \langle
\bX_a,\bX_b\right \rangle =\varepsilon_1 \delta _{ab}
\end{equation}
Putting this together with \eqref{invtsl3} defines the invariant form on $\reli$. Using \eqref{redefitsltwu1}, \eqref{gencont1} and taking the limit $c\rightarrow\infty$, one finds that the invariant tensor for the generators of the $\hsnwi$ algebra \eqref{HSNWalgebra1} is given by \eqref{itnw} plus
\be\label{invtcont1}
\left\langle \bmK_a  , \bSigma_b \right\rangle= -\gamma_0 \delta_{ab} ,\quad
\left\langle \bmK_{ab} ,  \bmK_{cd} \right\rangle= \gamma_0 \left( \delta _{a(c}\delta _{d)b}-\frac{2}{3}\delta _{ab}\delta _{cd} \right)  ,\quad 
\left\langle \bmK_a  ,\bmK_b \right\rangle= \gamma_1  \delta_{ab},
\ee
which is non-degenerate provided $\gamma_0\neq0$.

\item The invariant bilinear form on the $\sltwu$ algebra \eqref{sl2xu1v1} is given by 
\be
\left\langle \bmA,\bmA\right\rangle=\frac{1}{4}\sigma_0,
\qquad
\left\langle \bmB,\bmC\right\rangle= \frac{1}{2}\sigma_0,
\qquad
\left\langle \bmD,\bmD\right\rangle= \frac{1}{4}\sigma_1,
\ee
which, by redefining $\varepsilon_1=-\sigma_0$, $\varepsilon_2=-\sigma_0+\sigma_1$, can be rewritten in terms of the generators $\bX$ and $\bX_{ab}$ of \eqref{sl2xu1v2} as
\be\label{invtsl2u1}
 \left\langle \bX,  \bX \right\rangle= \varepsilon_1
,\quad 
\left\langle \bX_{ab} ,  \bX_{cd} \right\rangle= -\varepsilon_1  \delta _{a(c}\delta _{d)b}+\varepsilon_ 2 \delta _{ab}\delta _{cd}.
\ee
Thus, we will consider the invariant tensor on $\relii$ defined by \eqref{invtsl3} and \eqref{invtsl2u1}. Using the redefinitions \eqref{gencont2}, \eqref{redefitsltwu1} and $\gamma_2=\varepsilon_2-\frac23 \varepsilon_0$, the limit $c\rightarrow\infty$ yields an invariant bilinear form on the $\hsnwii$ algebra \eqref{HSNWalgebra2}, given by \eqref{itnw} plus the relations
\be\label{invtcont2}
\bal
&\left\langle \bmKt_{ab} , \bSigmat_{cd}\right\rangle= \gamma_0 \left( \delta _{a(c}\delta _{d)b}-\frac{2}{3}\delta _{ab}\delta _{cd}\right)
,\\
&\left\langle \bmKt_{ab} ,  \bmKt_{cd} \right\rangle= -\gamma_1 \delta _{a(c}\delta _{d)b}+\gamma_2 \delta _{ab}\delta _{cd}  ,\quad 
\left\langle \bmKt_a , \bmKt_b \right\rangle=- \gamma_0  \delta_{ab}.
\eal
\ee
Again, this invariant tensor has non-vanishing determinant as long as $\gamma_0\neq0$. If, instead of $\bSigmat_{ab}$, one uses the generator $\bSigmat^{\prime}_{ab}$ defined in \eqref{redefSigma}, the first relation in \eqref{invtcont2} is replaced by
\be\label{invtcont2prime}
\left\langle \bmKt_{ab} , \bSigmat^{\prime}_{cd}\right\rangle= \gamma_0 \delta _{a(c}\delta _{d)b}.
\ee 

\end{itemize}
The next step is to use these results to define non-degenerate invariant bilinear forms on the expanded algebras $S\times\hsnwi$ and  $S\times\hsnwii$ given by \eqref{HSNWalgebra1exp} and \eqref{HSNWalgebra2exp}, respectively. Following \cite{Boulanger:2002bt,Izaurieta:2006zz}, the procedure is summarized as follows: given a set of Lie algebra generators $T_\alpha$ together with an invertible invariant bilinear form $\langle T_\alpha T_\beta \rangle = g_{\alpha\beta}$, the expanded Lie algebra spanned by the generators $T^{(i)}_\alpha= \lambda_i \times T_\alpha$ admits the non-degenerate invariant tensor  $\langle T^{(i)}_\alpha T^{(j)}_\beta \rangle = h_{ij} g_{\alpha\beta}$, where the components of $h_{ij}=\langle \lambda_i \lambda_j \rangle$ include a set of arbitrary constants. When considering the $S\times \nw$ algebra. \eqref{expNW}, we find \cite{Penafiel:2019czp}
\be\label{itnwexp}
\left \langle \bk^{(i)},\bsigma^{(j)} \right \rangle = - \mu_{i\diamond j} ,
\quad \left \langle \bk_a^{(i)},
\bk_b^{(j)} \right \rangle = \mu_{i\diamond j}\delta_{ab} ,
\quad \left \langle \bk^{(i)},\bk^{(j)} \right
\rangle = \nu_{i\diamond j},
\ee
where $\mu_{i}$ and $\nu_{i}$ stand for arbitrary constants. The procedure can be generalized straightforwardly to the case of the $S\times \hsnwi$ and $S\times \hsnwii$ algebras:
\begin{itemize}

\item The generators of $S\times \hsnwi$ are given by \eqref{expandedgen} and \eqref{expandedgenhs1}. Thus, we can use the results \eqref{itnw} and \eqref{invtcont1} to obtain the invariant tensor \eqref{itnwexp} together with the extension
\be\label{itexp1}
\bal
&\left\langle \bmK_a^{(i)}  , \bSigma_b^{(j)} \right\rangle= -\mu_{i\diamond j} \delta_{ab} ,
\\
&\left\langle \bmK_{ab}^{(i)} ,  \bmK_{cd}^{(j)} \right\rangle= \mu_{i\diamond j} \left( \delta _{a(c}\delta _{d)b}-\frac{2}{3}\delta _{ab}\delta _{cd} \right)  ,
\\
&\left\langle \bmK_a^{(i)}  ,\bmK_b^{(j)} \right\rangle= \nu_{i\diamond j}  \delta_{ab}.
\eal
\ee

\item For $S\times \hsnwii$ we use the invariant form defined by \eqref{itnw} and \eqref{invtcont2} to compute an invariant bilinear form for the expanded generators \eqref{expandedgen} and \eqref{expandedgenhs2}. After the expansion, the invariant bilinear form \eqref{itnwexp} is extended to include the relations
\be\label{itexp2}
\bal
&\left\langle \bmKt_{ab}^{(i)} , \bSigmat_{cd}^{(j)}\right\rangle= \mu_{i\diamond j} \left( \delta _{a(c}\delta _{d)b}-\frac{2}{3}\delta _{ab}\delta _{cd}\right)
,\\
&\left\langle \bmKt_{ab}^{(i)} ,  \bmKt_{cd}^{(j)} \right\rangle= -\nu_{i\diamond j}  \delta _{a(c}\delta _{d)b}+\rho_{i\diamond j}  \delta _{ab}\delta _{cd} ,
\\
&\left\langle \bmKt_a^{(i)} , \bmKt_b^{(j)} \right\rangle=- \mu_{i\diamond j}  \delta_{ab}.
\eal
\ee
In complete analogy to \eqref{invtcont2prime}, one can simplify the first relation in \eqref{itexp2} to
\be\label{invtcont2primeexp}
\left\langle \bmKt_{ab}^{(i)} , \bSigmat^{\prime(j)}_{cd}\right\rangle= \mu_{i\diamond j} \delta _{a(c}\delta _{d)b}
\ee 
by redefining the expanded generator $\bSigmat_{ab}^{(i)}$ as in \eqref{redefSigmaexp}.

\end{itemize}

\subsection{Chern-Simons actions}
CS forms provide actions for topological field theories in $2+1$ dimensions. In the case of NR symmetries, CS theories have been used to define three-dimensional versions of Newton-Cartan gravity and extensions thereof \cite{Papageorgiou:2009zc,Bergshoeff:2016lwr,Hartong:2016yrf,Matulich:2019cdo}. In our case, after having defined an non-degenerate invariant bilinear form on the expanded spin-$3$ algebras $S\times \hsnwi$ and $S\times \hsnwii$, it is natural to construct topological higher-spin gravity actions by considering the integral of a CS three-form over a three-dimensional manifold
\be\label{CSform}
I=\int_{M_3} \mathcal L,\qquad \mathcal L= \left\langle \bA\wed\left( \bF -\frac{1}{6} \left[\bA, \bA\right] \right)\right\rangle,
\ee
where $\bA$ is a connection one-form taking values on the Lie algebra of interest and $\bF=\ed \bA + \bA^2$ is the corresponding curvature two-form\footnote{Throughout the analysis, wedge product between forms is understood.}. Up to boundary terms, the theory is invariant under infinitesimal gauge transformations $\delta \bA= d\lambda + [\bA,\lambda]$, with $\lambda$ a local parameter taking values on the Lie algebra. As done in previous sections, we start by briefly reviewing the pure Nappi-Witten case, without higher-spin fields. The gauge connection one-form on the $\nw$ algebra \eqref{nwalg} and its curvature read
\be\label{NWAF}
\bA^{\nw}= \theta^a \bk_a + \alpha \bk +\beta \bsigma,\quad \bF^{\nw}= \bar R^a \bk_a + \bar R \bk + \bar F \bsigma,
\ee
where the curvature components are given by
\be\label{NWcurv}
\bar R^a=\ed\theta^a+\epsilon^{ab}\,\theta^b\wed\alpha,\quad
\bar R= \ed \alpha,\quad
\bar F= \ed \beta -\frac12\epsilon_{ab}\,\theta^a\wed\theta^b.
\ee
Using the invariant form \eqref{itnw}, the CS action invariant under the $\nw$ algebra takes the form
\be\label{CSNW}
\mathcal{L}_{\nw}=\gamma_0\left(  \theta^a \wed \bar R^a-\alpha\wed \ed\beta-\beta\wed \ed\alpha\right) +\gamma_1\,\alpha \wed \ed\alpha .
\ee
The next step is to generalize this action to the case of the expanded Nappi-Wiiten algebra $S\times\nw$ given in Eq. \eqref{expNW} to introduce the semigroup elements \eqref{slaw} through the expanded generators \eqref{expandedgen}. The connection and the curvature for $S\times\nw$ are
\be\label{expNWAF}
\bA^{S\times\nw}= \theta^a_{(i)} \bk_a^{(i)} + \alpha_{(i)} \bk^{(i)} +\beta_{(i)} \bsigma^{(i)},\quad \bF^{S\times\nw}= \bar R^a_{(i)} \bk_a^{(i)} + \bar R_{(i)} \bk^{(i)} + \bar F_{(i)} \bsigma^{(i)}\\
\ee
where we have defined the expanded curvature components
\be\label{expNWFcomp}
\bar R^a_{(i)}=\ed\theta^a_{(i)}+\epsilon^{ab}\,\theta^b_{(j)}\wed\alpha_{(k)}\,\delta^{j\diamond k}_{i},\quad
\bar R_{(i)}= \ed \alpha_{(i)},\quad
\bar F_{(i)}= \ed \beta_{(i)} -\frac12\epsilon_{ab}\,\theta^a_{(j)}\wed\theta^b_{(k)}\,\delta^{j\diamond k}_{i},
\ee
and now the Einstein summation convention also holds for repeated semigroup indices. After implementing the expansion mechanism, the CS action \eqref{CSNW} is extended to \cite{Penafiel:2019czp}
\be\label{CSNWexp}
\mathcal{L}_{S\times\nw}=\mu_{i\diamond j}\left(  \theta^a_{(i)} \wed \bar R^a_{(j)}-\alpha_{(i)}\wed \ed\beta_{(j)}-\beta_{(i)}\wed \ed\alpha_{(j)}\right) +\nu_{i\diamond j}\,\alpha_{(i)} \wed \ed\alpha_{(j)} \,,
\ee
where the constants $\gamma_0$ and $\gamma_1$ are replaced by the corresponding ones entering in the expanded invariant tensor \eqref{itnwexp}, $\mu_{i}$ and $\nu_{i}$.
\\

\noindent\textbf{Spin-$3$ generalization}: The procedure previously presented can be further generalized to define CS actions invariant under the spin-$3$ extension of the Nappi-Witten algebras presented in Section \ref{HSNW}. In order to simplify the discussion, we make a parallel construction for both symmetries $\hsnwi$ and $\hsnwii$. We thus consider two connection one-forms, each one taking values on one of the spin-$3$ extensions of the Nappi-Witten algebra
\be\label{Aexpnw}
\bal
\bA^{\hsnwi}&= \bA^{\nw}+ \Theta^{ab}\bmK_{ab}+A^a \bmK_a +B^a \bSigma_a,
\\[6pt]
\bA^{\hsnwii}&=\bA^{\nw} + \Theta^a\bmKt_a+A^{ab} \bmKt_{ab} +B ^{ab} \bSigmat_{ab},
\eal
\ee
with associated curvatures
\be
\bal
\bF^{\hsnwi}&= R^a \bk_a+R \bk+ F \bsigma
+ \mathcal R^{ab} \bmK_{ab}
+\mathcal R^a \bmK_a +\mathcal F^a \bSigma_a,
\\[6pt]
\bF^{\hsnwii}&= \tilde{R}^a \bk_a+\tilde{R} \bk+ \tilde{F} \bsigma
+\tilde{\mathcal R}^a \bmKt_a  + \tilde{\mathcal R}^{ab} \bmKt_{ab}+\tilde{\mathcal F}^{ab} \bSigmat_{ab}.
\eal
\ee
Notice that, since the higher-spin generators in each expanded algebra have different index structure, the corresponding higher-spin gauge field content is different in each case as well. This manifests in the form of the components of the curvatures. Indeed, for $\bF^{\hsnwi}$ we find
\be
\bal
&R^a= \bar R^a
+ 2\epsilon^{ab} \Theta^{bc}\wed A^c,
\\
&R=\bar R +\frac 12 \epsilon_{ab} A^a\wed A^b,
\\
&F=\bar F +\epsilon_{ab}\left(A^a B^b-2\Theta^{ac}\wed\Theta^{cb}\right),
\\
&\mathcal R^{ab} = \ed \Theta^{ab}
+ 
\epsilon^{c(a} \left(\frac12
 A^{b)}\wed \theta^c
-\Theta^{b)c}\wed\alpha
\right)- \delta^{ab}\epsilon_{cd}\theta^c \wed A^d,
\\
&\mathcal R^a = \ed A^a + \epsilon^{ab} A^b\wed \alpha,
\\
&\mathcal F^a = \ed B^a +  \epsilon^{ab} \left(B^b\wed  
\alpha+ A^b\wed \beta\right)-2\epsilon_{bc}\Theta^{ab}\wed\theta^c,
\eal
\ee
whereas the components of the curvature $\bF^{\hsnwii}$ are
\be
\bal
&\tilde{R}^a= \bar R^a
+2\epsilon^{ab} A^{bc}\wed\Theta^c,
\\
&\tilde{R}=\bar R -2\epsilon_{ab} A^{ac}\wed A^{cb},
\\
&\tilde{F}=\bar F +\epsilon_{ab}\left(\frac12\Theta^a\wed\Theta^b-4A^{ac} B^{cb}\right),
\\
&\tilde{\mathcal R}^a = \ed \Theta^a
+\epsilon^{ab} \Theta^b\wed \alpha -2\epsilon_{bc} A^{ab} \wed \theta^c,
\\
&\tilde{\mathcal R}^{ab} = \ed A^{ab}
-
\epsilon^{c(a}  A^{b)c}\wed\alpha,
\\
&\tilde{\mathcal F}^{ab} = \ed B^{ab}  +\epsilon^{c(a}\left( \frac12
 \Theta^{b)}\wed \theta^c -A^{b)c}\wed\beta -B^{b)c}\wed\alpha\right)
- \delta^{ab}\epsilon_{cd}\theta^c \wed \Theta^d.
\eal
\ee
Naturally, both curvatures possess the same terms $\bar R^a$, $\bar R$ and $\bar F$, defined in Eq. \eqref{NWcurv}, associated to the Nappi-Witten subalgebra that both spin-$3$ symmetries share. The CS forms invariant under $\hsnwi$ and $\hsnwii$ can be computed by evaluating \eqref{CSform} for the corresponding connections and curvatures and later using the invariant tensors given in \eqref{invtcont1} and \eqref{invtcont2}. This yields two inequivalent spin-$3$ extensions of the Nappi-Witten CS action \eqref{CSNW}, 
\be\label{CShs}
\bal
\mathcal L_{\hsnwi}&= \mathcal L_{\nw}
+ \gamma_0\left(2\Theta^{ab} \wed \mathcal R^{ab}-\frac23 \Theta^{aa} \wed  \mathcal R^{bb} -A^a \wed \mathcal F^a-B^a \wed  \mathcal R^a \right)+\gamma_1 A^a \wed \mathcal R^a,
\\[6pt]
\mathcal L_{\hsnwii}&= \mathcal L_{\nw}
+ \gamma_0\left(2A^{ab} \wed  \tilde{\mathcal F}^{ab}+2B^{ab}  \wed \tilde{\mathcal R}^{ab}-
\frac23 A^{aa}  \wed \tilde{\mathcal F}^{bb}-\frac23 B^{aa} \wed  \tilde{\mathcal R}^{bb}
-\Theta^a \wed  \tilde{\mathcal R}^a\right)
\\&- 2\gamma_1 A^{ab} \wed \tilde{\mathcal R}^{ab}
+\gamma_2 A^{aa} \wed \tilde{\mathcal R}^{bb}.
\eal
\ee

\noindent\textbf{Expansion}: Now we turn our attention to the expanded algebras $S\times\hsnwi$ and $S\times\hsnwii$. In order to obtain the corresponding expanded CS actions, we follow the same procedure as in the $\nw$ case. Therefore, we define two expanded connections
\be\label{expAs}
\bal
\bA^{S\times\hsnwi}&=\bA^{S\times\nw}+ \Theta_{(i)}^{ab}\bmK_{ab}^{(i)}+A^a_{(i)}\bmK_a^{(i)}+B_{(i)}^a \bSigma_a^{(i)},
\\[6pt]
\bA^{S\times\hsnwii}&=\bA^{S\times\nw}+ \Theta_{(i)}^a\bmKt_a^{(i)}+A^{ab}_{(i)}\bmKt_{ab}^{(i)}+B_{(i)}^{ab} \bSigmat_{ab}^{(i)},
\eal
\ee
where $\bA^{S\times\nw}$ is the connection that takes values on the Expanded Nappi-Witten algebra \eqref{expNWAF}. The corresponding curvatures have the form
\be\label{expFs}
\bal
\bF^{S\times\hsnwi}&= R^a_{(i)} \bk_a^{(i)}+R_{(i)} \bk^{(i)}+ F_{(i)} \bsigma^{(i)}
+ \mathcal R^{ab}_{(i)} \bmK_{ab}^{(i)}
+\mathcal R^a_{(i)} \bmK_a^{(i)} +\mathcal F^a_{(i)} \bSigma_a^{(i)},
\\[6pt]
\bF^{S\times\hsnwii}&= \tilde{R}^a_{(i)} \bk_a^{(i)}+\tilde{R}_{(i)} \bk^{(i)}+ \tilde{F}_{(i)} \bsigma^{(i)}
+\tilde{\mathcal R}^a_{(i)} \bmKt_a^{(i)}  + \tilde{\mathcal R}^{ab}_{(i)} \bmKt_{ab}^{(i)}+\tilde{\mathcal F}^{ab}_{(i)} \bSigmat_{ab}^{(i)},
\eal
\ee
where the components of the curvature two-form on $S\times\hsnwi$ are given by
\be\label{expNWFcomp1}
\bal
&R^a_{(i)}= \bar R^a_{(i)}
+ 2\epsilon^{ab} \Theta^{bc}_{(j)} \wed A^c_{(k)} 
\,\delta^{j\diamond k}_{i},
\\
&R_{(i)}=\bar R_{(i)} +\frac 12 \epsilon_{ab} A^a_{(j)} \wed A^b_{(k)} 
\,\delta^{j\diamond k}_{i},
\\
&F_{(i)}=\bar F_{(i)} +\epsilon_{ab}\left(A^a_{(j)}  \wed B^b_{(k)} -2\Theta^{ac}_{(j)} \wed\Theta^{cb}_{(k)} \right)
\delta^{j\diamond k}_{i},
\\
&\mathcal R^{ab}_{(i)} = \ed \Theta^{ab}_{(i)}
+ 
\epsilon^{c(a} \left(\frac12
 A^{b)}_{(j)} \wed \theta^c_{(k)} 
-\Theta^{b)c}_{(j)} \wed\alpha_{(k)} 
\right)
\delta^{j\diamond k}_{i}
- \delta^{ab}\epsilon_{cd}\theta^c_{(j)} \wed A^d_{(k)} 
\,\delta^{j\diamond k}_{i},
\\
&\mathcal R^a_{(i)} = \ed A^a_{(i)} + \epsilon^{ab} A^b_{(j)} \wed \alpha_{(k)} 
\,\delta^{j\diamond k}_{i},
\\
&\mathcal F^a_{(i)} = \ed B^a_{(i)} +  \epsilon^{ab} \left(B^b_{(j)} \wed  
\alpha_{(k)} + A^b_{(j)} \wed \beta_{(k)} \right)\delta^{j\diamond k}_{i}
-2\epsilon_{bc}
\Theta^{ab}_{(j)} \wed\theta^c_{(k)} 
\,\delta^{j\diamond k}_{i},
\eal
\ee
while for the components of the curvature on the second expanded algebra $\bF^{S\times\hsnwii}$ we get
\be\label{expNWFcomp2}
\bal
&\tilde{R}^a_{(i)}= \bar R^a_{(i)}
+2\epsilon^{ab} A^{bc}_{(j)} \wed\Theta^c_{(k)} 
\,\delta^{j\diamond k}_{i},
\\
&\tilde{R}_{(i)}=\bar R_{(i)} -2\epsilon_{ab} A^{ac}_{(j)} \wed A^{cb}_{(k)} 
\,\delta^{j\diamond k}_{i},
\\
&\tilde{F}_{(i)}=\bar F_{(i)} +\epsilon_{ab}\left(\frac12\Theta^a_{(j)} \wed\Theta^b_{(k)} -4A^{ac}_{(j)}  B^{cb}_{(k)} \right)
\delta^{j\diamond k}_{i},
\\
&\tilde{\mathcal R}^a_{(i)} = \ed \Theta^a_{(i)}
+\epsilon^{ab} \Theta^b_{(j)} \wed \alpha_{(k)}  \,\delta^{j\diamond k}_{i}
-2\epsilon_{bc} A^{ab}_{(j)}  \wed \theta^c_{(k)} 
\,\delta^{j\diamond k}_{i},
\\
&\tilde{\mathcal R}^{ab}_{(i)} = \ed A^{ab}_{(i)}
-
\epsilon^{c(a}  A^{b)c}_{(j)} \wed\alpha_{(k)} 
\,\delta^{j\diamond k}_{i},
\\
&\tilde{\mathcal F}^{ab}_{(i)} = \ed B^{ab}_{(i)}  +\epsilon^{c(a}\left( \frac12
 \Theta^{b)}_{(j)} \wed \theta^c_{(k)}  -A^{b)c}_{(j)} \wed\beta_{(k)}  -B^{b)c}_{(j)} \wed\alpha_{(k)} \right)\delta^{j\diamond k}_{i}
- \delta^{ab}\epsilon_{cd}\theta^c_{(j)}  \wed 
\Theta^d_{(k)} 
\,\delta^{j\diamond k}_{i}.
\eal
\ee
Note that, as before, the components of the curvature two-forms along the expanded generators $\bk_a^{(i)}$, $\bk^{(i)}$ and $\bsigma^{(i)}$ contain the terms given in \eqref{expNWFcomp} plus contributions of the higher-spin fields that are different in each case due to the structure of the commutation relations involving the higher-spin generators. The expanded CS actions can be obtained in a straightforward way by applying the same procedure used for the action \eqref{CSNW} to obtain \eqref{CSNWexp}: i) take the actions given in \eqref{CShs} and replace $\gamma_0\rightarrow\mu_{i\diamond j}$, $\gamma_1\rightarrow\nu_{i\diamond j}$, $\gamma_2\rightarrow\rho_{i\diamond j}$, ii) insert the semigroup indices $i$ and $j$ in the gauge fields and curvatures, contracted with the invariant bilinear form constants in the proper way, iii) take care of using Kronecker deltas of the form $\delta^{j\diamond k}_{i}$ in the terms that come from the commutator $[\bA,\bA]$. We find the resulting actions 
\be\label{hsCS1}
\bal
\mathcal L_{S\times\hsnwi}=\; \mathcal L_{S\times\nw}
+ \mu_{i\diamond j}\left(2\Theta^{ab}_{(i)} \wed \mathcal R^{ab}_{(j)}-\dfrac23 \Theta^{aa}_{(i)} \wed \mathcal R^{bb}_{(j)} -A^a_{(i)} \wed \mathcal F^a_{(j)}-B^a_{(i)} \wed \mathcal R^a_{(j)} \right)
+ \nu_{i\diamond j} \;A^a_{(i)} \wed  \mathcal R^a_{(j)},
\eal
\ee
\be\label{hsCS2}
\bal
\mathcal L_{S\times\hsnwii}&=\; \mathcal L_{S\times\nw}
+  \mu_{i\diamond j} \left(2A^{ab}_{(i)}  \wed \tilde{\mathcal F}^{ab}_{(j)}+2B^{ab}_{(i)}  \wed \tilde{\mathcal R}^{ab}_{(j)}-
\dfrac23 A^{aa}_{(i)} \wed  \tilde{\mathcal F}^{bb}_{(j)}-\dfrac23 B^{aa}_{(i)} \wed  \tilde{\mathcal R}^{bb}_{(j)}
-\Theta^a_{(i)}  \wed \tilde{\mathcal R}^a_{(j)}\right)
\\[10pt]
&\hskip.5truecm- 2 \nu_{i\diamond j} \;A^{ab}_{(i)} \wed \tilde{\mathcal R}^{ab}_{(j)}
+ \rho_{i\diamond j} \;A^{aa}_{(i)} \wed \tilde{\mathcal R}^{bb}_{(j)},
\eal
\ee
where one should remark that, in the same way as for the spatial indices, there is an implicit sum over repeated semigroup indices $i$ and $j$. Naturally, these actions can alternatively be obtained by evaluating \eqref{CSform} using the connections and curvatures given in \eqref{expAs} and \eqref{expFs} together with the invariant bilinear forms given in Eqs. \eqref{itnwexp}-\eqref{itexp2}. As we have seen, however, once the CS action invariant under some to-be-expanded algebra is known, the expansion method provides a quite efficient and direct way to obtain the CS actions invariant under the corresponding expanded algebra.

\subsection{Non-relativistic limit of expanded $\sltr$ algebras}
\label{NRlim}
It is natural to expect that the NR spin-$3$ symmetries $S\times \hsnwi$ and $S\times \hsnwii$, respectively defined by \eqref{HSNWalgebra1exp} and \eqref{HSNWalgebra2exp}, are not only expansions of a NR algebra, but also NR contractions of some relativistic symmetry. In \cite{Caroca:2017izc}, spin-$3$ extensions of relativistic algebras in $2+1$ dimensions were obtained as expansions of $\sltr$. Combining the notation introduced in \cite{Caroca:2017izc} and the definition \eqref{slaw}, an expanded algebra of the form $S\times \sltr$ has the commutation relations 
\be\label{sl3exp}
\left[\bL_A^{(i)},\bL_B^{(j)}\right] =\epsilon_{AB}^{\quad C}\bL_{C}^{(i\diamond j)}\,\quad
\left[\bL_A^{(i)},\bmL_{BC}^{(j)}\right] =\epsilon_{A(B}^{\quad \;D}\bmL_{C)D}^{(i\diamond j)}\,\quad
\left[\bmL_{AB}^{(i)},\bmL_{CD}^{(j)}\right]  =- \eta_{(A(C}\epsilon_{D)B)}^{\quad \;E} \bL_E^{(i\diamond j)}.
\ee
Depending on the chosen semigroup, this structure can reproduce spin-$3$ extensions of the Poincar\'e, AdS, Maxwell and AdS-Lorentz algebras (see \eqref{SMSEhsrel}). The method was also extended to define spin-$3$ extensions of the Generalized Poincar\'e  (also known as $\mathfrak B_m$) \cite{Izaurieta:2009hz} and Generalized AdS (also known as $\mathfrak C_m$) \cite{Concha:2016kdz} algebras in three space-time dimensions.

It is interesting to point out that the contractions introduced in Section \ref{NRcontractions} reveal the way to extend \eqref{sl3exp} in order to define expanded relativistic symmetries whose NR limits are precisely the expanded algebras $S\times \hsnwi$ and $S\times \hsnwii$.

\begin{itemize}
\item By expanding the $\sotr$ algebra \eqref{so3alg} we find
\be\label{so3exp}
\bal
\left[\bX^{(i)},\bX_a^{(j)}\right]=&\epsilon_{ab}\bX_b^{(i\diamond j)}
,\qquad
\left[\bX_a^{(i)}, \bX_b^{(j)}\right]=&\epsilon_{ab} \bX^{(i\diamond j)},
\eal
\ee
which can be put together with \eqref{sl3exp} to define the contraction 
\be\label{expreli}
\left[S\times \sltr\right] \oplus \left[S\times \sotr\right]
\quad
\underset{c\rightarrow\infty}{\longrightarrow} \qquad S\times\hsnwi
\ee
by means of the generator redefinition
\be\label{redefsl3iexp}
\bal
&\bL_{0}^{(i)}= \dfrac{1}{2}\bk^{(i)}+c^{2}\bsigma^{(i)},
&\quad&
\bL_a^{(i)}=c\, \bk_a^{(i)},
&\quad&
\bX^{(i)} =\dfrac{1}{2}\bk-c^{2}\bsigma^{(i)},
\\
&\bmL_{a0}^{(i)} =\dfrac{1}{2}\bmK_a^{(i)}+c^{2}\bSigma_a^{(i)},
&\quad&
\bmL_{ab}^{(i)} = c\,\bmK_{ab}^{(i)},
&\quad&
\bX_a^{(i)}=\dfrac{1}{2}\bmK_a^{(i)}-c^{2}\bSigma_a^{(i)}.
\eal
\ee
In the limit $c\rightarrow \infty$ this contraction yields exactly the commutation relations \eqref{HSNWalgebra1exp} of the algebra $S\times \hsnwi$. What we have done here, oppositely to the \emph{first contract and later expand} procedure given in the previous sections, is to expand the relativistic relations \eqref{sl3exp} and \eqref{so3exp} before applying the contraction.

\item The expansion and the contraction also commute if we try to obtain the $S\times \hsnwii$ algebra. One can use \eqref{sl3exp} together with the expansion of the $\sltwu$ algebra \eqref{sl2xu1v2}
\be\label{sl2xu1exp}
\bal
\left[\bX^{(i)},\bX_{ab}^{(j)}\right]=&-\epsilon_{c(a}\bX_{b)c}^{(i\diamond j)}
,\qquad
\left[\bX_{ab}^{(i)},\bX_{cd}^{(j)}\right]=&\delta_{(a(c}\epsilon_{d)b)} \bX^{(i\diamond j)}
\eal
\ee
to define a NR limit 
\be\label{exprelii}
\left[S\times \sltr\right] \oplus \left[S\times \sltw\right]\oplus \left[S\times \uo \right]
\quad
\underset{c\rightarrow\infty}{\longrightarrow} \qquad S\times\hsnwii.
\ee
Thus, by defining relations analog to \eqref{redefrelii},
\be\label{redefsl3iiexp}
\bal
&\bL_{0}^{(i)}= \dfrac{1}{2}\bk^{(i)}+c^{2}\bsigma^{(i)},
&\quad&
\bL_a^{(i)}=c\, \bk_a^{(i)},
&\quad&
\bX^{(i)} =\dfrac{1}{2}\bk-c^{2}\bsigma^{(i)},
\\
&\bmL_{ab}^{(i)} =\dfrac{1}{2}\bmKt_{ab}^{(i)}+c^{2}\bSigmat_{ab}^{(i)},
&\quad&
\bmL_{a0}^{(i)} = c\,\bmKt_a^{(i)} ,
&\quad&
\bX_{ab}^{(i)}=\dfrac{1}{2}\bmKt_{ab}^{(i)}-c^{2}\bSigmat_{ab}^{(i)},
\eal
\ee
the limit $c\rightarrow\infty$ leads to the $S\times\hsnwii$ commutation relations \eqref{HSNWalgebra2exp}, as expected. This construction provides an alternative method to generate the expanded algebras that will be presented in the next section, which is based on the fact that (for the type of expansions here considered) the expansion and contraction procedures commute. The construction of the invariant bilinear forms \eqref{itnwexp}-\eqref{itexp2}, as well as the CS actions \eqref{hsCS1} and \eqref{hsCS2} that are invariant under these expanded NR symmetries can be obtained by means of this alternative method as well.

\end{itemize}

\subsection{Flat limit}
\label{flatlimit}
In the following section, we will use two particular semigroups to construct higher-spin NR symmetries, namely, the $S_E^{(n)}$ and $S_M^{(n)}$ semigroups \cite{Izaurieta:2006zz}, defined by the following multiplication rules \eqref{slaw}.
\be\label{sen}
\hskip-2.3truecm S_E^{(n)}=\left\{\lambda_0,\lambda_1,\dots,\lambda_{n+1} \right\},\qquad i\diamond j =\left\{
\begin{array}{lll}
i +j  \quad & \mathrm{if}\quad i+j \leq n\,,
&  \\
n+1 \quad & \mathrm{if}\quad i+j >n\,. &
\end{array}%
\right.
\ee
\be\label{smn}
S_M^{(n)}=\{\lambda_0 ,\lambda_1,\dots,\lambda_n\},\qquad
\hskip.5truecm
i\diamond j=\left\{
\begin{array}{lll}
i+j \,\,\,\, & \mathrm{if}\,\,\,\,\ i +j \leq n\,,
&  \\
i +j -2\left[ \frac{n+1}{2}\right]\,\,\, & \mathrm{if}%
\,\,\,\,i+j>n \,.&
\end{array}%
\right. 
\ee
Note that $S_E^{(n)}$ has an extra \emph{zeroth} element $\lambda_{n+1}$ ($\lambda_i\cdot\lambda_{n+1}=\lambda_{n+1}\;\forall\;\lambda_i\in S_E^{(n)}$), not present in $S_M^{(n)}$. When defining an expanded Lie algebra, the fact that the semigroup $S_E^{(n)}$ has a zero requires to impose the \emph{reduction condition} $T^{(n+1)}_\alpha\equiv0$ on the expanded generators, together with $ \mu_{n+1}\equiv0$ and  $\nu_{n+1}\equiv0$ (and additionally $\rho_{n+1}\equiv0$ for $S\times\hsnwii$) on the invariant tensor constants introduced in \eqref{itnwexp}-\eqref{itexp2}. This condition is not necessary when defining expanded algebras by means of the semigroup $S_M^{(n)}$.

In \cite{Penafiel:2019czp}, it was shown that the expanded Nappi-Witten algebras $S_E^{(n)}\times\nw$ define \emph{Generalized Bargmann algebras}, i.e. an infinite family of extensions of the Bargmann algebra in $2+$1 dimensions. On the other hand, the expanded algebras $S_M^{(n)}\times\nw$ were named  \emph{Generalized Newton-Hooke algebras} since they provide generalizations of the $S_E^{(n)}\times\nw$ algebras that include a cosmological constant. By using the same semigroups in the definition of the expanded higher-spin Nappi-Witten algebras $\hsnwi$ and $\hsnwii$, we can define the following families of spin-$3$ symmetries
\begin{align}
&\textnormal{Generalized spin-}3\;\textnormal{Bargmann algebras}:&\quad& 
S_E^{(n)}\times \hsnwi,\;
S_E^{(n)}\times \hsnwii, \label{genbarg}\\[9pt]
&\textnormal{Generalized spin-}3\;\textnormal{Newton-Hooke algebras}:&\quad&
S_M^{(n)}\times \hsnwi,\; \label{gennh}
S_M^{(n)}\times \hsnwii.
\end{align}
These symmetries are related by an In\"o\"u-Wigner contraction, which can be interpreted as a vanishing cosmological constant limit or, equivalently, as the limit when the AdS radius, say $\ell$, becomes infinite. The constant $\ell$ can be made explicit in the symmetry algebras $S^{(n)}_M\times \hsnwi$ and $S^{(n)}_M\times \hsnwii$ previously presented by rescaling the expanded generators \eqref{expandedgen}, \eqref{expandedgenhs1}, \eqref{expandedgenhs2} in the form
\be\label{recalingsell}
\bal
&\bk_a^{(i)}\rightarrow\ell^i\, \bk_a^{(i)},&\quad& 
\bk^{(i)}\rightarrow\ell^i \,\bk^{(i)},&\quad& 
\bsigma^{(i)}\rightarrow\ell^i\, \bsigma^{(i)},\\
&\bmK_{ab}^{(i)}\rightarrow\ell^i \,\bmK_{ab}^{(i)},&\quad& 
\bmK_a^{(i)}\rightarrow\ell^i \,\bmK_a^{(i)},&\quad& 
\bSigma_a^{(i)}\rightarrow\ell^i \,\bSigma_a^{(i)},\\
&\bmK_a^{(i)}\rightarrow\ell^i \,\bmK_a^{(i)},&\quad& 
\bmK_{ab}^{(i)}\rightarrow\ell^i \,\bmK_{ab}^{(i)},&\quad& 
\bSigma_{ab}^{(i)}\rightarrow\ell^i\, \bSigma_{ab}^{(i)},
\eal
\ee
and similarly for the invariant tensor constants given in \eqref{itnwexp}-\eqref{itexp2}, i.e.
\be
\mu_i\rightarrow\ell^i \mu_i,\quad \nu_i\rightarrow\ell^i \rho_i,\quad\rho_i\rightarrow\ell^i \rho_i.
\ee
For a fixed value of $n$, the contraction $\ell\rightarrow\infty$ takes an expanded algebra \eqref{gennh} into the corresponding expanded algebra \eqref{genbarg}, i.e.
\be\label{SMSEcontraction}
S_M^{(n)}\times \hsnwi \xrightarrow[\ell\rightarrow\infty]\, S_E^{(n)}\times \hsnwi,\qquad
S_M^{(n)}\times \hsnwii \xrightarrow[\ell\rightarrow\infty]\, S_E^{(n)}\times \hsnwii.
\ee
The rescaling \eqref{recalingsell} implies that the gauge fields in the connection one forms \eqref{Aexpnw} and \eqref{expAs} should be rescaled with the opposite powers of $\ell$,
\be\label{rescNRfields}
\bal
&\theta^a_{(i)}\rightarrow\dfrac{1}{\ell^i}\theta^a_{(i)},&\quad&
\alpha_{(i)}\rightarrow\dfrac{1}{\ell^i} \alpha_{(i)},&\quad&
\beta_{(i)}\rightarrow\dfrac{1}{\ell^i} \beta_{(i)},\\
&\Theta^{ab}_{(i)}\rightarrow\dfrac{1}{\ell^i}\Theta^{ab}_{(i)},&\quad&
A^a_{(i)}\rightarrow\dfrac{1}{\ell^i} A^a_{(i)},&\quad&
B^a_{(i)}\rightarrow\dfrac{1}{\ell^i} B^a_{(i)},\\
&\Theta^a_{(i)}\rightarrow\dfrac{1}{\ell^i}\Theta^a_{(i)},&\quad&
A^{ab}_{(i)}\rightarrow\dfrac{1}{\ell^i} A^{ab}_{(i)},&\quad&
B^{ab}_{(i)}\rightarrow\dfrac{1}{\ell^i} B^a_{(i)},\\
\eal
\ee
and similarly for the corresponding curvatures \eqref{expFs}. In this way, one can apply the flat limit $\ell\rightarrow\infty$ to the expanded CS actions \eqref{hsCS1} and \eqref{hsCS2}, reducing higher-spin gravity actions with $S^{(n)}_M$-expanded invariance to the corresponding $S^{(n)}_E$-expanded ones.
\\

\noindent\textbf{NR and flat limits}: As a final remark, the flat limit $\ell\rightarrow\infty$ can also be used at the level of the relativistic symmetries introduced in Section \ref{NRlim}. In particular, one can relate the NR limit given in \eqref{expreli} for $S=S^{(n)}_M$ with the corresponding ones for $S=S^{(n)}_E$.

\begin{itemize}
\item We first look at the $\reli$ case \eqref{expreli}, where one can extend the diagram \eqref{SMSEcontraction} as
\be\label{fulldiagram1}
\bal
&\left[S^{(m)}_M\times \sltr\right] \oplus \left[S^{(m)}_M\times \sotr\right]
\;
&\underset{c\rightarrow\infty}{\longrightarrow}& \qquad S^{(m)}_M\times\hsnwi
\\
&\hskip2.3truecm {\scriptstyle \ell\rightarrow\infty}\;\Big\downarrow  & &\hskip.9truecm {\scriptstyle \ell\rightarrow\infty}\;\Big\downarrow\\
&\left[S^{(m)}_E\times \sltr\right] \oplus \left[S^{(m)}_E\times \sotr\right]
\;
&\underset{c\rightarrow\infty}{\longrightarrow}& \qquad S^{(m)}_E\times\hsnwi.
\eal
\ee
To see this, it is necessary to extend the rescaling \eqref{recalingsellRel} to the case of the relativistic algebras \eqref{expreli} and \eqref{exprelii} for $S=S^{(n)}_M$,
\be\label{recalingsellRel}
\bL_A^{(i)}\rightarrow\ell^i\, \bL_A^{(i)},\quad 
\bmL_{AB}^{(i)}\rightarrow\ell^i \,\bmL_{AB}^{(i)},
\ee
and establish the following relations between relativistic spin-$3$ symmetries \cite{Caroca:2017izc}
\be\label{SMSEhsrel}
\bal
&S^{(1)}_M\times\sltr\simeq\hsrela &\xrightarrow[\ell\rightarrow\infty]\, &\qquad S^{(1)}_E\times\sltr\simeq\hsrelp\\
&S^{(2)}_M\times\sltr\simeq\hsrelal &\xrightarrow[\ell\rightarrow\infty]\, &\qquad S^{(2)}_E\times\sltr\simeq\hsrelm\\
&&\vdots&\\
&S^{(n)}_M\times\sltr&\xrightarrow[\ell\rightarrow\infty]\, &\qquad S^{(n)}_E\times\sltr ,\\
\eal
\ee
where the first line in \eqref{SMSEhsrel} relates the spin-$3$ extensions of the AdS and the Poincar\'e algebras in $2+1$ dimensions, whereas the second line does so for the spin-$3$ extensions of the Maxwell and the AdS-Lorentz algebras. However, in order to make contact with \eqref{fulldiagram1}, one has to introduce the expanded algebra \eqref{so3exp} and similarly rescale the generators of $S^{(n)}_M\times \sotr$ as
\be\label{rescellX1}
\bX^{(i)}\rightarrow\ell^i\, \bX^{(i)},\quad 
\bX_a^{(i)}\rightarrow\ell^i\, \bX_a^{(i)},\\
\ee
which leads to the contractions
\be\label{diagexpso3}
\bal
&S^{(1)}_M\times\sotr\simeq\mathfrak{so}(4) &\xrightarrow[\ell\rightarrow\infty]\, &\qquad S^{(1)}_E\times\sotr\simeq\mathfrak{iso}(3)\\
&S^{(2)}_M\times\sotr\simeq\mathfrak{e{\textnormal -}adslor_3} &\xrightarrow[\ell\rightarrow\infty]\, &\qquad S^{(2)}_E\times\sotr\simeq\mathfrak{e{\textnormal -}max_3}\\
&&\vdots&
\\
&S^{(m)}_M\times \mathfrak{so}(3) &\xrightarrow[\ell\rightarrow\infty]\, &\qquad S^{(m)}_E\times \mathfrak{so}(3) ,\\
\eal
\ee
where $\mathfrak{e{\textnormal -}max_3}$ and $\mathfrak{e{\textnormal -}adslor_3}$ denote the Euclidean versions of the Maxwell algebra and the AdS-Lorentz algebra in three spatial dimensions, respectively. We have related the first two examples in \eqref{diagexpso3} to known algebras since these cases will be explicitly used in the next section.

\item When considering the expansion of $\relii$ introduced in \eqref{exprelii}, one can also combine the $c\rightarrow\infty$ and $\ell\rightarrow\infty$ limits and extend the diagram \eqref{SMSEcontraction}in the form
\be\label{fulldiagram2}
\bal
&\left[S^{(m)}_M\times \sltr\right] \oplus \left[S^{(m)}_M\times \sltw\right]\oplus \left[S^{(m)}_M\times \uo \right]
\;
&\underset{c\rightarrow\infty}{\longrightarrow}& \qquad S^{(m)}_M\times\hsnwii
\\
&\hskip3.8truecm {\scriptstyle \ell\rightarrow\infty}\;\Big\downarrow & &\hskip.9truecm {\scriptstyle \ell\rightarrow\infty}\;\Big\downarrow
\\
&\left[S^{(m)}_E\times \sltr\right] \oplus \left[S^{(m)}_E\times \sltw\right]\oplus \left[S^{(m)}_E\times \uo \right]
\;
&\underset{c\rightarrow\infty}{\longrightarrow}& \qquad S^{(m)}_E\times\hsnwii.
\eal
\ee
The expansion of $\sltr$ in \eqref{fulldiagram2} is again \eqref{SMSEhsrel}, which now is to be supplemented with expansions of $\sltwu$. Expanding the generators of $\sltw$ introduced in \eqref{sl2xu1v1} and using the rescaling
\be\label{rescABC}
\bmA^{(i)}\rightarrow\ell^i\, \bmA^{(i)},\quad 
\bmB^{(i)}\rightarrow\ell^i\, \bmB^{(i)},\quad 
\bmC^{(i)}\rightarrow\ell^i\, \bmC^{(i)},
\ee
allows one to relate the following expanded symmetries in the flat limit $\ell\rightarrow\infty$
\be\label{diagexpsltw}
\bal
&S^{(1)}_M\times\sltw\simeq\mathfrak{so}(2,2) &\xrightarrow[\ell\rightarrow\infty]\, &\qquad S^{(1)}_E\times\sltw\simeq\mathfrak{isl}(2,\mathbb R)\\
&S^{(2)}_M\times\sltw\simeq\mathfrak{adslor_{2+1}} &\xrightarrow[\ell\rightarrow\infty]\,&\qquad S^{(2)}_E\times\sltw\simeq\mathfrak{max_{2+1}}\\
&&\vdots&
\\
&S^{(m)}_M\times \sltw &\xrightarrow[\ell\rightarrow\infty]\, &\qquad S^{(m)}_E\times \sltw.
\eal
\ee
Due to the isomorphism $\mathfrak{so}(2,1)\simeq \sltw$, these relations correspond to the Euclidean continuation of the ones given in \eqref{diagexpso3}. In the first case, the expanded algebras are the AdS and the Poincar\'e algebras in three space-time dimensions, whereas in the second case one finds the Maxwell and the AdS-Lorentz algebras in $2+1$ dimensions, which are known to be related by an In\"o\"u-Wigner contraction. Note that \eqref{rescABC} can be equivalently defined in the basis \eqref{sl2xu1v2} where the expanded generators \eqref{sl2xu1exp} are rescaled as
\be\label{rescellX2}
\bX^{(i)}\rightarrow\ell^i\, \bX^{(i)},\quad 
\bX_{ab}^{(i)}\rightarrow\ell^i\, \bX_{ab}^{(i)}.
\ee
Finally, to complete the expansion of \eqref{sl2xu1v1}, we need to expand the $\uo$ generator $\bmD$, which simply produces $n$ copies of $\uo$. In this case, the corresponding rescaling of $\bmD^{(i)}\rightarrow\ell^i\, \bmD^{(i)}$ and flat limit does not modify the algebra, i.e.
\be\label{expuo}
S^{(n)}_M\times\uo\simeq S^{(n)}_E\times\uo \simeq \uo \oplus \underset{n\;{\rm times}}{\cdots} \oplus\uo.
\ee
Putting it all together allows one to construct the left-hand side of \eqref{fulldiagram2}.
\end{itemize}
The contractions here presented also relate the CS gravity theories invariant under the corresponding symmetry algebras. It is thus a direct but involved computation to obtain the CS actions invariant under the spin-$3$ NR algebras \eqref{hsCS1} and \eqref{hsCS1}, from the corresponding CS theories based on the relativistic symmetries considered in \eqref{expreli} and \eqref{expreli}, respectively, which requires one to introduce rescalings in the relativistic gauge fields in complete analogy with Eq. \eqref{rescNRfields} in the NR case.
\\

\noindent\textbf{Pure Nappi-Witten case}: It is worth remarking that, since the generators $\bL_A^{(i)}$ in \eqref{sl3exp} define an expansion of the $\sltw$ subalgebra of $\sltr$, a diagram of the form \eqref{diagexpsltw} is also obtained from \eqref{SMSEhsrel} when eliminating the higher-spin generators $\bmL_{AB}^{(i)}$, reducing the analysis to the pure gravity case. Dropping the higher-spin generators $\bX_a^{(i)}$ and $\bX_{ab}^{(i)}$ in \eqref{so3exp} and \eqref{sl2xu1exp} leaves in each case only the expanded generator $\bX^{(i)}$ for which the expansion can be summarized in a diagram equivalent to \eqref{expuo}. Thus, in the pure gravity case, both diagrams \eqref{fulldiagram1} and \eqref{fulldiagram2} coincide and reduce to.
\be\label{fulldiagram2}
\bal
&\left[S^{(m)}_M\times \sltw\right] \oplus \left[S^{(m)}_M\times \uo \right]
\;
&\underset{c\rightarrow\infty}{\longrightarrow}& \qquad S^{(m)}_M\times\nw
\\
&\hskip3.8truecm {\scriptstyle \ell\rightarrow\infty}\;\Big\downarrow & &\hskip.9truecm {\scriptstyle \ell\rightarrow\infty}\;\Big\downarrow
\\
&\left[S^{(m)}_E\times \sltw\right] \oplus \left[S^{(m)}_E\times \uo \right]
\;
&\underset{c\rightarrow\infty}{\longrightarrow}& \qquad S^{(m)}_E\times\nw.
\eal
\ee
This means that for a given $n$, an expanded Nappi-Witten algebra \eqref{expnw} is the NR contractions of the semidirect sum of the corresponding expanded $\sltw$ algebra and $n$ Abelian generators. As we will see in the following section, this reproduces the known contractions that lead to the Extended Bargmann algebra \cite{Bergshoeff:2016lwr} and the Extended Electric NR Maxwell algebra \cite{Aviles:2018jzw}, and extends the result to expansions with arbitrary $n$.

\section{Examples}
\label{Examples}
Let us consider explicit examples of the higher-spin symmetries presented in the previous section.  
In the following, we explicitly evaluate expanded symmetries with the semigroups $S_E^{(n)}$ and $S_M^{(n)}$ for $n=1$ and $n=2$. The $n=1$ case reproduces the higher-spin extensions of the extended Bargmann and Newton-Hooke algebras in $2+1$ dimensions presented in \cite{Bergshoeff:2016soe}. The $n=2$ case yields novel higher-spin extensions of the NR Maxwell algebra and the NR AdS-Lorentz algebra in $2+1$ dimensions. 
The explicit construction of the expanded algebras for $n\geq3$ can be straightforwardly done along the same lines and they will not be given here.

\subsection{Spin-$3$ Bargmann algebras}
\label{HSB}
The Bargmann algebra is the universal central extension of the Galilei algebra \cite{Bacry:1968zf} (for a recent review on non-Lorentzian symmetries, see also \cite{Bergshoeff:2022eog}). In $2+1$ dimensions, it admits a second non-trivial central extension\footnote{In fact, the Galilei algebra in three space-time dimensions admits a third central extension, which is non-physical \cite{Brihaye:1995nv} and will not be considered here.}, that allows defining a non-degenerate invariant bilinear form \cite{Papageorgiou:2009zc}. As a first example, we reconstruct the spin-$3$ extensions of the Extended Bargmann algebra in $2+1$ dimensions  \cite{Bergshoeff:2016soe}, as particular expansions of our spin-$3$ Nappi-Witten algebras. As shown in \cite{Penafiel:2019czp}, the extended Bargmann algebra in $2+1$ dimensions
\be\label{ExtBargAlg}
\bal
&\left[\bJ,\bG_a\right]=\epsilon_{ab}\bG_b,
&
&\left[\bG_a,\bG_b \right]=-\epsilon_{ab}\bS,\\[5pt]
&\left[\bJ,\bP_a \right]=  [\bH,\bG_a ] =\epsilon_{ab}\bP_b,
&
&\left[\bG_a,\bP_b \right]=-\epsilon_{ab}\bM,
\eal
\ee
is the expanded Nappi-Witten algebra $S^{(1)}_E\times \nw$, generated by 
\be\label{BargRenameGen}
\bal
&\bJ=\bk^{(0)},& &\bG_a=\bk_a^{(0)},& &\bS=\bsigma^{(0)},\\
&\bH=\bk^{(1)},& &\bP_a=\bk_a^{(1)}, & &\bM=\bsigma^{(1)},
\eal
\ee
where the semigroup $S^{(1)}_E$ is obtained by replacing $n=1$ in \eqref{sen}. Here $\bJ$, $\bH$, $\bG_a$ and $\bP_a$ stand respectively for the generators of spatial rotations, temporal translations, Galilean boosts and spatial translations, whereas $\bS$ and $\bM$ are central extensions associated to spin and mass\footnote{While the mass generator $\bM$ is present in higher-dimensional versions of the Bargmann algebra and its physical meaning is understood (see, for example, \cite{Bacry:1968zf}), the generator $\bS$ is particular to the three-dimensional case and its interpretation as a spin angular momentum generator has not yet been fully established \cite{Duval:2002cw,Hagen:2002pg}.}. In the same way, the spin-$3$ extensions of the Bargmann algebra can be defined by evaluating the expanded higher-spin symmetries \eqref{HSNWalgebra1exp} and \eqref{HSNWalgebra2exp} for $S=S_E^{(1)}$. 

\begin{itemize}
\item The first spin-$3$ extension of the Extended Bargmann algebra in $2+1$ dimensions, $\hsbi$, is given by $S_E^{(1)}\times\hsnwi$. It can be obtained by defining the higher-spin generators
\be\label{hsBarg1RenameGen}
\bal
&\bmJ_a=\bmK_a^{(0)},&
&\bmG_{ab}=\bmK_{ab}^{(0)},&
&\bmS_a=\bSigma_a^{(0)}, \\
&\bmH_a=\bmK_a^{(1)},&
&\bmP_{ab}=\bmK_{ab}^{(1)},&
&\bmM_a=\bSigma_a^{(1)},
\eal
\ee
and using the relations \eqref{HSNWalgebra1exp} and \eqref{sen} for $n=1$. The commutation relations of the expanded generators \eqref{BargRenameGen} and \eqref{hsBarg1RenameGen} are given by \eqref{ExtBargAlg} together with 
\be\label{BargHS1}
\bal
&\left[\bJ,\bmJ_a\right]  = \epsilon_{ab}\bmJ_b,
&
&\left[\bJ,\bmG_{ab}\right] =  -\epsilon_{c(a}\bmG_{b)c},&
&\left[\bJ,\bmS_a\right] =\left[\bS,\bmJ_a\right]  = \epsilon_{ab}\bmS_b,
\\[5pt]
&\left[\bG_a,\bmJ_b\right] = -\epsilon_{a(c}\bmG_{b)c},
&
&\left[\bG_a,\bmG_{bc}\right]  =  -\epsilon_{a(b}\bmS_{c)}, &
&\left[\bmJ_a,\bmJ_b\right]  =  \epsilon_{ab}\bJ,
\\[5pt]
&\left[\bmJ_a, \bmS_b\right] = \epsilon_{ab}\bS,&
&\left[\bmJ_a,\bmG_{bc}\right]  =\delta_{a(b}\epsilon_{c)d}\bG_d,&
&\left[\bmG_{ab},\bmG_{cd}\right] = \delta_{(a(c}\epsilon_{d)b)}\bS,
\eal
\ee
together with
\be\label{BargHS2}
\bal
&\left[\bJ,\bmH_a\right]=\left[\bH,\bmJ_a\right]   = \epsilon_{ab}\bmH_b,&
&\left[\bG_a,\bmH_b\right] = \left[\bP_a,\bmJ_b\right] = -\epsilon_{a(c}\bmP_{b)c},
\\[5pt]
&\left[\bJ,\bmP_{ab}\right]=\left[\bH,\bmG_{ab}\right]  =  -\epsilon_{c(a}\bmP_{b)c},&
&\left[\bmJ_a,\bmP_{bc}\right]  =\left[\bmH_a,\bmG_{bc}\right]   =\delta_{a(b}\epsilon_{c)d}\bP_d,
\\[5pt]
&\left[\bmJ_a,\bmH_b\right]  =  \epsilon_{ab}\bH,&
&\left[\bG_a,\bmP_{bc}\right]  =  \left[\bP_a,\bmG_{bc}\right]  =-\epsilon_{a(b}\bmM_{c)},
\\[5pt]
&\left[\bmJ_a, \bmM_b\right] =\left[\bmH_a, \bmS_b\right]= \epsilon_{ab}\bM,&
&\left[\bJ,\bmM_a\right] =\left[\bS,\bmH_a\right]  = \left[\bH,\bmS_a\right]=\left[\bM,\bmJ_a\right]  =\epsilon_{ab}\bmM_b.
\\[5pt]
&\left[\bmG_{ab},\bmP_{cd}\right] = \delta_{(a(c}\epsilon_{d)b)}\bM,
\eal
\ee
Notice that, when including the higher-spin extension, the generators $\bS$ and $\bM$ are not central anymore. The $\hsbi$ algebra was obtained in \cite{Bergshoeff:2016soe} from a double contraction (large-speed-of-light and large-AdS-raduis limit) of the spin-$3$ extension of the AdS algebra in three dimensions $\hsrela$.

\item One can define a second spin-$3$ extension of the Extended Bargman algebra in $2+1$ dimensions, $\hsbii$, as the expanded algebra $S_E^{(1)}\times \hsnwii$. Now, instead of \eqref{hsBarg1RenameGen}, the alternative generators
\be\label{hsBarg2RenameGen}
\bal
&\bmJ_{ab}=\bmKt_{ab}^{(0)},& 
& \bmG_a=\bmKt_a^{(0)},&
&\bmS_{ab}=\bSigmat_{ab}^{(0)},& \\
&\bmH_{ab}=\bmKt_{ab}^{(1)},&
&\bmP_a=\bmKt_a^{(1)},&
&\bmM_{ab}=\bSigmat_{ab}^{(1)},
\eal
\ee
are introduced. Putting together the Extended Bargmann generators \eqref{BargRenameGen} with \eqref{hsBarg2RenameGen}, and using the commutation relations of the $\hsnwii$ algebra \eqref{HSNWalgebra2exp} for the $S_E^{(1)}$ semigroup \eqref{sen}, one finds the commutation relations \eqref{ExtBargAlg} plus
\be\label{BargHS1T}
\bal
&[\bJ,\bmG_a ] =\epsilon_{ab}\bmG_b,&
&[\bJ,\bmJ_{ab}] =-\epsilon_{c(a}\bmJ_{b)c},&
&[\bJ,\bmS_{ab}] =[\bS,\bmJ_{ab}] =-\epsilon_{c(a}\bmS_{b)c},
\\[5pt]
&[\bG_a ,\bmG_b ]=-\epsilon_{a(c}\bmS_{b)c},&
&[\bG_a ,\bmJ_{bc}]=-\epsilon_{a(b}\bmG_{c)},&
&[\bmJ_{ab},\bmJ_{cd}] =\delta_{(a(c}\epsilon_{d)b)} \bJ,
\\[5pt]
&[\bmG_a ,\bmG_b ] =\epsilon_{ab}\bS,&
&[\bmG_a ,\bmJ_{bc}] =\delta_{a(b}\epsilon_{c)d}\bG_d,&
&[\bmJ_{ab},\bmS_{cd}] =\delta_{(a(c}\epsilon_{d)b)} \bS,
\eal
\ee
and
\be\label{BargHS2T}
\bal
&[\bJ,\bmP_a ]=[\bH,\bmG_a ] =\epsilon_{ab}\bmP_b,&
&[\bG_a ,\bmH_{bc}]=[\bP_a ,\bmJ_{bc}]=-\epsilon_{a(b}\bmP_{c)},
\\[5pt]
&[\bJ,\bmH_{ab}] =[\bH,\bmJ_{ab}] =-\epsilon_{c(a}\bmH_{b)c},&
&[\bmG_a ,\bmH_{bc}] =[\bmP_a ,\bmJ_{bc}] =\delta_{a(b}\epsilon_{c)d}\bP_d,
\\[5pt]
&[\bmJ_{ab},\bmH_{cd}] =\delta_{(a(c}\epsilon_{d)b)} \bH,&
&[\bmJ_{ab},\bmM_{cd}] =[\bmH_{ab},\bmS_{cd}] =\delta_{(a(c}\epsilon_{d)b)} \bM,
\\[5pt]
&[\bmG_a ,\bmP_b ] =\epsilon_{ab}\bM,&
&[\bJ,\bmM_{ab}] =[\bH,\bmS_{ab}] =[\bS,\bmH_{ab}] 
\\[5pt]
&[\bG_a ,\bmP_b ]=[\bP_a ,\bmG_b ]=-\epsilon_{a(c}\bmM_{b)c},&
&\hskip1.6truecm=[\bM,\bmJ_{ab}] =-\epsilon_{c(a}\bmM_{b)c}.
\eal
\ee
Even though this algebra is not explicitly equivalent to the $\hsbii$ algebra obtained in \cite{Bergshoeff:2016soe} by means of the Medina-Revoy theorem, one can show they are isomorphic by using the redefinition \eqref{redefSigmaexp} for the higher-spin generators $\bmS_{ab}$ and $\bmM_{ab}$. 
\end{itemize}
Thus, we have explicitly constructed the spin-$3$ NR algebras
\be\label{hsbiandii}
\hsbi=S^{(1)}_E\times \hsnwi,\qquad 
\hsbii=S^{(1)}_E\times \hsnwii.
\ee
As argued in \cite{Bergshoeff:2016soe}, two additional higher-spin extensions of the Extended Bargmann algebra can be obtained by interchanging 
\be\label{swap1}
\bal
&\bmJ_a\leftrightarrow \bmH_a,&\quad&
\bmG_{ab}\leftrightarrow \bmP_{ab},&\quad&
\bmS_a\leftrightarrow \bmM_a,\\
&\bmJ_{ab}\leftrightarrow \bmH_{ab},&\quad&
\bmG_a\leftrightarrow \bmP_b,&\quad&
\bmS_{ab}\leftrightarrow \bmM_{ab},
\eal
\ee
in \eqref{BargHS1} and  \eqref{BargHS2}, and in \eqref{BargHS1T} and \eqref{BargHS2T}, respectively.
\\

\noindent\textbf{Semidirect sum structure}: The algebras $\hsbi$ and $\hsbii$ are semidirect sums. This is to be expected since they are higher-spin extensions of the NR limit of the Poincar\'e algebra, which is a semidirect sum itself. Indeed, the first line of \eqref{ExtBargAlg} together with Eq. \eqref{BargHS1} (Eq. \eqref{BargHS1T}) is a subalgebra isomorphic to $\hsnwi$ ($\hsnwii$). The remaining generators form an Abelian ideal that can be identified with $\hsnwi$ ($\hsnwii$) seen as an Abelian vector space, which can be denoted by $\hsnwi^{({\rm ab})}$ ($\hsnwii^{({\rm ab})}$), and on which $\hsnwi$ ($\hsnwii$) acts with the adjoint representation \cite{Barnich:2014kra}. Thus, the $\hsbi$ ($\hsbii$) algebra can be written in a compact form as
\be
\left[(X,X^\ast),(Y,Y^\ast)\right]=\left([X,Y],[X,Y^\ast]-[Y,X^\ast]\right), 
\ee
where $X$ is an element on the $\hsnwi$ ($\hsnwii$) algebra and $X^\ast$ is an element on $\hsnwi^{({\rm ab})}$ ($\hsnwii^{({\rm ab})}$). Therefore, the spin-$3$ extensions of the Extended Bargmann algebra in $2+1$ dimensions can be defined as the semidirect sums
\be
\bal
\hsbi& \simeq \hsnwi \oright _{\rm ad}  \hsnwi^{({\rm ab})},\\
\hsbii&\simeq \hsnwii \oright _{\rm ad}  \hsnwii^{({\rm ab})}.
\eal 
\ee
This is the NR analog of the semidirect sum \cite{Campoleoni:2016vsh}
\be\label{hspoin}
\hsrelp\simeq\sltr\oright _{\rm ad}  \sltr^{({\rm ab})},
\ee
which corresponds to the spin-$3$ extension of the Poincar\'e algebra in three space-time dimensions \cite{Afshar:2013vka,Gonzalez:2013oaa}.
\\

\noindent\textbf{NR limit}: The spin-$3$ extensions of the Bargman algebra \eqref{hsbiandii} can alternatively be obtained as In\"o\"u-Wigner contractions of relativistic symmetries. In \cite{Caroca:2017izc}, it was shown that the spin-$3$ extension of the Poincar\'e algebra in three space-time dimensions \eqref{hspoin} is the expansion
\be\label{hspoin2}
\hsrelp\simeq S^{(1)}_E\times \sltr,
\ee
with expanded generators
\be\label{hspoingen}
\bJ_A^\r = \bL^{(0)}_A,\quad
\bP_A^\r = \bL^{(1)}_A,\quad
\bmJ_{AB}^\r = \bmL^{(0)}_{AB},\quad
\bmP_{AB}^\r = \bmL^{(1)}_{AB},
\ee
and commutation relations given by \eqref{sl3exp} evaluated for the semigroup product law of $S^{(1)}_E$ given in \eqref{sen}. In order to define the NR limits \eqref{expreli} and \eqref{exprelii}, we need to introduce the expanded algebras \eqref{so3exp} and \eqref{sl2xu1exp}. For $S=S^{(1)}_E$, these are respectively isomorphic to the Euclidean algebra in three spatial dimensions $\mathfrak{iso}(3)$ (spanned by $\{\bX^{(0)},\bX^{(1)},\bX_a^{(0)},\bX_a^{(1)}\}$ in Eq. \eqref{so3exp}) and the Poincar\'e algebra in three space-time dimensions, $\mathfrak{isl}(2,\mathbb R)$, supplemented with two Abelian generators\footnote{The isomorphism $S^{(1)}_E\times\Big( \sltwu\Big) \simeq \mathfrak{isl}(2,\mathbb R)\oplus \uo\oplus\uo$ is more transparent at the level of the commutation relations when expanding $\sltwu$ in the basis \eqref{sl2xu1v1} rather than \eqref{sl2xu1v2}.} (spanned by $\{\bX^{(0)},\bX^{(1)},\bX_{ab}^{(0)},\bX_{ab}^{(1)}\}$ in Eq. \eqref{sl2xu1exp}), i.e.
\be\label{otherexpspoin}
\bal
&S^{(1)}_E\times \sotr \simeq \mathfrak{iso}(3),\\
&S^{(1)}_E\times\Big( \sltwu\Big) \simeq \mathfrak{isl}(2,\mathbb R)\oplus \uo\oplus\uo.
\eal
\ee
These isomorphisms also follow from the diagrams  \eqref{diagexpso3}, \eqref{diagexpsltw} and \eqref{expuo}. Inverting \eqref{redefsl3iexp}, one finds the relation between the Poincar\'e and Bargman generators
\be\label{IWpoin1}
\bal
\bJ & =  \bJ_{0}^\r+\bX^{(0)}, &
\bJ_a & =  \frac{1}{c}\bJ_a^\r, &
\bS & =  \frac{1}{2c^2}\left(\bJ_{0}^\r-\bX^{(0)}\right),\\
\bH & =  \bP_{0}^\r+\bX^{(1)}, &
\bP_a & =  \frac{1}{c}\bP_a^\r, &
\bM & =  \frac{1}{2c^2}\left(\bP_{0}^\r-\bX^{(1)}\right),
\eal
\ee
and the corresponding relations between NR and relativistic spin-$3$ generators
\be\label{IWpoin2}
\bal
\bmJ_a&= \bmJ_{a0}^\r+\bX_a^{(0)},
&
\bmG_{ab} &=  \frac{1}{c}\bmJ_{ab}^\r,
&
\bmS_a&=  \frac{1}{2c^{2}}\left(\bmJ_{a0}^\r-\bX_a^{(0)}\right),\\
\bmH_a&= \bmP_{a0}^\r+\bX_a^{(1)},
&
\bmP_{ab} &=  \frac{1}{c}\bmP_{ab}^\r,
&
\bmM_a&=  \frac{1}{2c^{2}}\left(\bmP_{a0}^\r-\bX_a^{(1)}\right).
\eal
\ee
Similarly, using \eqref{redefsl3iiexp} one gets \eqref{IWpoin1} and
\be\label{IWpoin3}
\bal
\bmJ_{ab}&= \bmJ_{ab}^\r+\bX_{ab}^{(0)},
&
\bmG_a &=  \frac{1}{c}\bmJ_{a0}^\r,
&
\bmS_{ab}&=  \frac{1}{2c^{2}}\left(\bmJ_{ab}^\r-\bX_{ab}^{(0)}\right),\\
\bmH_{ab}&= \bmP_{ab}^\r+\bX_{ab}^{(1)},
&
\bmP_a &=  \frac{1}{c}\bmP_{a0}^\r,
&
\bmM_{ab}&=  \frac{1}{2c^{2}}\left(\bmP_{ab}^\r-\bX_{ab}^{(1)}\right).
\eal
\ee
Thus, putting it all together one finds that $\hsbi$ can be obtained from the contraction defined by \eqref{IWpoin1} and \eqref{IWpoin2}, while $\hsbii$ follows from taking $c\rightarrow\infty$ in \eqref{IWpoin1} and \eqref{IWpoin3}. Schematically, the NR limits \eqref{expreli} and \eqref{exprelii} for $S=S^{(1)}_E$ can be written as
\be\label{contractionshspoin}
\bal
&\hsrelp\oplus \mathfrak{iso}(3)\;&
\underset{c\rightarrow\infty}{\longrightarrow}& \qquad \hsbi
\\[6pt]
&\hsrelp \oplus \mathfrak{isl}(2,\mathbb R)\oplus \uo\oplus\uo\;&\underset{c\rightarrow\infty}{\longrightarrow}&\qquad \hsbii
\eal
\ee
Moreover, one can see from \eqref{IWpoin1} that dropping the higher-spin generators precisely corresponds to the contraction relating the Poincar\'e and the Extended Bargmann algebras in three space-time dimensions \cite{Bergshoeff:2016lwr}. In other words, when removing the spin-$3$ generators, the left-hand sides of the contractions in \eqref{contractionshspoin} boil down to $\mathfrak{iso}(2,1)\oplus \uo \oplus\uo$ whereas the right-hand sides reduce to the Extended Bargman algebra in $2+1$ dimensions.

\subsection{Spin-$3$ non-relativistic Maxwell algebras}
\label{NRMax}
The Maxwell algebra is an extension of the Poincar\'e algebra that describes particle systems in the presence of a constant electromagnetic field \cite{Bacry:1970ye,Schrader:1972zd}. In $2+1$ dimensions, the electric NR limit of the Maxwell algebra \cite{Gonzalez:2016xwo,Rubio:2018itx,Gomis:2019fdh} admits three central extensions \cite{Aviles:2018jzw} (there dubbed \emph{exotic}), in complete analogy with the two central extensions appearing in the Extended Bargmann algebra. This particular extended NR limit\footnote{Other NR limits of the Maxwell algebra have been considered in \cite{Bacry:1970du,Beckers:1983gp,negro1990local2,Gomis:2019fdh}.}, which from now on in this paper will be simply referred to as NR Maxwell algebra in $2+1$ dimensions, is isomorphic to the expanded Nappi-Witten algebra $S_E^{(2)}\times\nw$ \cite{Penafiel:2019czp}, and its commutation relations are \eqref{ExtBargAlg} plus
\be\label{ExtNRMaxAlg}
\bal
\left[\bJ,\bZ_a\right]=\left[\bH,\bP_a\right]=
\left[\bY,\bG_a\right]=\epsilon_a^{\;\;b}\bZ_b,\qquad
\left[\bG_a,\bZ_b\right]=\left[\bP_a,\bP_b\right]=-\epsilon_{ab}\bT,
\eal
\ee
where the expanded Nappi-Witten generators are the ones given in \eqref{BargRenameGen} and
\be\label{MaxwellRenameGen}
\bY=\bk^{(2)},\quad \bZ_a=\bk_b^{(2)}, \quad \bT=\bsigma^{(2)}.
\ee
In the NR Maxwell algebra, the generators $\bY$ and $\bZ_a$ are associated to the magnetic and the electric field that define the constant electromagnetic background, while $\bT$ is a third central term (which, as in the previous case, ceases being central when the higher-spin generators are introduced). One can define two inequivalent spin-$3$ extensions of the NR Maxwell algebra in three dimensions by evaluating the expanded algebras \eqref{HSNWalgebra1exp} and \eqref{HSNWalgebra2exp} for $S=S^{(2)}_E$, i.e. 
\be\label{hsmiandii}
\hsmi=S^{(2)}_E\times \hsnwi,\qquad 
\hsmii=S^{(2)}_E\times \hsnwii.
\ee
\begin{itemize}
\item In order to define the first higher-spin extension $\hsmi$, we evaluate the commutation relations of $S^{(2)}_E\times\hsnwi$ using \eqref{HSNWalgebra1exp} and the semigroup product law \eqref{sen} for $n=2$. We define
\begin{equation}\label{hsMaxwell1RenameGen}
\bmY_a=\bmK_a^{(2)},\quad \bmZ_{ab}=\bmK_{ab}^{(2)},\quad \bmT_a=\bSigma_a^{(2)},
\end{equation}
which, together with the set of expanded generators appearing in Eqs. \eqref{BargRenameGen}, \eqref{hsBarg1RenameGen} and \eqref{MaxwellRenameGen}, satisfy the commutation relations \eqref{ExtBargAlg}, \eqref{BargHS1}, \eqref{BargHS2} \eqref{ExtNRMaxAlg} and
\be\label{MaxHS}
\bal
&\left[\bJ,\bmY_a\right]=\left[\bY,\bmJ_a\right] = \left[\bH,\bmH_a\right] = \epsilon_{ab}\bmY_b,\\[5pt]
&\left[\bJ,\bmZ_{ab}\right]=\left[\bY,\bmG_{ab}\right]  =  \left[\bH,\bmP_{ab}\right]  = -\epsilon_{c(a}\bmZ_{b)c},\\[5pt]
&\left[\bmJ_a,\bmY_b\right]  =  \left[\bmH_a,\bmH_b\right]  =\epsilon_{ab}\bY,\\[5pt]
&\left[\bmJ_a, \bmT_b\right] =\left[\bmY_a, \bmS_b\right]=\left[\bmH_a, \bmM_b\right]= \epsilon_{ab}\bT,\\[5pt]
&\left[\bmG_{ab},\bmZ_{cd}\right] = \left[\bmP_{ab},\bmP_{cd}\right] =\delta_{(a(c}\epsilon_{d)b)}\bT,\\[5pt]
&\left[\bG_a,\bmY_b\right] = \left[\bZ_a,\bmJ_b\right]  = \left[\bP_a,\bmH_b\right] =-\epsilon_{a(c}\bmZ_{b)c},
\\[5pt]
&\left[\bmJ_a,\bmZ_{bc}\right]  =\left[\bmY_a,\bmG_{bc}\right]   =\left[\bmH_a,\bmP_{bc}\right]  =\delta_{a(b}\epsilon_{c)d}\bZ_d,
\\[5pt]
&\left[\bG_a,\bmZ_{bc}\right]  =  \left[\bZ_a,\bmG_{bc}\right]  = \left[\bP_a,\bmP_{bc}\right]  =-\epsilon_{a(b}\bmT_{c)},
\\[5pt]
&\left[\bJ,\bmT_a\right] =\left[\bS,\bmY_a\right]  = \left[\bY,\bmS_a\right]=\left[\bT,\bmJ_a\right]  =\left[\bH,\bmM_a\right]=\left[\bM,\bmH_a\right]    =\epsilon_{ab}\bmT_b.
\\[5pt]
\eal
\ee
\item The second spin-$3$ extension of the NR Maxwell symmetry, $\hsmii$, follows from defining
\begin{equation}\label{hsMaxwell2RenameGen}
\bmY_{ab}=\bmKt_{ab}^{(2)},\quad \bmZ_a=\bmKt_a^{(2)},\quad \bmT_{ab}=\bSigmat_{ab}^{(2)},
\end{equation}
in addition to \eqref{BargRenameGen}, \eqref{hsBarg2RenameGen} and \eqref{MaxwellRenameGen}. Using again the multiplication law of the semigroup $S^{(2)}_E$ the algebra $\hsmii$ can be read off from \eqref{HSNWalgebra2exp} and is given by \eqref{ExtBargAlg}, \eqref{BargHS1T}, \eqref{BargHS2T}, \eqref{ExtNRMaxAlg} together with the commutation realations
\be\label{MaxHST}
\bal
&[\bJ,\bmZ_a ]=[\bY,\bmG_a ] =[\bH,\bmP_a ] =\epsilon_{ab}\bmZ_b,\\[5pt]
&[\bJ,\bmY_{ab}] =[\bY,\bmJ_{ab}]=[\bH,\bmH_{ab}]  =-\epsilon_{c(a}\bmY_{b)c},\\[5pt]
&[\bmJ_{ab},\bmY_{cd}]=[\bmH_{ab},\bmH_{cd}]  =\delta_{(a(c}\epsilon_{d)b)} \bY,\\[5pt]
&[\bmG_a ,\bmZ_b ]=[\bmP_a ,\bmP_b ]  =\epsilon_{ab}\bT,\\[5pt]
&[\bG_a ,\bmZ_b ]=[\bZ_a ,\bmG_b ]=[\bP_a ,\bmP_b ]=-\epsilon_{a(c}\bmT_{b)c},\\[5pt]
&[\bG_a ,\bmY_{bc}]=[\bZ_a ,\bmJ_{bc}]=[\bP_a ,\bmH_{bc}]=-\epsilon_{a(b}\bmZ_{c)},
\\[5pt]
&[\bmG_a ,\bmY_{bc}] =[\bmZ_a ,\bmJ_{bc}] =[\bmP_a ,\bmH_{bc}] =\delta_{a(b}\epsilon_{c)d}\bZ_d,
\\[5pt]
&[\bmJ_{ab},\bmT_{cd}] =[\bmY_{ab},\bmS_{cd}] =[\bmH_{ab},\bmM_{cd}] =\delta_{(a(c}\epsilon_{d)b)} \bT,
\\[5pt]
&[\bJ,\bmT_{ab}] =[\bY,\bmS_{ab}] =[\bS,\bmY_{ab}]=[\bT,\bmJ_{ab}] =[\bH,\bmM_{ab}] =[\bM,\bmH_{ab}] =-\epsilon_{c(a}\bmT_{b)c}.
\eal
\ee
\end{itemize}
In the same way as in the Bargmann case, one can rename the higher-spin generators in the algebras given above either as
\be\label{swap2}
\bal
&\bmH_a\leftrightarrow \bmY_a,&\quad&
\bmP_{ab}\leftrightarrow \bmZ_{ab},&\quad&
\bmM_a\leftrightarrow \bmT_a,\\
&\bmH_{ab}\leftrightarrow \bmY_{ab},&\quad&
\bmP_a\leftrightarrow \bmZ_b,&\quad&
\bmM_{ab}\leftrightarrow \bmT_{ab},
\eal
\ee
or as
\be\label{swap3}
\bal
&\bmJ_a\leftrightarrow \bmY_a,&\quad&
\bmG_{ab}\leftrightarrow \bmZ_{ab},&\quad&
\bmS_a\leftrightarrow \bmT_a,\\
&\bmJ_{ab}\leftrightarrow \bmY_{ab},&\quad&
\bmG_a\leftrightarrow \bmZ_b,&\quad&
\bmS_{ab}\leftrightarrow \bmT_{ab},
\eal
\ee
giving isomorphic spin-$3$ extensions of the NR Maxwell algebra in $2+1$ dimensions, which can be physically different depending on how the higher-spin generators are physically interpreted. Similarly one can try the combinations \eqref{swap1}, \eqref{swap2} and \eqref{swap1}, \eqref{swap3}.

It is interesting to note that the expanded algebras $S^{(2)}_E\times\hsnwi$ and $S^{(2)}_E\times\hsnwii$ also define spin-$3$ extensions of a NR version of the Hietarinta algebra in $2+1$ dimensions \cite{Bansal:2018qyz,Chernyavsky:2020fqs} if one interchanges the generators
\be
\bH\leftrightarrow \bY,\quad
\bP_a\leftrightarrow \bZ_a,\quad
\bM\leftrightarrow \bT,\\
\ee
in the NR Maxwell sublagebra defined by Eqs. \eqref{ExtBargAlg} and \eqref{ExtNRMaxAlg}.
\\

\noindent\textbf{Extended semidirect sum structure}: The relativistic Maxwell algebra has the structure of an extended semidirect sum, where the Poincar\'e translations become non-Abelian. On the other hand, the higher-spin symmetry $\hsmi$ ($\hsmii$) extends the algebra $\hsbi$ ($\hsbii$) found in Section \ref{HSB} by rendering its Abelian ideal of space and time translations non-Abelian. Since the extra generators in this extension form, in turn, a new Abelian ideal that can be identified with $\hsnwi^{({\rm ab})}$ ($\hsnwi^{({\rm ab})}$), one can conclude that $\hsmi$ ($\hsmii$) is an extended semidirect sum algebra of the same type as the original Maxwell algebra in three space-time dimensions. Indeed, the commutation relations of the $\hsmi$ ($\hsmii$) algebra can be written in a compact form as \cite{Salgado-Rebolledo:2019kft}
\be
\left[(X,X^\ast,X^\prime),(Y,Y^\ast,Y^\prime)\right]=\left([X,Y],[X,Y^\ast]-[Y,X^\ast],[X,Y^\prime]-[Y,X^\prime]+[X^\ast,Y^\ast]\right), 
\ee
with $X$ an element of $\hsnwi$ ($\hsnwii$), and $X^\ast$ and $X^\prime$ belonging to $\hsnwi^{({\rm ab})}$ ($\hsnwii^{({\rm ab})}$). Therefore, the spin-$3$ extensions of the NR Maxwell algebra in $2+1$ dimensions can be expressed as
\be\label{extsds}
\bal
\hsmi& = \hsnwi \oright _{\rm ad}  \hsnwi^{({\rm ab})}_{\rm ext},\\
\hsmii&= \hsnwii \oright _{\rm ad}  \hsnwii^{({\rm ab})}_{\rm ext},
\eal 
\ee
where $\hsnwi^{({\rm ab})}_{\rm ext}$ ($\hsnwii^{({\rm ab})}_{\rm ext}$) denotes the extended group $\hsnwi \otimes\hsnwi$ ($\hsnwii \otimes\hsnwii$) endowed with the product
\be\label{exttranslations}
\left(X^\ast,X^\prime\right)\hat{+}\left(Y^\ast,Y^\prime \right)=\left(X^\ast+Y^\ast\,,\; X^\prime+Y^\prime +\frac{1}{2}\,[X^\ast,Y^\ast] \right).
\ee
The relativistic analog of \eqref{extsds} is the extended semidirect sum that defines the spin-$3$ extension of the Maxwell algebra\footnote{A higher-spin extension of the Maxwell algebra in higher dimensions has been considered in \cite{Fedoruk:2013sna}.} in $2+1$ dimensions \cite{Salgado-Rebolledo:2019kft}
\be\label{hsmax}
\hsrelm\simeq\sltr  \oright _{\rm ad}  \sltr^{({\rm ab})}_{\rm ext}.
\ee

\noindent\textbf{NR limit}: As shown in the general case treated in Section \ref{NRlim}, the higher-spin NR symmetries \eqref{hsmiandii} can be obtained as NR limits applied to relativistic symmetries. The algebra \eqref{hsmax} was constructed in \cite{Caroca:2017izc} as the expansion
\be\label{hsmax2}
\hsrelm\simeq S^{(2)}_E\times \sltr,
\ee
with the expanded generators \eqref{hspoingen} plus
\be\label{hsmaxgen}
\bZ_A^\r = \bL^{(2)}_A,\quad
\bmZ_{AB}^\r = \bmL^{(2)}_{AB},
\ee
and commutation relations \eqref{sl3exp} for $S=S^{(2)}_E$. Now we look at the expanded algebras \eqref{so3exp} and \eqref{sl2xu1exp}, necessary to define the NR limits \eqref{redefsl3iexp} and \eqref{redefsl3iiexp}. When using the semigroup product law of $S^{(2)}_E$ given in \eqref{sen}, these can be shown to be isomorphic to the Euclidean Maxwell algebra in three dimensions $\mathfrak{e\textnormal{-}max_{3}}$ (spanned by $\{\bX^{(0)},\bX^{(1)},\bX_a^{(0)},\bX_a^{(1)},\bX_a^{(2)}\}$ in Eq. \eqref{so3exp}) and of the direct sum of the Maxwell algebra in $2+1$ dimensions and three Abelian generators (spanned by $\{\bX^{(0)},\bX^{(1)},\bX_{ab}^{(0)},\bX_{ab}^{(1)},\bX_{ab}^{(2)}\}$ in Eq. \eqref{sl2xu1exp}),
\be
\bal
&S^{(2)}_E\times \sotr \simeq \mathfrak{e\textnormal{-}max_{3}},\\
&S^{(2)}_E\times\Big( \sltwu\Big) \simeq \mathfrak{max_{2+1}}\oplus \uo\oplus\uo\oplus\uo.
\eal
\ee
Using the relation \eqref{redefsl3iexp} to relate relativistic and NR generators, we find for the NR Maxwell generators the relations \eqref{IWpoin1} plus
\be\label{IWmax1}
\bal
\bY & =  \bZ_{0}^\r+\bX^{(2)}, &
\bZ_a & =  \frac{1}{c}\bZ_a^\r, &
\bT & =  \frac{1}{2c^2}\left(\bZ_{0}^\r-\bX^{(2)}\right),
\eal
\ee
whereas the higher-spin generators of $\hsmi$ satisfy \eqref{IWpoin2} and
\be\label{IWmax2}
\bal
\bmY_a&= \bmZ_{a0}^\r+\bX_a^{(2)},
&
\bmZ_{ab} &=  \frac{1}{c}\bmZ_{ab}^\r,
&
\bmT_a&=  \frac{1}{2c^{2}}\left(\bmZ_{a0}^\r-\bX_a^{(2)}\right).\\
\eal
\ee
The second possible contraction \eqref{redefsl3iiexp}, yields once again \eqref{IWpoin1} and \eqref{IWmax1}, but the higher-spin extension is given by \eqref{IWpoin3} together with
\be\label{IWmax3}
\bal
\bmY_{ab}&= \bmZ_{ab}^\r+\bX_{ab}^{(2)},
&
\bmZ_a &=  \frac{1}{c}\bmZ_{a0}^\r,
&
\bmT_{ab}&=  \frac{1}{2c^{2}}\left(\bmZ_{ab}^\r-\bX_{ab}^{(2)}\right),
\eal
\ee
and leads to $\hsmii$ in the limit $c\rightarrow\infty$. Therefore, the spin-$3$ algebras \eqref{hsmiandii} follow from the NR limits
\be\label{contractionshsmax}
\bal
&\hsrelm\oplus \mathfrak{e\textnormal{-}max_{3}}\;&
\underset{c\rightarrow\infty}{\longrightarrow}& \qquad \hsmi
\\[6pt]
&\hsrelm \oplus \mathfrak{max_{2+1}}\oplus \uo\oplus\uo\oplus\uo\;&\underset{c\rightarrow\infty}{\longrightarrow}&\qquad \hsmii.
\eal
\ee
The relations \eqref{IWpoin1} and \eqref{IWmax1} show that, in absence of spin-$3$ generators, the standard contraction of the Maxwell algebra in three space-time dimensions and three Abelian generators that yields the NR Maxwell algebra \cite{Aviles:2018jzw} is recovered. 

\subsection{Spin-$3$ Newton-Hooke algebras}
\label{sectionNH}

One can include a cosmological constant in the Bargman algebra by considering a NR contraction of the (anti-)deSitter algebra \cite{Bacry:1968zf}. This leads to the Newton-Hooke algebra, which in three space-time dimensions admits two central extensions \cite{Hartong:2016yrf} and is given by the extended Bargmann commutation relations \eqref{ExtBargAlg} plus
\be\label{generatorsnh}
\left[\bH,\bP_a\right]= \epsilon_{ab}\bG_b,\quad
\left[\bP_a,\bP_b\right]=- \epsilon_{ab}\;\bS.
\ee
The Extended Newton-Hooke algebra is isomorphic to the expanded Nappi-Witten algebra $S_M ^{(1)}\times \nw$. Thus, we can construct the corresponding spin-$3$ extensions by evaluating expanded algebras
\be\label{hsniandii}
\hsni=S^{(1)}_M \times \hsnwi,\qquad
\hsnii = S^{(1)}_M \times \hsnwii.
\ee
We find the explicit form of their commutation relations by using the product law of the semigroup $S^{(1)}_M\backsimeq\mathbb{Z}_2$ given in \eqref{smn} for $n=1$.
\begin{itemize}
\item We define the first higher-spin extension of the Extended Newton-Hooke algebra in $2+1$ dimensions algebra, $\hsni$, by defining the higher-spin generators \eqref{hsBarg1RenameGen}, which  yields the commutation relations \eqref{ExtBargAlg}, \eqref{BargHS1}, \eqref{BargHS2}, \eqref{generatorsnh} and
\be\label{NewtonHookeHS}
\bal
&\left[\bH,\bmH_a\right]  =  \epsilon_{ab}\bmJ_b,&
&\left[\bH,\bmP_{ab}\right] =  -  \epsilon_{c(a}\bmG_{b)c},&
&\left[\bH,\bmM_a\right] =\left[\bM,\bmH_a\right]  =  \epsilon_{ab}\bmS_b,
\\[5pt]
&\left[\bP_a,\bmH_b\right] = -  \epsilon_{a(c}\bmG_{b)c},&
&\left[\bP_a,\bmP_{bc}\right]  =  -  \epsilon_{a(b}\bmS_{c)}, &
&\left[\bmH_a,\bmH_b\right]  =   \epsilon_{ab}\bJ,
\\[5pt]
&\left[\bmH_a, \bmM_b\right] =  \epsilon_{ab}\bS,&
&\left[\bmH_a,\bmP_{bc}\right]  = \delta_{a(b}\epsilon_{c)d}\bG_d,&
&\left[\bmP_{ab},\bmP_{cd}\right] =  \delta_{(a(c}\epsilon_{d)b)}\bS.
\eal
\ee
\item The commutation relations of the second higher-spin extension $\hsnii$ are obtained by considering the expanded generators \eqref{hsBarg2RenameGen}. The result is the relations \eqref{ExtBargAlg}, \eqref{BargHS1T}, \eqref{BargHS2T}, \eqref{generatorsnh} plus
\be\label{NewtonHookeHST}
\bal
&[\bH,\bmP_a ] =\epsilon_{ab}\bmG_b,&
&[\bH,\bmH_{ab}] =-\epsilon_{c(a}\bmJ_{b)c},&
&[\bH,\bmM_{ab}] =[\bM,\bmH_{ab}] =-\epsilon_{c(a}\bmS_{b)c},
\\[5pt]
&[\bP_a ,\bmP_b ]=-\epsilon_{a(c}\bmS_{b)c},&
&[\bP_a ,\bmH_{bc}]=-\epsilon_{a(b}\bmG_{c)},&
&[\bmH_{ab},\bmH_{cd}] =\delta_{(a(c}\epsilon_{d)b)} \bJ,
\\[5pt]
&[\bmP_a ,\bmP_b ] =\epsilon_{ab}\bS,&
&[\bmP_a ,\bmH_{bc}] =\delta_{a(b}\epsilon_{c)d}\bG_d,&
&[\bmH_{ab},\bmM_{cd}] =\delta_{(a(c}\epsilon_{d)b)} \bS.
\eal
\ee
\end{itemize}

\noindent\textbf{Direct sum structure}: The Extended Newton-Hooke algebra defined by \eqref{ExtBargAlg} and \eqref{generatorsnh} is isomorphic to a direct product of two copies of the Nappi-Witten algebra \cite{Alvarez:2007ys,Hartong:2017bwq,Joung:2018frr}, which is a NR analog of the isomorphism $\mathfrak{so}(2,2)\backsimeq\sltw\oplus\sltw$. Similarly, in the spin-$3$ case, the following isomorphisms hold
\be\bal\label{hsnhiso}
&\hsni \backsimeq \hsnwi\oplus \hsnwi,\\
&\hsnii \backsimeq \hsnwii\oplus \hsnwii.
\eal
\ee
Indeed, the combinations
\be\label{2nw}
\bk^{\pm}=\frac{1}{2}\left( \bJ \pm  \bH \right),\quad
\bk^{\pm}_a=\frac{1}{2}\left( \bG_a \pm  \bP_a \right),\quad
\bsigma^{\pm}=\frac{1}{2}\left( \bS \pm \bM \right),
\ee
satisfy the $\nw$ commutation relations \eqref{nwalg}. The isomorphism \eqref{hsnhiso} can be easily proven by using change of basis \eqref{2nw} and the higher-spin generalization
\be\label{2hsnw1}
\bmK^{\pm}_a=\frac{1}{2}\left( \bmJ_a \pm  \bmH_a \right),\quad
\bmK^{\pm}_{ab}=\frac{1}{2}\left( \bmG_{ab} \pm  \bmP_{ab} \right),\quad
\bSigma^{\pm}_a=\frac{1}{2}\left( \bmS_a \pm \bmM_a \right),
\ee
in the $\hsni$ algebra. Instead, supplementing \eqref{nwalg} with
\be\label{2hsnw2}
\bmKt^{\pm}_{ab}=\frac{1}{2}\left( \bmJ_{ab} \pm  \bmH_{ab} \right),\quad
\bmKt^{\pm}_a=\frac{1}{2}\left( \bmG_a \pm  \bmP_a\right),\quad
\bSigmat^{\pm}_{ab}=\frac{1}{2}\left( \bmS_{ab} \pm \bmM_{ab} \right),
\ee
does the job for the $\hsnii$ algebra. It is important to remark that all the previous results consider a negative cosmological constant (the Newton-Hooke symmetry in our analysis is strictly speaking Newton-Hooke$^-$). The results can be generalized to the case of a positive cosmological constant by redefining the generators in the form
\be\label{dscase1}
\bal
&\bH\rightarrow i\bH,&\quad &\bP_a\rightarrow i\bP_a,&\quad&\bM\rightarrow i\bM,\\
&\bmH_a\rightarrow i\bmH_a,&\quad &\bmP_{ab}\rightarrow i\bmP_{ab},&\quad&\bmM_a\rightarrow i\bmM_a,\\
&\bmH_{ab}\rightarrow i\bmH_{ab},&\quad &\bmP_a\rightarrow i\bmP_a,&\quad&\bmM_{ab}\rightarrow i\bmM_{ab}.
\eal
\ee

\noindent\textbf{NR and flat limits}: As in the Bargmann case, we can obtain \eqref{hsniandii} by taking a NR limit of suitable extensions of the spin-$3$ AdS algebra \cite{Campoleoni:2010zq}
\be\label{hsadsiso}
\hsrela\simeq \sltr\oplus\sltr.
\ee
Indeed, it is clear that one can obtain \eqref{hsnhiso} from \eqref{hsadsiso} using two copies of the contractions developed in Section \ref{HSNW} that take $\sltr$ into the NR algebras $\hsnwi$ and $\hsnwii$. Moreover, \eqref{hsadsiso} is equivalent to the expansion
\be\label{hsadsiso2}
\hsrela\simeq S^{(1)}_M\times \sltr,
\ee
introduced in \cite{Caroca:2017izc}, where the commutation relations can be read off from \eqref{sl3exp} using the product law of the semigroup $S^{(1)}_M$ and expanded generators \eqref{hspoingen} defined in the Poincar\'e case. Thus, one can define the expanded algebras \eqref{expreli} and \eqref{exprelii} for $S=S^{(1)}_M$ and take the NR limit to find \eqref{hsnhiso}. This requires to extend \eqref{hsadsiso2} with the algebras \eqref{so3exp} and \eqref{sl2xu1exp}, which we can identify using the following isomorphisms (see \eqref{diagexpso3}, \eqref{diagexpsltw} and \eqref{expuo}):
\be\label{otherexpsads}
\bal
&S^{(1)}_M\times \sotr  \simeq \sotr\oplus\sotr  \simeq\mathfrak{so}(4),\\
&S^{(1)}_M\times\Big( \sltwu\Big) \simeq \mathfrak{so}(2,2)\oplus \uo\oplus\uo.
\eal
\ee

As shown in Section \ref{flatlimit}, one can define a vanishing cosmological constant limit that takes \eqref{hsniandii} into \eqref{hsbiandii} by means of the rescaling \eqref{recalingsell}, i.e. to redefine
\eqref{BargRenameGen}, \eqref{hsBarg1RenameGen} and \eqref{hsBarg2RenameGen}
as
\be\label{rescaleP's}
\bal
&\bH\rightarrow\ell\,\bH,& &\bP_a\rightarrow\ell\,\bP_a, & &\bM\rightarrow\ell\,\bM,\\
&\bmH_a\rightarrow\ell\,\bmH_a,&
&\bmP_{ab}\rightarrow\ell\,\bmP_{ab},&
&\bmM_a\rightarrow\ell\,\bmM_a,\\
&\bmH_{ab}\rightarrow\ell\,\bmH_{ab},&  
&\bmP_a\rightarrow\ell\,\bmP_a,& & \bmM_{ab}\rightarrow\ell\,\bmM_{ab}.
\eal
\ee
This has the effect of adding a factor $1/\ell^2$ on the right-hand side of the commutation relations given in Eqs. \eqref{generatorsnh}, \eqref{NewtonHookeHS} and \eqref{NewtonHookeHST}, so that they dissapear in the limit $\ell\rightarrow\infty$. Similarly, the contraction defined by \eqref{recalingsellRel}, \eqref{rescellX1} and \eqref{rescellX2} in this case takes the form
\be\bal\label{reslcaleP'sRel}
&\bP_A^{\rm rel}\rightarrow \ell\,\bP_A^{\rm rel},&\quad&
\bmP_{AB}^{\rm rel}\rightarrow \ell\,\bmP_{AB}^{\rm rel}\\
&\bX^{(1)}\rightarrow \ell\,\bX^{(1)},&\quad&
\bX_a^{(1)}\rightarrow \ell\,\bX_a^{(1)},\quad
\bX_{ab}^{(1)}\rightarrow \ell\,\bX_{ab}^{(1)}
\eal\ee
and defines an analog contraction at the relativistic level, taking \eqref{hsadsiso2} and \eqref{otherexpspoin} into \eqref{hspoin2} and \eqref{otherexpsads}, respectively. Thus, the NR contraction leading to $\hsni$ together with the corresponding flat limit is
\be\label{fulldiagram1ads}
\bal
&\hsrela\oplus \mathfrak{so}(4)
\;
&\underset{c\rightarrow\infty}{\longrightarrow}& \qquad \hsni
\\
&\hskip1.0truecm {\scriptstyle \ell\rightarrow\infty}\;\Big\downarrow  & &\hskip.9truecm {\scriptstyle \ell\rightarrow\infty}\;\Big\downarrow\\
&\hsrelp\oplus \mathfrak{iso}(3)
\;
&\underset{c\rightarrow\infty}{\longrightarrow}& \qquad \hsbi,
\eal
\ee
whereas the corresponding diagram for $\hsnii$ has the form
\be\label{fulldiagram2ads}
\bal
&\hsrela \oplus \mathfrak{so}(2,2)\oplus\uo\oplus\uo
\;
&\underset{c\rightarrow\infty}{\longrightarrow}& \qquad \hsnii
\\
&\hskip2.3truecm {\scriptstyle \ell\rightarrow\infty}\;\Big\downarrow  & &\hskip.9truecm {\scriptstyle \ell\rightarrow\infty}\;\Big\downarrow\\
&\hsrelp \oplus \mathfrak{isl}(2,\mathbb R)\oplus \uo\oplus\uo
\;
&\underset{c\rightarrow\infty}{\longrightarrow}& \qquad \hsbii.
\eal
\ee
In both cases, the explicit relations of the $c\rightarrow\infty$ contractions are the same as the ones given in the Poincar\'e case \eqref{IWpoin1}-\eqref{IWpoin3}.


\subsection{Spin-$3$ non-relativistic AdS-Lorentz algebras}
As the last example, we turn our attention to the AdS-Lorentz symmetry. Based on the fact that NR AdS-Lorentz algebra is the expanded Nappi-Witten algebra \eqref{expNW} for $S=S^{(2)}_M$, we define the corresponding spin-$3$ extensions as
\be\label{hsaliandii}
\hsali= S^{(2)}_M\times \hsnwi, \qquad
\hsalii= S^{(2)}_M \times \hsnwii.
\ee
The relativistic AdS-Lorentz algebra is a generalization of the Maxwell algebra that provides a semi-simple extension of the Poincar\'e algebra \cite{Soroka:2006aj,Gomis:2009dm}, and whose NR limit has been considered in the context of higher-dimensional (extended) CS theories \cite{Gonzalez:2016xwo,Rubio:2018itx}. In $2+1$dimensions it admits three central extensions and its commutation relations are the ones given in \eqref{ExtBargAlg} and \eqref{ExtNRMaxAlg}, plus
\be\label{nonreladslalg}
\bal
&\left[\bY,\bZ_a\right]= \epsilon_{a}^{\;\;b} \bZ_b\,,&\quad&
\left[\bP_a,\bZ_b\right]=- \epsilon_{ab} \bM,\\[5pt]
&\left[\bH,\bZ_a\right]= \left[\bY,\bP_a\right]= \epsilon_{a}^{\;\;b}\bP_b,&\quad&
 \left[\bZ_a,\bZ_b\right]=- \epsilon_{ab}\bT.
\eal
\ee
As said before, this algebra follows from evaluating the expanded Nappi-Witten algebra \eqref{expNW} for the semigroup $S^{(n)}_M$ defined in \eqref{smn}, and setting $n=2$. The same procedure allows one to find the commutation relations of the higher-spin extensions \eqref{hsaliandii}.
\begin{itemize}
\item The first higher-spin generalization of the NR AdS-Lorentz algebra $\hsali$  includes the higher-spin generators \eqref{hsBarg1RenameGen} and \eqref{hsMaxwell1RenameGen}. In this case the commutation relations are the ones of the $\hsmi$ algebra,  \eqref{ExtBargAlg}, \eqref{BargHS1}, \eqref{BargHS2}, \eqref{ExtNRMaxAlg}, \eqref{MaxHS}, together with \eqref{nonreladslalg} plus
\be\label{AdSLHS1}
\bal
&\left[\bY,\bmY_a\right]  =   \epsilon_{ab}\bmY_b,
&
&\left[\bY,\bmZ_{ab}\right] =  -  \epsilon_{c(a}\bmZ_{b)c},&
&\left[\bY,\bmT_a\right] =\left[\bT,\bmY_a\right]  =   \epsilon_{ab}\bmT_b,
\\[5pt]
&\left[\bZ_a,\bmY_b\right] = -  \epsilon_{a(c}\bmZ_{b)c},
&
&\left[\bZ_a,\bmZ_{bc}\right]  =  -  \epsilon_{a(b}\bmT_{c)}, &
&\left[\bmY_a,\bmY_b\right]  =    \epsilon_{ab}\bY,
\\[5pt]
&\left[\bmY_a, \bmT_b\right] =   \epsilon_{ab}\bT,&
&\left[\bmY_a,\bmZ_{bc}\right]  =  \delta_{a(b}\epsilon_{c)d}\bZ_d,&
&\left[\bmZ_{ab},\bmZ_{cd}\right] =   \delta_{(a(c}\epsilon_{d)b)}\bT,
\eal
\ee
and
\be\label{AdSLHS2}
\bal
&\left[\bY,\bmH_a\right]=\left[\bH,\bmY_a\right]   = \epsilon_{ab}\bmH_b,&
&\left[\bZ_a,\bmH_b\right] = \left[\bP_a,\bmY_b\right] = -\epsilon_{a(c}\bmP_{b)c},
\\[5pt]
&\left[\bY,\bmP_{ab}\right]=\left[\bH,\bmZ_{ab}\right]  =  -\epsilon_{c(a}\bmP_{b)c},&
&\left[\bmY_a,\bmP_{bc}\right]  =\left[\bmH_a,\bmZ_{bc}\right]   =\delta_{a(b}\epsilon_{c)d}\bP_d,
\\[5pt]
&\left[\bmY_a,\bmH_b\right]  =  \epsilon_{ab}\bH,&
&\left[\bZ_a,\bmP_{bc}\right]  =  \left[\bP_a,\bmZ_{bc}\right]  =-\epsilon_{a(b}\bmM_{c)},
\\[5pt]
&\left[\bmY_a, \bmM_b\right] =\left[\bmH_a, \bmT_b\right]= \epsilon_{ab}\bM,&
&\left[\bY,\bmM_a\right] =\left[\bT,\bmH_a\right]  = \left[\bH,\bmT_a\right]=\left[\bM,\bmY_a\right]  =\epsilon_{ab}\bmM_b.
\\[5pt]
&\left[\bmZ_{ab},\bmP_{cd}\right] = \delta_{(a(c}\epsilon_{d)b)}\bM,
\eal
\ee

\item The second higher-spin extension of the NR AdS-Lorentz algebra $\hsalii$  includes the higher-spin generators \eqref{hsBarg2RenameGen} and \eqref{hsMaxwell2RenameGen} and has the commutation relations of the $\hsmii$ algebra,  \eqref{ExtBargAlg}, \eqref{BargHS1T}, \eqref{BargHS2T}, \eqref{ExtNRMaxAlg}, \eqref{MaxHST}, together with \eqref{nonreladslalg} plus
\be\label{AdSLHS1T}
\bal
&[\bY,\bmZ_a ] =\epsilon_{ab}\bmZ_b,&
&[\bY,\bmY_{ab}] =-\epsilon_{c(a}\bmY_{b)c},&
&[\bY,\bmT_{ab}] =[\bT,\bmY_{ab}] =-\epsilon_{c(a}\bmT_{b)c},
\\[5pt]
&[\bZ_a ,\bmZ_b ]=-\epsilon_{a(c}\bmT_{b)c},&
&[\bZ_a ,\bmY_{bc}]=-\epsilon_{a(b}\bmZ_{c)},&
&[\bmY_{ab},\bmY_{cd}] =\delta_{(a(c}\epsilon_{d)b)} \bY.
\\[5pt]
&[\bmZ_a ,\bmZ_b ] =\epsilon_{ab}\bT,&
&[\bmZ_a ,\bmY_{bc}] =\delta_{a(b}\epsilon_{c)d}\bZ_d,&
&[\bmY_{ab},\bmT_{cd}] =\delta_{(a(c}\epsilon_{d)b)} \bT,
\eal
\ee
and
\be\label{AdSLHS2T}
\bal
&[\bY,\bmP_a ]=[\bH,\bmZ_a ] =\epsilon_{ab}\bmP_b,&
&[\bZ_a ,\bmH_{bc}]=[\bP_a ,\bmY_{bc}]=-\epsilon_{a(b}\bmP_{c)},
\\[5pt]
&[\bY,\bmH_{ab}] =[\bH,\bmY_{ab}] =-\epsilon_{c(a}\bmH_{b)c},&
&[\bmZ_a ,\bmH_{bc}] =[\bmP_a ,\bmY_{bc}] =\delta_{a(b}\epsilon_{c)d}\bP_d,
\\[5pt]
&[\bmY_{ab},\bmH_{cd}] =\delta_{(a(c}\epsilon_{d)b)} \bH,&
&[\bmY_{ab},\bmM_{cd}] =[\bmH_{ab},\bmT_{cd}] =\delta_{(a(c}\epsilon_{d)b)} \bM,
\\[5pt]
&[\bmZ_a ,\bmP_b ] =\epsilon_{ab}\bM,&
&[\bY,\bmM_{ab}] =[\bH,\bmT_{ab}] =[\bT,\bmH_{ab}] 
\\[5pt]
&[\bZ_a ,\bmP_b ]=[\bP_a ,\bmZ_b ]=-\epsilon_{a(c}\bmM_{b)c},&
&\hskip1.7truecm =[\bM,\bmY_{ab}] =-\epsilon_{c(a}\bmM_{b)c}.
\eal
\ee
\end{itemize}

\noindent\textbf{Direct sum structure}: The AdS-Lorentz symmetry in $d+1$ dimensions was so dubbed because it is isomorphic to the direct $\mathfrak{so}(d,2)\oplus\mathfrak{so}(d,1)$ \cite{Gomis:2009dm}. Along the same lines, the NR AdS-Lorentz algebra is isomorphic to the direct sum of the Extended Newton-Hooke algebra and the Nappi-Witten algebra, which in turn is equivalent to three copies of the $\nw$ algebra. The spin-$3$ generalization of this statement reads
\be\bal
&\hsali \backsimeq \hsni\oplus \hsnwi \backsimeq \hsnwi\oplus \hsnwi\oplus \hsnwi,\\
&\hsalii \backsimeq \hsnii\oplus \hsnwii \backsimeq \hsnwii\oplus \hsnwii\oplus \hsnwii,
\eal
\ee
where, on the right-hand side of this relation, we have used \eqref{hsnhiso} (This is the NR analog of the fact that, in $2+1$  dimensions, the AdS-Lorentz algebra is isomorphic to the direct sum of three copies of $\sltw$ \cite{Concha:2018jjj}). This can be shown by considering the change of basis
\begin{align}
& \hat{\bJ}= \bY ,&\quad&\hat{\bG}_a= \bZ_a,&\quad&\hat{\bS}= \bT,\nonumber \\ 
& \hat{\bmJ}_a= \bmY_a ,&\quad&\hat{\bmG}_{ab}= \bmZ_{ab},&\quad&\hat{\bmS}_a= \bmT_a, \label{redefadsl1}\\
& \hat{\bmJ}_{ab}= \bmY_{ab} ,&\quad&\hat{\bmG}_a= \bmZ_a,&\quad&\hat{\bmS}_{ab}= \bmT_{ab},\nonumber 
\\[9pt]
& \hat{\bH}= \bH ,&\quad&\hat{\bP}_a= \bP_a,&\quad& \hat{\bM}= \bM,\nonumber \\ 
& \hat{\bmH}_a= \bmH_a ,&\quad&\hat{\bmP}_{ab}= \bmP_{ab},&\quad& \hat{\bmM}_a= \bmM_a, \label{redefadsl2}\\
& \hat{\bmH}_{ab}= \bmH_{ab} ,&\quad&\hat{\bmP}_a= \bmP_a,&\quad& \hat{\bmM}_{ab}= \bmM_{ab},\nonumber 
\\[9pt]
&  \hat{\bk}= \bJ - \bY  ,&\quad&\hat{\bk}_a= \bG_a- \bZ_a ,&\quad&   \hat{\bsigma}=\bS- \bT, \nonumber \\ 
&  \hat{\bmK}_a= \bmJ_a - \bmY_a  ,&\quad&\hat{\bmK}_{ab}= \bmG_{ab}- \bmZ_{ab} ,&\quad&   \hat{\bSigma}_a=\bmS_a- \bmT_a, \label{redefadsl3} \\ 
&  \hat{\bmKt}_{ab}= \bmJ_{ab} - \bmY_{ab}  ,&\quad&\hat{\bmKt}_a= \bmG_a- \bmZ_a ,&\quad&   \hat{\bSigmat}_{ab}=\bmS_{ab}- \bmT_{ab}.  \nonumber
\end{align}
The first and second lines in Eqs. \eqref{redefadsl1}-\eqref{redefadsl3} take the $\hsali$ algebra  into the direct sum of $\hsni$ and $\hsnwi$, whereas the first and third lines in Eqs. \eqref{redefadsl1}-\eqref{redefadsl3} do the analog job for $\hsalii$. Clearly, one can apply the change of basis \eqref{2nw}-\eqref{2hsnw2} to the hatted generators to end up with three copies of a spin-$3$ Nappi-Witten algebra in each case. The previous results can be extended to the case of a positive cosmological constant by using the redefinition \eqref{dscase1} together with 
\be
\bal
&\bY\rightarrow  -\bY,&\quad &\bZ_a\rightarrow  -\bZ_a,&\quad&\bT\rightarrow-  \bT,\\
&\bmY_a\rightarrow -\bmY_a,&\quad &\bmZ_{ab}\rightarrow -\bmZ_{ab},&\quad&\bmT_a\rightarrow -\bT_a,\\
&\bmY_{ab}\rightarrow -\bmY_{ab},&\quad &\bmZ_a\rightarrow -\bmZ_a,&\quad&\bmT_{ab}\rightarrow -\bT_{ab}.
\eal
\ee

\noindent\textbf{NR and flat limits}: We can construct relativistic symmetries that reproduce \eqref{hsaliandii} in the limit $c\rightarrow\infty$ starting from the higher-spin extension of the AdS-Lorentz symmetry in $2+1$ dimensions \cite{Caroca:2017izc}
\be\label{hsadsL}
\hsrelal=S^{(2)}_M\times \sltr.
\ee
The commutation relations of \eqref{hsadsL} can be read off from Eq. \eqref{sl3exp} with the same definition \eqref{hspoingen} and \eqref{hsmaxgen} for the expanded generators as the Maxwellian case. We extend this symmetry to define the algebras \eqref{expreli} and \eqref{exprelii} for $S=S^{(2)}_M$, which requires introducing
\be\label{otherexpsadslor}
\bal
&S^{(2)}_M\times \sotr \simeq \mathfrak{e\textnormal{-}adslor_{3}}\simeq \sotr\oplus\sotr\oplus\sotr,\\
&S^{(2)}_M\times\Big( \sltwu\Big) \simeq \mathfrak{adslor_{2+1}}\oplus \uo\oplus\uo\oplus\uo.
\eal
\ee
The NR contractions have the same form as the one found in the Maxwell case, which follows from the general expressions \eqref{redefsl3iexp} and \eqref{redefsl3iiexp}, and are explicitly given by \eqref{IWpoin1}, \eqref{IWpoin2}, \eqref{IWmax1}, \eqref{IWmax2}, and \eqref{IWpoin1}, \eqref{IWpoin3}, \eqref{IWmax1}, \eqref{IWmax3}, respectively.

The previous results can be connected with the ones found in Section \ref{NRMax} for the Maxwell case by means of a contraction analog to the vanishing cosmological limit that relates the Extended Newton-Hooke and the Bargmann symmetry. The flat limit in this case extends the relations \eqref{rescaleP's} by introducing the extra rescalings
\be\label{rescaleZ's}
\bal
&\bY\rightarrow\ell^2\,\bY,& &\bZ_a\rightarrow\ell^2\,\bZ_a, & &\bT\rightarrow\ell^2\,\bT,\\
&\bmY_a\rightarrow\ell^2\,\bmY_a,&
&\bmZ_{ab}\rightarrow\ell^2\,\bmZ_{ab},&
&\bmT_a\rightarrow\ell^2\,\bmT_a,\\
&\bmY_{ab}\rightarrow\ell^2\,\bmY_{ab},&  
&\bmZ_a\rightarrow\ell^2\,\bmZ_a,& 
& \bmT_{ab}\rightarrow\ell^2\,\bmT_{ab},
\eal
\ee
as defined in \eqref{recalingsell}, which sets all the commutators given in Eqs. \eqref{nonreladslalg}-\eqref{AdSLHS2T} to zero in the limit $\ell^2\rightarrow\infty$. For the relativistic generators we use the rescaling \eqref{recalingsellRel}, \eqref{rescellX1} and \eqref{rescellX2}, which lead to \eqref{reslcaleP'sRel} plus
\be\bal\label{reslcaleZ'sRel}
&\bZ_A^{\rm rel}\rightarrow \ell^2\,\bZ_A^{\rm rel},&\quad&
\bmZ_{AB}^{\rm rel}\rightarrow \ell^2\,\bmZ_{AB}^{\rm rel},\\
&\bX^{(2)}\rightarrow \ell^2\,\bX^{(2)},&\quad&
\bX_a^{(2)}\rightarrow \ell^2\,\bX_a^{(2)},\quad
\bX_{ab}^{(2)}\rightarrow \ell^2\,\bX_{ab}^{(2)}.
\eal\ee
Thus, in the case of the $\hsali$ algebra we find the following diagram analog to \eqref{fulldiagram1ads}\be\label{fulldiagram1adslor}
\bal
&\hsrelal\oplus\sotr\oplus\sotr\oplus\sotr
\;
&\underset{c\rightarrow\infty}{\longrightarrow}& \qquad \hsali
\\
&\hskip1.5truecm {\scriptstyle \ell\rightarrow\infty}\;\Big\downarrow  & &\hskip.9truecm {\scriptstyle \ell\rightarrow\infty}\;\Big\downarrow\\
&\hsrelm\oplus \mathfrak{e\textnormal{-}max_{3}}
\;
&\underset{c\rightarrow\infty}{\longrightarrow}& \qquad \hsmi,
\eal
\ee
while in the case of $\hsalii$ the corresponding diagram, analog to \eqref{fulldiagram2ads}, has the form
\be\label{fulldiagram2adslor}
\bal
&\hsrelal \oplus \mathfrak{adslor_{2+1}}\oplus \uo\oplus\uo\oplus\uo
\;
&\underset{c\rightarrow\infty}{\longrightarrow}& \qquad \hsalii
\\
&\hskip2.3truecm {\scriptstyle \ell\rightarrow\infty}\;\Big\downarrow  & &\hskip.9truecm {\scriptstyle \ell\rightarrow\infty}\;\Big\downarrow\\
&\hsrelm \oplus \mathfrak{max_{2+1}}\oplus \uo\oplus\uo\oplus\uo
\;
&\underset{c\rightarrow\infty}{\longrightarrow}& \qquad \hsmii.
\eal
\ee
In both cases, the explicit relations of the $c\rightarrow\infty$ contractions are the same as the ones given in the Poincar\'e case \eqref{IWpoin1}-\eqref{IWpoin3}.



\subsection{Higher-spin gravity theories}
\label{HSG}

Higher-spin gravity actions invariant under the spin-$3$ symmetries presented above can be constructed using the definitions \eqref{expAs} and \eqref{expFs} in \eqref{CSform}, together with the invariant tensors defined by \eqref{itnwexp}, \eqref{itexp1} and \eqref{itnwexp}, \eqref{itexp2}, which correspond to the expansions $S\times\hsnwi$ and $S\times\hsnwii$, respectively. A considerable easier task is to use the results \eqref{hsCS1} and \eqref{hsCS2} to simply evaluate the semigroup product law $i\diamond j$ for the semigroups under consideration. Here we adopt the latter strategy and describe the steps to be followed in order to find the result. Since the form of the resulting CS forms turns out to be rather cumbersome, the explicit form of the Lagrangians and the details of the derivations are given in Appendix \ref{appA}.

\begin{itemize}
\item \textbf{Spin-$3$ Bargmann algebras}: the action \eqref{CSNWexp} for $S=S_E^{(1)}$ is the one of Extended Bargmann gravity \cite{Papageorgiou:2009zc,Bergshoeff:2016lwr,Hartong:2016yrf,Penafiel:2016ufo} with gauge symmetry \eqref{ExtBargAlg}, and thus, the following gauge fields are identified
\be\label{changeFieldsb}
\bal
 \alpha_{(0)}= w, &\quad&&\theta_{(0)}^a= \omega^a, &\quad&
\beta_{(0)} =s,\\
\alpha_{(1)}= \tau, &\quad& &\theta_{(1)}^a= e^a, &\quad&\beta_{(1)}= m,\\
\eal
\ee 
where $w$ and $\omega^a$ are the spin connections associated with rotations and Galilean boosts, respectively, $\tau$ is the clock form, $e^a$ is the spatial dreibein, and $m$ and $s$ are the gauge fields that go along the mass and the spin generators.
The higher-spin extensions \eqref{hsCS1} and \eqref{hsCS2} introduce a spin-$3$ version of each generator given in Eq. \eqref{changeFieldsb}. It is therefore natural to adopt the notation
\be\label{changeFieldsb1}
\bal
& A^a_{(0)}= W^a,&\quad&\Theta_{(0)}^{ab}= \Omega^{ab},&\quad&
B^a_{(0)} =S^a,\\
& A^a_{(1)}= V^a,&\quad&\Theta_{(1)}^{ab}= E^{ab},&\quad& B^a_{(1)}= M^a,
\eal
\ee
when defining the gauge fields associated to the $\hsbi$ algebra. Similarly, when considering the $\hsbii$ symmetry, one can define the spin-$3$ extension of \eqref{changeFieldsb} as
\be\label{changeFieldsb2}
\bal
&A^{ab}_{(0)}= W^{ab},&\quad &\Theta_{(0)}^a= \Omega^a,&\quad&
B^{ab}_{(0)} =S^{ab},\\
&A^{ab}_{(1)}= V^{ab},&\quad&\Theta_{(1)}^a= E^a, &\quad& B^{ab}_{(1)}= M^{ab}.
\eal
\ee 
The explicit form of the spin-$3$ CS gravity Lagrangians \eqref{hsCS1} and \eqref{hsCS2} for $S^{(1)}_E$ follows from Eqs. \eqref{hsCSNH1} and \eqref{hsCSNH2} of Appendix \ref{appNHB} after taking the limit $\ell\rightarrow\infty$.

\item \textbf{Spin-$3$ Maxwell algebras}: considering $S^{(2)}_E$ in the CS form \eqref{CSNWexp} yields a NR Maxwellian CS gravity theory \cite{Aviles:2018jzw,Penafiel:2019czp} for the gravitational gauge fields \eqref{changeFieldsb} and the Maxwellian gauge fields
\be\label{changeFieldsm}
\alpha_{(2)}= y,\quad \theta_{(2)}^a= k^a,\quad\beta_{(2)}= t. 
\ee
One can then extend such theory by evaluating the Lagrangian \eqref{hsCS2} for $S^{(2)}_E$, which yields a CS action with $\hsmi$ gauge symmetry that includes the spin-$3$ gauge fields \eqref{changeFieldsb2} plus 
\be\label{changeFieldsm1}
A^a_{(2)}= Y^a,\quad \Theta_{(2)}^{ab}= K^{ab},\quad B^a_{(2)}= T^a.
\ee
Similarly, one can write the $\hsmii$-invariant CS theory coming from \eqref{hsCS2} in terms of \eqref{changeFieldsb2} and
\be\label{changeFieldsm2}
A^{ab}_{(2)}= Y^{ab},\quad \Theta_{(2)}^a= K^a,\quad B^{ab}_{(2)}= T^{ab}.
\ee
These are respectively given in Eqs. \eqref{hsCSNRAdSL1} and \eqref{hsCSNRAdSL2} of Appendix \ref{appAdSLM} when the limit $\ell\rightarrow\infty$ is taken.
\item \textbf{Spin-$3$ Newton-Hooke algebras}: replacing $S=S^{(1)}_M$ in the CS form \eqref{CSNWexp} yields Extended Newton-Hooke CS gravity \cite{Papageorgiou:2010ud,Hartong:2016yrf,Penafiel:2016ufo}. 
When considering the extensions $\hsni$ and $\hsnii$, the CS forms \eqref{hsCS1} and \eqref{hsCS2} provide actions involving the same fields \eqref{changeFieldsb}-\eqref{changeFieldsb2} as the Bargmann case. One can implement the rescaling \eqref{rescNRfields}, which now looks like
\be\label{changeFieldsn}
\bal
& \tau\rightarrow\ell^{-1}\, \tau,&\quad& e^a\rightarrow\ell^{-1}\, e^a,&\quad& m\rightarrow\ell^{-1}\, m,
\\
&V^a\rightarrow\ell^{-1}\, V^a,&\quad& E^{ab}\rightarrow\ell^{-1}\, E^{ab}, &\quad& M^a\rightarrow\ell^{-1}\, M^a,
\\
&V^{ab}\rightarrow\ell^{-1}\, V^{ab},&\quad& E^a\rightarrow\ell^{-1}\, E^a,&\quad& M^{ab}\rightarrow\ell^{-1}\, M^{ab},
\eal
\ee 
to find the CS forms given in Eqs. \eqref{hsCSNH1} and \eqref{hsCSNH2} of Appendix \ref{appNHB}.

\item \textbf{Spin-$3$ AdS-Lorentz algebras}: evaluating the action \eqref{CSNWexp} for $S^{(2)}_M$ yields the NR AdS-Lorentz CS gravity theory
first considered in Refs. \cite{Concha:2019lhn,Penafiel:2019czp}. The spin-$3$ extensions defined by and \eqref{hsCS1} and \eqref{hsCS2} lead to CS theories invariant under $\hsali$ and $\hsalii$, respectively. These involve the gauge fields \eqref{changeFieldsb}-\eqref{changeFieldsm2} appearing in the Maxwellian case that we previously considered, and reduces to it when applying the flat limit $\ell\rightarrow\infty$, provided the fields are rescaled as in Eq. \eqref{rescNRfields}, i.e.
\be\label{changeFieldsal}
\bal
&k^a\rightarrow\ell^{-2}\, k^a,&\quad& y\rightarrow\ell^{-2}\, y,&\quad& t\rightarrow\ell^{-2}\, t,
\\
&K^{ab}\rightarrow\ell^{-2}\, K^{ab},&\quad& Y^a\rightarrow\ell^{-2}\, Y^a,&\quad& T^a\rightarrow\ell^{-2}\, T^a,
\\
&K^a\rightarrow\ell^{-2}\, K^a,&\quad& Y^{ab}\rightarrow\ell^{-2}\, Y^{ab},&\quad& T^{ab}\rightarrow\ell^{-2}\, T^{ab}.
\eal
\ee
The explicit form of the CS forms is given in Eqs. \eqref{hsCSNRAdSL1} and \eqref{hsCSNRAdSL2} of Appendix \ref{appAdSLM}.
\end{itemize}

\noindent Naturally, all the CS forms here considered can be obtained from the corresponding relativistic CS forms constructed out of the expanded/extended $\sltr$ algebras appearing in \eqref{diagexpsltw}, where the rescaling of the relativistic gauge fields in terms of $c$ can be easily read off from \eqref{redefsl3iexp} and \eqref{redefsl3iiexp} by demanding that the relativistic gauge connections match the NR ones given in Eq. \eqref{expAs} .

\section{Concluding Remarks}
We have constructed NR higher-spin theories of Galilean type in $2+1$ space-time dimensions as expansions of particular higher-spin extensions of the Nappi-Witten algebras, namely, $\hsnwi$ and $\hsnwii$. These extended Nappi-Witten symmetries were obtained after applying In\"on\"u-Wigner contractions to suitable extensions of $\sltr$:
\be\label{discussion1}
\bal
&\reli\;&
\underset{c\rightarrow\infty}{\longrightarrow}& \qquad\hsnwi
\\[6pt]
&\relii\;&\underset{c\rightarrow\infty}{\longrightarrow}& \qquad\hsnwii.
\eal
\ee
The expansion method applied to these algebras gave rise to two infinite families of NR spin-$3$ symmetries, which are labeled by the coefficient $n$ in the semigroups $S^{(n)}_E$ and $S^{(n)}_M$ used in the expansion mechanism. These are
\begin{align}
&\textnormal{Generalized spin-}3\;\textnormal{Bargmann algebras}:& & 
S_E^{(n)}\times \hsnwiii \label{genbarg}\\[9pt]
&\textnormal{Generalized spin-}3\;\textnormal{Newton-Hooke algebras}:&&
S_M^{(n)}\times \hsnwiii.\; \label{gennh}
\end{align}
The first level of these families of algebras reproduces the higher-spin extensions of the Extended Bargmann and Newton-Hooke algebras in three space-time dimensions previously found in \cite{Bergshoeff:2016soe}, whereas the $n=2$ case yields novel spin-$3$ NR symmetries in $2+1$-dimensions, which come to extend the (exotic) NR Maxwell algebra \cite{Aviles:2018jzw} and the NR AdS-Lorentz algebra \cite{Penafiel:2019czp}. Greater values of $n$ correspond to spin-$3$ extensions of Generalized Galilean symmetries algebras \cite{Gonzalez:2016xwo,Penafiel:2019czp}. We have shown that these two families of NR spin-$3$ algebras are related by a vanishing cosmological constant limit, i.e.
\be
\bal
&S^{(1)}_M\times\hsnwiii\simeq\hsniii &\xrightarrow[\ell\rightarrow\infty]\, &\qquad S^{(1)}_E\times\hsnwiii\simeq\hsbiii\\
&S^{(2)}_M\times\hsnwiii\simeq\hsaliii&\xrightarrow[\ell\rightarrow\infty]\, &\qquad S^{(2)}_E\times\hsnwiii\simeq\hsmiii\\
&&\vdots&\\
&S^{(n)}_M\times\hsnwiii&\xrightarrow[\ell\rightarrow\infty]\, &\qquad S^{(n)}_E\times\hsnwiii.\\
\eal
\ee
Moreover, we have shown that all the algebras of this form define a NR version of an expanded $\sltr$ symmetry. The relation between relativistic and non-relativistic symmetries is however non-trivial. Indeed, the relativistic algebras needed to define a $c\rightarrow\infty$ contraction leading to a non-degenerate invariant tensor are semidirect sum extensions of the expanded $\sltr$ algebras considered in \cite{Caroca:2017izc}. Unlike the known contractions giving rise to double and triple central extension in the spin-$2$ case (see for example \cite{Bergshoeff:2016lwr,Aviles:2018jzw,Concha:2019lhn}), these extensions do not add only Abelian generators to the relativistic symmetries under consideration, but include Euclidean and Lorentzian symmetries in three dimensions (see \eqref{fulldiagram1ads}, \eqref{fulldiagram2ads} and \eqref{fulldiagram1adslor}, \eqref{fulldiagram2adslor}), which are expansions of the $\sotr$ or the $\sltwu$ algebras. The addition of such non-Abelian generators to the relativistic spin-$3$ algebras constructed in \cite{Caroca:2017izc} was not to be expected and provides a non-trivial generalization of the known spin-$2$ constructions. In this way, the expansion mechanism produces a non-degenerate invariant bilinear form for every NR expanded algebra, together with a CS higher-spin gravity action. The particular cases $n=1,2$ were explicitly worked out.
\\\\
\noindent An interesting question for future research is the generalization of our results to the case of NR spin-$N$ symmetries for $N>3$. Indeed the fact that CS forms based on the $\mathfrak{sl}(N)\times \mathfrak{sl}(N)$ algebra define more general higher-spin gravity theories in $2+1$ dimensions \cite{Campoleoni:2010zq} admitting a flat limit \cite{Campoleoni:2016vsh}, together with the structure of the algebras $\reli$ and $\relii$
suggest that spin-$N$ extensions of the Nappi-Witten algebra $\nw$ could be obtained as contractions of suitable extensions of $\mathfrak{sl}(N)$. A direct generalization of the structure \eqref{discussion1} signals the symmetries
\be\label{discussion2}
\mathfrak{sl}(N,\mathbb R)\oplus \mathfrak{so}(N)
,\qquad
\mathfrak{sl}(N,\mathbb R)
\oplus
\mathfrak{sl}(N-1,\mathbb R)
\oplus\cdots\oplus
\mathfrak{sl}(3,\mathbb R)\oplus
\mathfrak{sl}(2,\mathbb R)\oplus \mathfrak{so}(2),
\ee
as potential candidates to be expanded. One could as well explore extensions of $\mathfrak{sl}(N,\mathbb R)$ that look like
\be
\bal
\mathfrak{sl}(N,\mathbb R)
\oplus
\mathfrak{sl}(N-1,\mathbb R)
\oplus\cdots\oplus
\mathfrak{sl}(N-\mathfrak{m},\mathbb R)\oplus \mathfrak{so}(N-\mathfrak{m})
\;&\underset{c\rightarrow\infty}{\longrightarrow}& \quad\mathfrak{hs}_N \mathfrak{nw}(\mathfrak{1+m}),
\eal
\ee
and thus interpolate between the ones given in \eqref{discussion2}, which could lead to several possible higher-spin extensions of the Nappi-Witten algebra. We hope to explore this problem elsewhere.\\

\noindent Due to the recently discovered relations between NR geometry and condensed matter systems \cite{Duval:2001hu,Son:2008ye,Geracie:2014nka,Gromov:2014vla}, and in particular, the recent application of CS theories invariant under extended Nappi-Witten algebras to construct effective descriptions of the fractional quantum Hall effect \cite{Salgado-Rebolledo:2021wtf,Salgado-Rebolledo:2021nfj}, it would be interesting to consider the NR spin-$3$ CS theories here obtained in the description of collective excitations in quantum Hall systems along the lines of \cite{Cappelli:2015ocj}, where NR higher-spin fields in three space-time dimensions appear in the multipole expansion of the low-energy excitations of a quantum Hall fluid. Moreover, in \cite{Salgado-Rebolledo:2021nfj}, a supersymmetric extension of the Nappi-Witten algebra was considered in the description of the Moore-Read fractional quantum Hall state. Thus, supersymmetric extensions of the $\hsnwi$ and $\hsnwii$ algebras defined along the same lines can be of interest in the exploration of collective excitations of the fractional quantum Hall effect in terms of superpartners \cite{Gromov:2019cgu}. Furthermore, a natural generalization of our results is the inclusion of a space-time boundary and the construction of (Hamiltonian reductions of) higher-spin WZW theories (along the lines of, for example, \cite{Coussaert:1995zp}), which could lead to interesting edge dynamics and could be useful both in the context of the holographic duality in 3D gravity as well as as in condensed matter physics. We expect to consider this possibility in a future work.

\section*{Acknowledgements}
We are grateful to Andrea Cappelli, Giandomenico Palumbo, Simon Pekar, Konstantinos Sfestos and Utku Zorba for useful comments. P.S-R. thanks Joaquim Gomis for pointing out Ref. \cite{Boulanger:2002bt}, where Lie algebra expansions based on associative commutative algebras were considered. We specially thank Michelle Lacoste Adunka for carefully reading the manuscript. R.C. is partially funded by the National Agency for Research and Development ANID (ex-CONICYT) - FONDECYT grant N$^\circ$1211077 and by the Research Project Code DIREG$\_$09/2020 of the Universidad Cat\'olica de la Sant\'isima Concepci\'on, Chile. R.C. would like to thank the Direcci\'on de Investigaci\'on and the Vicerrector\'ia de Investigaci\'on of the Universidad Cat\'olica de la Sant\'isima Concepci\'on, Chile, for their constant support. P.S-R. has received funding from the Norwegian Financial Mechanism 2014-2021 via the National Science Centre (NCN) POLS grant 2020/37/K/ST3/03390.

\appendix

\section{Non-relativistic spin-$3$ Chern-Simons theories}
\label{appA}

In this appendix, we explicitly construct the CS forms associated with the expanded algebras $S\times \hsnwi$ and $S\times \hsnwii$ for the semigroups $S^{(1)}_M$ and $S^{(2)}_M$. The corresponding forms for the semigroups $S^{(1)}_E$ and $S^{(2)}_E$ can be obtained by considering the flat limit $\ell\rightarrow\infty$.

\subsection{Newton-Hooke and Bargmann cases}
\label{appNHB}

The CS forms (quasi)invariant under the spin-$3$ extensions of the Extended Newton-Hooke algebra, $\hsni$ and $\hsnii$, defined in \eqref{hsniandii} are obtained by evaluating Eqs. \eqref{hsCS1} and \eqref{hsCS2} for the semigroup $S^{(1)}_M$. Using the product law \eqref{smn} for $n=1$, we find the following values for the pairs $(i,j)$
\be\label{pairs1}
\bal
&i\diamond j=0\quad\rightarrow &\quad& (i,j)=\{(0,0),(1,1)\},\\
&i\diamond j=1\quad\rightarrow &\quad& (i,j)=\{(0,1),(1,0)\},  
\eal
\ee
which can be used to read off the terms that multiply the constants $\mu_i$, $\nu_i$ and $\rho_i$ in the expanded actions \eqref{hsCS1} and \eqref{hsCS2}. Using \eqref{changeFieldsb}-\eqref{changeFieldsb2} and applying the rescaling \eqref{changeFieldsn}, we find for \eqref{CSNWexp}
\be\label{CSNH}
\bal
\mathcal{L}_{S^{(1)}_M\times\nw}&=\mu_{0}\left(\omega^a \wed \bar R^a_{(0)}-2w \wed \ed s\right) +\nu_{0} \,w\wed \ed w+\mu_{1}\left(  \omega^a \wed \bar R^a_{(1)}+e^a \wed \bar R^a_{(0)}-2w \wed \ed m-2\tau \wed \ed s\right) 
\\
&
+2\nu_{1} \,w \wed \ed \tau+\frac{\mu_{0}}{\ell^2}\left(e^a \wed \bar R^a_{(1)}-2\tau \wed \ed m\right)
+\frac{\nu_{0} }{\ell^2}\,\tau \wed \ed\tau,
\eal
\ee
where the curvature components that appear in the action can be derived from \eqref{expNWFcomp} and read
\be
\bar R^a_{(0)}=\ed\omega^a+\epsilon^{ab} \;\omega^b \wed w
+\frac{1}{\ell^2}\epsilon^{ab} \;e^b \wed\tau,
\quad
\bar R^a_{(1)}=\ed e^a +\epsilon^{ab}\left(\omega^b \wed\tau+e^b \wed w\right),
\ee
Eq. \eqref{CSNH} defines Extended Newton-Hooke CS gravity {\cite{Papageorgiou:2010ud,Hartong:2016yrf,Penafiel:2016ufo}. In the limit $\ell\rightarrow\infty$, the second line in \eqref{CSNH} disappears, leading to Extended Bargmann CS gravity \cite{Papageorgiou:2009zc,Bergshoeff:2016lwr,Hartong:2016yrf}. Let us look now at the $\hsni$ extension of this  Lagrangian. Applying the same steps to the CS form \eqref{hsCS1} yields
\be\label{hsCSNH1}
\bal
&\mathcal L_{\hsni}= \mathcal L_{S^{(1)}_M\times\nw}
    + \mu_{0}\left(2\Omega^{ab} \wed \mathcal R^{ab}_{(0)}-\dfrac23 \Omega^{aa} \wed \mathcal R^{bb}_{(0)} -W^a \wed \mathcal F^a_{(0)}-S^a \wed \mathcal R^a_{(0)} \right)
+ \nu_{0} \;W^a \wed  \mathcal R^a_{(0)}
\\
&+ \mu_{1}\bigg(
2 E^{ab}  \wed \mathcal R^{ab}_{(0)}-\dfrac23 E^{aa} \wed \mathcal R^{bb}_{(0)} -V^a  \wed \mathcal F^a_{(0)}-M^a \wed \mathcal R^a_{(0)}
+
2\Omega^{ab}  \wed \mathcal R^{ab}_{(1)}-\dfrac23 \Omega^{aa} \wed \mathcal R^{bb}_{(1)} -W^a  \wed \mathcal F^a_{(1)}-S^a \wed \mathcal R^a_{(1)} 
\bigg)
\\
&
+ \nu_{1} \left(W^a \wed  \mathcal R^a_{(1)}
+V^a \wed  \mathcal R^a_{(0)}\right)
+\frac{\mu_{0}}{\ell^2}\left(2E^{ab} \wed \mathcal R^{ab}_{(1)}-\dfrac23 E^{aa} \wed \mathcal R^{bb}_{(1)} -V^a \wed \mathcal F^a_{(1)}-M^a \wed \mathcal R^a_{(1)} \right)
+ \frac{\nu_{0}}{\ell^2} V^a \wed  \mathcal R^a_{(1)},
\eal
\ee
where, as it happens for \eqref{CSNH}, we do not need to evaluate all the curvatures in \eqref{expNWFcomp1}, but only
\be
\bal
\mathcal R^{ab}_{(0)} &= \ed \Omega^{ab}
+ 
\epsilon^{c(a} \left(\frac12
 W^{b)} \wed \omega^c 
-\Omega^{b)c} \wed w 
+\frac{1}{2\ell^2}
 V^{b)} \wed e^c
-\frac{1}{\ell^2}E^{b)c} \wed\tau
\right)
- \delta^{ab}\epsilon_{cd}\left(\omega^c \wed W^d+\frac{1}{\ell^2} e^c \wed V^d \right)
,\\
\mathcal R^a_{(0)} &= \ed W^a + \epsilon^{ab} W^b \wed w + \frac{1}{\ell^2}\epsilon^{ab} V^b \wed \tau 
,\\
\mathcal F^a_{(0)} &= \ed S^a +  \epsilon^{ab} \left(S^b \wed  
w + W^b \wed s \right)
-2\epsilon_{bc}
\Omega^{ab} \wed\omega^c\\
&+\frac{1}{\ell^2}\epsilon^{ab} \left(M^b \wed  
\tau + V^b \wed m \right)
-\frac{2}{\ell^2}\epsilon_{bc}
E^{ab} \wed e^c
,\\
\mathcal R^{ab}_{(1)} &= \ed E^{ab}
+ 
\epsilon^{c(a} \left(\frac12
 W^{b)} \wed e^c 
+V^{b)} \wed \omega^c 
-\Omega^{b)c} \wed\tau -E^{b)c} \wed\alpha_{(0)} 
\right)
- \delta^{ab}\epsilon_{cd}\left(\omega^c \wed + e^c \wed W^d \right)
,\\
\mathcal R^a_{(1)} &= \ed V^a + \epsilon^{ab} \left(W^b \wed \tau 
+  V^b \wed w\right)
,\\
\mathcal F^a_{(1)} &= \ed M^a +  \epsilon^{ab} \left(S^b \wed  
\tau+M^b \wed  
w + W^b \wed m + V^b \wed s \right)
-2\epsilon_{bc}\left(
\Omega^{ab} \wed e^c
+
E^{ab} \wed\omega^c 
\right),
\eal
\ee
These expressions also follow from using the values \eqref{pairs1} for the pairs $(i,j)$ to quickly find the terms appearing in each component of the curvature two-form. One can similarly evaluate the spin-$3$ extension \eqref{hsCS2} corresponding to the expanded algebra $\hsnii$, which is given by
\be\label{hsCSNH2}
\bal
&\mathcal L_{\hsnii}=\mathcal L_{S^{(1)}_M\times\nw}
+  \mu_{0} \left(2W^{ab}  \wed \tilde{\mathcal F}^{ab}_{(0)}+2S^{ab}  \wed \tilde{\mathcal R}^{ab}_{(0)}-
\dfrac23 W^{aa} \wed  \tilde{\mathcal F}^{bb}_{(0)}-\dfrac23 S^{aa} \wed  \tilde{\mathcal R}^{bb}_{(0)}
-\Omega^a  \wed \tilde{\mathcal R}^a_{(0)}\right)
\\
&+  \mu_{1} \bigg(
+2V^{ab}  \wed \tilde{\mathcal F}^{ab}_{(0)}+2M^{ab}  \wed \tilde{\mathcal R}^{ab}_{(0)}-
\dfrac23 V^{aa}  \wed  \tilde{\mathcal F}^{bb}_{(0)}-\dfrac23 M^{aa}  \wed  \tilde{\mathcal R}^{bb}_{(0)}
-E^a  \wed \tilde{\mathcal R}^a_{(0)}
\\
&
+2W^{ab} \wed \tilde{\mathcal F}^{ab}_{(1)}+2S^{ab}  \wed \tilde{\mathcal R}^{ab}_{(1)}-
\dfrac23 W^{aa} \wed  \tilde{\mathcal F}^{bb}_{(1)}-\dfrac23 S^{aa} \wed  \tilde{\mathcal R}^{bb}_{(1)}
-\Omega^a  \wed \tilde{\mathcal R}^a_{(1)}
\bigg)
\\
&- 2 \nu_{0} \;W^{ab} \wed \tilde{\mathcal R}^{ab}_{(0)}
+  \rho_{0}\; W^{aa} \wed \tilde{\mathcal R}^{bb}_{(0)}
- 2\nu_{1} \left(W^{ab} \wed \tilde{\mathcal R}^{ab}_{(1)}+V^{ab} \wed \tilde{\mathcal R}^{ab}_{(0)}\right)
+ \rho_{1} \left(W^{aa} \wed \tilde{\mathcal R}^{bb}_{(1)}+V^{aa} \wed \tilde{\mathcal R}^{bb}_{(0)}\right)
\\
&+  \frac{\mu_{0}}{\ell^2} \left(2V^{ab}  \wed \tilde{\mathcal F}^{ab}_{(1)}+2M^{ab}  \wed \tilde{\mathcal R}^{ab}_{(1)}-
\dfrac23 V^{aa} \wed  \tilde{\mathcal F}^{bb}_{(1)}-\dfrac23 M^{aa} \wed  \tilde{\mathcal R}^{bb}_{(1)}
-E^a  \wed \tilde{\mathcal R}^a_{(1)}\right)
- \frac{2\nu_{0}}{\ell^2} V^{ab} \wed \tilde{\mathcal R}^{ab}_{(1)}
+ \frac{ \rho_{0}}{\ell^2} V^{aa} \wed \tilde{\mathcal R}^{bb}_{(1)}.
\eal
\ee
In this expression, the relevant curvature components follow from evaluating \eqref{expNWFcomp2} for $S^{(1)}_M$. One then finds
\be
\bal
\tilde{\mathcal R}^a_{(0)} &= \ed \Omega^a
+\epsilon^{ab} \Omega^b \wed w 
-2\epsilon_{bc} W^{ab}  \wed \omega^c 
+\frac{1}{\ell^2}\left(\epsilon^{ab} E^b \wed \tau
-2\epsilon_{bc} V^{ab}  \wed e^c
 \right)
,\\
\tilde{\mathcal R}^{ab}_{(0)} &= \ed W^{ab}
-
\epsilon^{c(a}  W^{b)c} \wed w
-
\frac{1}{\ell^2}\epsilon^{c(a}  V^{b)c} \wed\tau
,\\
\tilde{\mathcal F}^{ab}_{(0)} &= \ed S^{ab} +\epsilon^{c(a}\left( \frac12
 \Omega^{b)} \wed \omega^c  -W^{b)c} \wed s  -S^{b)c} \wed w \right) 
- \delta^{ab}\epsilon_{cd}\omega^c  \wed 
\Omega^d \\
&+
\frac{1}{\ell^2}\epsilon^{c(a}\left( \frac12
 E^{b)} \wed e^c  -V^{b)c} \wed m  -M^{b)c} \wed\tau \right) 
-\frac{1}{\ell^2} \delta^{ab}\epsilon_{cd} e^c  \wed 
E^d ,\\
\tilde{\mathcal R}^a_{(1)} &= \ed E^a
+\epsilon^{ab} \left(\Omega^b \wed \tau+E^b \wed w  \right)
-2\epsilon_{bc} \left( W^{ab}  \wed e^c+V^{ab}  \wed \omega^c \right)
,\\
\tilde{\mathcal R}^{ab}_{(1)}& = \ed V^{ab}
-
\epsilon^{c(a}\left(  W^{b)c} \wed\tau 
+V^{b)c} \wed w \right)
,\\
\tilde{\mathcal F}^{ab}_{(1)}& = \ed M^{ab}  +\epsilon^{c(a}\left( \frac12
 \Omega^{b)} \wed e^c  -W^{b)c} \wed m  -S^{b)c} \wed\tau 
 +\frac12
 E^{b)} \wed \omega^c  -V^{b)c} \wed s  -M^{b)c} \wed w \right) 
\\&- \delta^{ab}\epsilon_{cd}\left(\omega^c  \wed 
E^d+ e^c  \wed 
\Omega^d \right).
\eal
\ee
Due to the relation \eqref{SMSEcontraction}, it is clear that in the flat limit $\ell\rightarrow\infty$ the forms \eqref{hsCSNH1} and \eqref{hsCSNH2} reduce to two spin-$3$ Bargmann gravity Lagrangians
\be\label{SMSEcontraction1}
\mathcal L_{\hsni} \xrightarrow[\ell\rightarrow\infty]\, \mathcal L_{\hsbi},\qquad
\mathcal L_{\hsnii} \xrightarrow[\ell\rightarrow\infty]\, \mathcal L_{\hsbii},
\ee
which could have been independently obtained by using the $S^{(1)}_E$ product law \eqref{sen}
\be\label{pairs11}
\bal
&i\diamond j=0\quad\rightarrow &\quad& (i,j)=\{(0,0)\},\\
&i\diamond j=1\quad\rightarrow &\quad& (i,j)=\{(0,1),(1,0)\},  
\eal
\ee
when evaluating the expanded CS actions \eqref{hsCS1} and \eqref{hsCS2}.  It comes without saying that \eqref{SMSEcontraction1} can be extended to include the relativistic case, in analogy with \eqref{fulldiagram1ads} and \eqref{fulldiagram2ads}, which requires to introduce CS actions invariant under the relativistic algebra \eqref{hsadsiso}, such as the one considered in \cite{Campoleoni:2010zq}, together with the CS forms associated to the extensions \eqref{otherexpsads}. 

As stated before, the CS Lagrangians \eqref{hsCSNH1} and \eqref{hsCSNH2} could also be constructed by replacing the expanded connections \eqref{expAs} and curvatures \eqref{expFs} in \eqref{CSform}, and using the invariant bilinear forms \eqref{itnwexp}-\eqref{itexp1} for $S=S^{(1)}_M$. However, here we have chosen the alternative path of evaluating the non-vanishing combinations \eqref{pairs1} directly in \eqref{hsCS1} and \eqref{hsCS2}.

\subsection{Non-relativistic AdS-Lorentz and Maxwell cases}
\label{appAdSLM}

Following the same procedure adopted in the previous example, we derive the $S^{(2)}_M$-expanded CS forms. In this case we use the semigroup product law \eqref{smn} for $n=2$, for which the relevant pairs of semigroup indices are
\be\label{pairs2}
\bal
&i\diamond j=0\quad\rightarrow &\quad& (i,j)=\{(0,0)\},\\
&i\diamond j=1\quad\rightarrow &\quad& (i,j)=\{(0,1),(1,0),(1,2),(2,1)\}  ,\\
&i\diamond j=2\quad\rightarrow &\quad& (i,j)=\{(0,2),(2,0),(1,1),(2,2)\}.  
\eal
\ee
Using the gauge field definitions \eqref{changeFieldsb}-\eqref{changeFieldsb2} and \eqref{changeFieldsm}-\eqref{changeFieldsm2}, together with the rescaling \eqref{changeFieldsal}, one can evaluate the CS forms \eqref{hsCS1} and \eqref{hsCS2}, which define higher-spin gravity theories invariant under the gauge symmetries $\hsali$ and $\hsalii$. We first look at the Lagrangian \eqref{CSNWexp}, which define NR AdS-Lorentz CS gravity \cite{Concha:2019lhn,Penafiel:2019czp}
\be\label{CSNRAdSL}
\bal
&\mathcal{L}_{S^{(2)}_M\times\nw}=\mu_{0}\left(  \omega^a \wed \bar R^a_{(0)}-2 w \wed \ed s \right) +\nu_{0}\, w  \wed \ed w  +\mu_{1}\left(  \omega^a \wed \bar R^a_{(1)}+e^a \wed \bar R^a_{(0)}
-2 w \wed \ed m -2 \tau \wed \ed s 
\right) \\
&+2\nu_{1}\, w  \wed \ed \tau 
+\mu_{2}\left(  \omega^a \wed \bar R^a_{(2)}+e^a \wed \bar R^a_{(1)}+k^a \wed \bar R^a_{(0)}
-2 w \wed \ed t -2 \tau \wed \ed m -2 y \wed \ed s \right)
+\nu_{2}\left(2 w  \wed \ed y + \tau  \wed \ed \tau  \right)
\\
&+\frac{\mu_{1}}{\ell^2}\left(  e^a \wed \bar R^a_{(2)}+k^a \wed \bar R^a_{(1)}
-2 \tau \wed \ed t -2 y \wed \ed m \right) 
+\frac{2\nu_{1}}{\ell^2}\, \tau  \wed \ed y 
+\frac{\mu_{2}}{\ell^2}\left(k^a \wed \bar R^a_{(2)}
-2 y \wed \ed t \right)
+\frac{\nu_{2}}{\ell^2}\, y  \wed \ed y .
\eal
\ee
Here, the curvature components \eqref{expNWFcomp} take the form
\be
\bal
\bar R^a_{(0)}&=\ed\omega^a+\epsilon^{ab}\omega^b \wed w,\\
\bar R^a_{(1)}&=\ed e^a +\epsilon^{ab}\left(\omega^b \wed\tau+e^b \wed w \wed \right)
+\frac{1}{\ell^2}\epsilon^{ab}\left( e^b \wed y+ k^b \wed \tau\right),\\
\bar R^a_{(2)}&=\ed k^a +\epsilon^{ab}\left(\omega^b \wed y+k^b \wed w+e^b \wed \tau \right)+\frac{1}{\ell^2}\epsilon^{ab}\; k^b \wed y.\\
\eal
\ee
In the limit $\ell\rightarrow\infty$, the last line in Eq. \eqref{CSNH} disappears. The first two lines correspond to the CS form $\mathcal{L}_{S^{(2)}_E\times\nw}$, which is precisely the (exotic) Maxwellian NR gravity theory first found in \cite{Aviles:2018jzw}. The first higher-spin extension \eqref{hsCS1} takes the form
\be\label{hsCSNRAdSL1}
\bal
&\mathcal L_{\hsali}= \mathcal L_{S^{(1)}_M\times\nw}
+ \mu_0 \bigg(2\Omega^{ab} \wed \mathcal R^{ab}_{(0)}
-\dfrac23 \Omega^{aa} \wed \mathcal R^{bb}_{(0)}
 -W^a \wed \mathcal F^a_{(0)}-S^a \wed \mathcal R^a_{(0)}\bigg)
\\
&+ \mu_1 \bigg(2\Omega^{ab} \wed \mathcal R^{ab}_{(1)}+2E^{ab} \wed \mathcal R^{ab}_{(0)}
-\dfrac23 \Omega^{aa} \wed \mathcal R^{bb}_{(1)}
-\dfrac23 E^{aa} \wed \mathcal R^{bb}_{(0)}
 -W^a \wed \mathcal F^a_{(1)}-V^a \wed \mathcal F^a_{(0)}-S^a \wed \mathcal R^a_{(1)}-M^a \wed \mathcal R^a_{(0)}\bigg)
\\&
+ \mu_{2}\bigg(2\Omega^{ab} \wed \mathcal R^{ab}_{(2)}
+2K^{ab} \wed \mathcal R^{ab}_{(0)}
+2E^{ab} \wed \mathcal R^{ab}_{(1)}
-\dfrac23 \Omega^{aa} \wed \mathcal R^{bb}_{(2)} 
-\dfrac23 K^{aa} \wed \mathcal R^{bb}_{(0)} 
-\dfrac23 E^{aa} \wed \mathcal R^{bb}_{(1)}
\\
&-W^a \wed \mathcal F^a_{(2)}
-Y^a \wed \mathcal F^a_{(0)} 
-V^a \wed \mathcal F^a_{(1)}
-S^a \wed \mathcal R^a_{(2)} -T^a \wed \mathcal R^a_{(0)}
 -M^a \wed \mathcal R^a_{(1)}
\bigg)
\\
&
+ \nu_{0} \;W^a \wed  \mathcal R^a_{(0)}
+ \nu_{1} \left(W^a \wed  \mathcal R^a_{(1)}
+V^a \wed  \mathcal R^a_{(0)}\right)
+ \nu_{2} \left(W^a \wed  \mathcal R^a_{(2)}
+Y^a \wed  \mathcal R^a_{(0)}+V^a \wed  \mathcal R^a_{(1)}\right)
\\
&+\frac{\mu_1}{\ell^2}\bigg(2E^{ab} \wed \mathcal R^{ab}_{(2)}+2K^{ab} \wed \mathcal R^{ab}_{(1)}
-\dfrac23 E^{aa} \wed \mathcal R^{bb}_{(2)}
-\dfrac23 K^{aa} \wed \mathcal R^{bb}_{(1)}
 -V^a \wed \mathcal F^a_{(2)}-A^a_{(2)} \wed \mathcal F^a_{(1)}-M^a \wed \mathcal R^a_{(2)}-T^a \wed \mathcal R^a_{(1)}\bigg)
\\
&
+ \frac{\nu_{1}}{\ell^2} \left(V^a \wed  \mathcal R^a_{(2)}
+Y^a \wed  \mathcal R^a_{(1)}\right)
+\frac{\mu_2}{\ell^2}\bigg(2K^{ab} \wed \mathcal R^{ab}_{(2)}-\dfrac23 K^{aa} \wed \mathcal R^{bb}_{(2)} -Y^a \wed \mathcal F^a_{(2)}-T^a \wed \mathcal R^a_{(2)}
\bigg)
+ \frac{\nu_{2}}{\ell^2} Y^a \wed  \mathcal R^a_{(2)},
\eal
\ee
where the higher-spin curvatures appearing in the action can be found by evaluating Eq. \eqref{expNWFcomp1} and are given by
\be
\bal
\mathcal R^{ab}_{(0)} &= \ed \Omega^{ab}
+ 
\epsilon^{c(a} \left(\frac12
 W^{b)} \wed \omega^c
-\Omega^{b)c} \wed w 
\right)
- \delta^{ab}\epsilon_{cd}\omega^c \wed W^d ,\\
\mathcal R^a_{(0)} &= \ed W^a + \epsilon^{ab} W^b \wed  w 
,\\
\mathcal F^a_{(0)} &= \ed S^a +  \epsilon^{ab} \left(S^b \wed  
 w + W^b \wed s \right)
-2\epsilon_{bc}
\Omega^{ab} \wed\omega^c
,\\
\mathcal R^{ab}_{(1)} &= \ed E^{ab}
+ 
\epsilon^{c(a} \left(\frac12
 W^{b)} \wed e^c 
 +\frac12
 V^{b)} \wed \omega^c 
-\Omega^{b)c} \wed\tau
-E^{b)c} \wed w 
\right)
- \delta^{ab}\epsilon_{cd}\left(\omega^c \wed V^d+e^c \wed W^d\right) 
\\&
+ 
\frac{1}{\ell^2}\epsilon^{c(a} \left(\frac12
 V^{b)} \wed k^c 
 +\frac12
 Y^{b)} \wed e^c
-E^{b)c} \wed y
-K^{b)c} \wed\tau 
\right)
-\frac{1}{\ell^2} \delta^{ab}\epsilon_{cd}\left(e^c \wed Y^d+k^c \wed V^d\right),
\\
\mathcal R^a_{(1)} &= \ed V^a + \epsilon^{ab} W^b\left( \wed \tau 
+ V^b \wed w\right)
+\frac{1}{\ell^2}\epsilon^{ab} \left( V^b \wed \alpha_{(2)}+Y^b \wed \tau\right)
,\\
\mathcal F^a_{(1)} &= \ed M^a +  \epsilon^{ab} \left(S^b \wed  
\tau+M^b \wed  
w + W^b \wed m + V^b \wed s \right)
-2\epsilon_{bc}\left(
\Omega^{ab} \wed e^c
+
E^{ab} \wed\omega^c 
\right)
\\
& +\frac{1}{\ell^2}  \epsilon^{ab} \left(M^b \wed  
y+T^b \wed  
\tau + V^b \wed t + Y^b \wed m \right)
-\frac{2}{\ell^2}\epsilon_{bc}\left(
E^{ab} \wed k^c 
+
K^{ab} \wed e^c
\right)
,\\
\mathcal R^{ab}_{(2)} &= \ed K^{ab}
+ 
\frac12\epsilon^{c(a} \left(
 W^{b)} \wed k^c 
+
 Y^{b)} \wed \omega^c 
+
 V^{b)} \wed e^c 
 \right)
 -\epsilon^{c(a}\left(
\Omega^{b)c} \wed y
+K^{b)c} \wed w 
+E^{b)c} \wed\tau
\right)\\
&- \delta^{ab}\epsilon_{cd}\left(\omega^c \wed Y^d+k^c \wed W^d
+e^c \wed V^d\right) 
+ \frac{1}{\ell^2}
\epsilon^{c(a} \left(\frac12
 Y^{b)} \wed k^c
-K^{b)c} \wed y
\right)
-\frac{1}{\ell^2} \delta^{ab}\epsilon_{cd} k^c \wed Y^d
,\\
\mathcal R^a_{(2)} &= \ed Y^a + \epsilon^{ab}\left( W^b \wed y+Y^b \wed w+V^b \wed \tau \right)
+ \frac{1}{\ell^2}\epsilon^{ab} Y^b \wed y
,\\
\mathcal F^a_{(2)} &= \ed T^a +  \epsilon^{ab} \left(S^b \wed  
y+T^b \wed  
w +M^b \wed  
\tau + W^b \wed t + Y^b \wed s
+V^b \wed m\right)
\\
&-2\epsilon_{bc}\left(
\Omega^{ab} \wed k^c
+
K^{ab} \wed\omega^c 
+
E^{ab} \wed e^c 
\right)
+\frac{1}{\ell^2}\epsilon^{ab} \left(T^b \wed  
y + Y^b \wed t \right)
-\frac{2}{\ell^2}\epsilon_{bc}
K^{ab} \wed k^c.
\eal
\ee
Following the same steps to find a second spin-$3$ extension of the NR AdS-Lorentz CS form leads to
\be\label{hsCSNRAdSL2}
\bal
&\mathcal L_{\hsalii}= \mathcal L_{S^{(1)}_M\times\nw}
+  \mu_{0} \left(2W^{ab}  \wed \tilde{\mathcal F}^{ab}_{(0)}+2S^{ab}  \wed \tilde{\mathcal R}^{ab}_{(0)}-
\dfrac23 W^{aa} \wed  \tilde{\mathcal F}^{bb}_{(0)}-\dfrac23 S^{aa} \wed  \tilde{\mathcal R}^{bb}_{(0)}
-\Omega^a \wed \tilde{\mathcal R}^a_{(0)}\right)
\\
&+  \mu_{1} \bigg(
2W^{ab}  \wed \tilde{\mathcal F}^{ab}_{(1)}
+2V^{ab}  \wed \tilde{\mathcal F}^{ab}_{(0)}
+2S^{ab}  \wed \tilde{\mathcal R}^{ab}_{(1)}
+2M^{ab}  \wed \tilde{\mathcal R}^{ab}_{(0)}
-\dfrac23 W^{aa} \wed  \tilde{\mathcal F}^{bb}_{(1)}
\\
&
-\dfrac23 V^{aa} \wed  \tilde{\mathcal F}^{bb}_{(0)}
-\dfrac23 S^{aa} \wed  \tilde{\mathcal R}^{bb}_{(1)}
-\dfrac23 M^{aa} \wed  \tilde{\mathcal R}^{bb}_{(0)}
-\Omega^a \wed \tilde{\mathcal R}^a_{(1)}
-E^a  \wed \tilde{\mathcal R}^a_{(0)}
\bigg)
\\
&+  \mu_{2} \bigg(
2W^{ab}  \wed \tilde{\mathcal F}^{ab}_{(2)}
+2Y^{ab} \wed \tilde{\mathcal F}^{ab}_{(0)}
+2V^{ab} \wed \tilde{\mathcal F}^{ab}_{(1)}
+2S^{ab} \wed \tilde{\mathcal R}^{ab}_{(2)}
+2T^{ab} \wed \tilde{\mathcal R}^{ab}_{(0)}
+2M^{ab} \wed \tilde{\mathcal R}^{ab}_{(1)}
-\dfrac23 W^{aa} \wed  \tilde{\mathcal F}^{bb}_{(2)}
\\
&
-\dfrac23 Y^{aa} \wed  \tilde{\mathcal F}^{bb}_{(0)}
-\dfrac23 V^{aa} \wed  \tilde{\mathcal F}^{bb}_{(1)}
-\dfrac23 S^{aa} \wed  \tilde{\mathcal R}^{bb}_{(2)}
-\dfrac23 T^{aa} \wed  \tilde{\mathcal R}^{bb}_{(0)}
-\dfrac23 M^{aa} \wed  \tilde{\mathcal R}^{bb}_{(1)}
-\Omega^a \wed \tilde{\mathcal R}^a_{(2)}
-K^a  \wed \tilde{\mathcal R}^a_{(0)}
-E^a \wed \tilde{\mathcal R}^a_{(1)}
\bigg)
\\
& -2 \nu_{0} \;W^{ab} \wed \tilde{\mathcal R}^{ab}_{(0)}
 -2 \nu_{1} \left(W^{ab} \wed \tilde{\mathcal R}^{ab}_{(1)}+V^{ab} \wed \tilde{\mathcal R}^{ab}_{(0)}\right)
 -2 \nu_{2} \left(W^{ab} \wed \tilde{\mathcal R}^{ab}_{(2)}+Y^{ab} \wed \tilde{\mathcal R}^{ab}_{(0)}+V^{ab} \wed \tilde{\mathcal R}^{ab}_{(1)}\right)
\\
&
+ \rho_{0} \;W^{aa} \wed \tilde{\mathcal R}^{bb}_{(0)}
+ \rho_{1} \left(W^{aa} \wed \tilde{\mathcal R}^{bb}_{(1)}+V^{aa} \wed \tilde{\mathcal R}^{bb}_{(0)}\right)
+ \rho_{2} \left(W^{aa} \wed \tilde{\mathcal R}^{bb}_{(2)}+A^{aa}_{(2)} \wed \tilde{\mathcal R}^{bb}_{(0)}+V^{aa} \wed \tilde{\mathcal R}^{bb}_{(1)}\right)
\\
&
+  \frac{\mu_{1}}{\ell^2} \bigg(
2V^{ab}  \wed \tilde{\mathcal F}^{ab}_{(2)}
+2Y^{ab}  \wed \tilde{\mathcal F}^{ab}_{(1)}
+2M^{ab}  \wed \tilde{\mathcal R}^{ab}_{(2)}
+2T^{ab}  \wed \tilde{\mathcal R}^{ab}_{(1)}
-\dfrac23 V^{aa} \wed  \tilde{\mathcal F}^{bb}_{(2)}
\\
&
-\dfrac23 Y^{aa} \wed  \tilde{\mathcal F}^{bb}_{(1)}
-\dfrac23 M^{aa} \wed  \tilde{\mathcal R}^{bb}_{(2)}
-\dfrac23 T^{aa} \wed  \tilde{\mathcal R}^{bb}_{(1)}
-E^a  \wed \tilde{\mathcal R}^a_{(2)}
-K^a  \wed \tilde{\mathcal R}^a_{(1)}
\bigg)
\eal
\ee

\be\nonumber
\bal
&
+\frac{\mu_2}{\ell^2}\bigg(2Y^{ab}  \wed \tilde{\mathcal F}^{ab}_{(2)}+2T^{ab}  \wed \tilde{\mathcal R}^{ab}_{(2)}-
\dfrac23 Y^{aa} \wed  \tilde{\mathcal F}^{bb}_{(2)}-\dfrac23 T^{aa} \wed  \tilde{\mathcal R}^{bb}_{(2)}
-K^a  \wed \tilde{\mathcal R}^a_{(2)}
\bigg)
\\
&
-\frac{2 \nu_{1}}{\ell^2} \left(V^{ab} \wed \tilde{\mathcal R}^{ab}_{(2)}+Y^{ab} \wed \tilde{\mathcal R}^{ab}_{(1)}\right)
-\frac{2 \nu_{2}}{\ell^2} Y^{ab} \wed \tilde{\mathcal R}^{ab}_{(2)}
+\frac{  \rho_{1} }{\ell^2} \left(V^{aa} \wed \tilde{\mathcal R}^{bb}_{(2)}+Y^{aa} \wed \tilde{\mathcal R}^{bb}_{(1)}\right)
+\frac{ \rho_{2} }{\ell^2} Y^{aa} \wed \tilde{\mathcal R}^{bb}_{(2)},
\eal
\ee
where the curvature components relevant here follow from \eqref{expNWFcomp2} and read
\be
\bal
\tilde{\mathcal R}^a_{(0)} &= \ed \Omega^a
+\epsilon^{ab} \Omega^b \wed w 
-2\epsilon_{bc} W^{ab}  \wed \omega^c 
\\
\tilde{\mathcal R}^{ab}_{(0)} &= \ed W^{ab}
-
\epsilon^{c(a}  W^{b)c} \wed w
\\
\tilde{\mathcal F}^{ab}_{(0)} &= \ed S^{ab} +\epsilon^{c(a}\left( \frac12
 \Omega^{b)} \wed \omega^c  -W^{b)c} \wed s  -S^{b)c} \wed w \right) 
- \delta^{ab}\epsilon_{cd}\omega^c  \wed 
\Omega^d \\
\tilde{\mathcal R}^a_{(1)} &= \ed E^a
+\epsilon^{ab} \left(\Omega^b \wed \tau+E^b \wed w  \right)
-2\epsilon_{bc} \left( W^{ab}  \wed e^c+V^{ab} \wed \omega^c \right)
\\
&
+\frac{1}{\ell^2}\epsilon^{ab} \left(E^b \wed y+K^b \wed \tau  \right)
-\frac{2}{\ell^2}\epsilon_{bc} \left( V^{ab}  \wed k^c+Y^{ab}  \wed e^c \right)
\\
\tilde{\mathcal R}^{ab}_{(1)}& = \ed V^{ab}
-
\epsilon^{c(a}\left(  W^{b)c} \wed\tau 
+V^{b)c} \wed w \right)
-
\frac{1}{\ell^2}\epsilon^{c(a}\left(  V^{b)c} \wed y 
+Y^{b)c} \wed\tau \right)
\\
\tilde{\mathcal F}^{ab}_{(1)}& = \ed M^{ab}  +\epsilon^{c(a}\left( \frac12
 \Omega^{b)} \wed e^c 
 +\frac12
 E^{b)} \wed \omega^c -W^{b)c} \wed m  -V^{b)c} \wed s  -S^{b)c} \wed\tau  -M^{b)c} \wed w \right) 
\\&- \delta^{ab}\epsilon_{cd}\left(\omega^c  \wed 
E^d+e^c  \wed 
\Omega^d \right)- \frac{1}{\ell^2}\delta^{ab}\epsilon_{cd}\left(e^c  \wed 
K^d+k^c  \wed 
E^d \right)\\
&
+\frac{1}{\ell^2}\epsilon^{c(a}\left( \frac12
 E^{b)} \wed k^c  
 +\frac12
 K^{b)} \wed e^c -V^{b)c} \wed t  -Y^{b)c} \wed m  -M^{b)c} \wed y  -T^{b)c} \wed\tau \right) \\
\tilde{\mathcal R}^a_{(2)} &= \ed K^a
+\epsilon^{ab} \left(\Omega^b \wed \alpha_{(2)}+K^b \wed w  +E^b \wed \tau \right)
-2\epsilon_{bc} \left( W^{ab}  \wed k^c+Y^{ab} \wed \omega^c +V^{ab}  \wed e^c \right)\\
&
+\frac{1}{\ell^2}\epsilon^{ab} K^b \wed y 
-\frac{2}{\ell^2}\epsilon_{bc} Y^{ab}  \wed k^c,
\\
\tilde{\mathcal R}^{ab}_{(2)}& = \ed Y^{ab}
-
\epsilon^{c(a}\left(  W^{b)c} \wed y
+Y^{b)c} \wed w +V^{b)c} \wed\tau \right)
-\frac{1}{\ell^2}
\epsilon^{c(a} Y^{b)c} \wed y
\\
\tilde{\mathcal F}^{ab}_{(2)}& = \ed T^{ab} - \delta^{ab}\epsilon_{cd}\left(\omega^c  \wed 
K^d+k^c  \wed 
\Omega^d +e^c  \wed 
E^d \right)
\\
&
+\epsilon^{c(a}\left( \frac12
 \Omega^{b)} \wed k^c  -W^{b)c} \wed t  -S^{b)c} \wed y
 +\frac12
 K^{b)} \wed \omega^c  -Y^{b)c} \wed s  -T^{b)c} \wed w
+\frac12
 E^{b)} \wed e^c  -V^{b)c} \wed m  -M^{b)c} \wed\tau  \right) 
\\
&
+\frac{1}{\ell^2}\epsilon^{c(a}\left( \frac12
 K^{b)} \wed k^c -Y^{b)c} \wed t  -T^{b)c} \wed y 
\right) - \frac{1}{\ell^2}\delta^{ab}\epsilon_{cd}\;k^c  \wed
K^d.
\eal
\ee
Using \eqref{SMSEcontraction}, one finds that applying the flat limit $\ell\rightarrow\infty$ to \eqref{hsCSNRAdSL1} and \eqref{hsCSNRAdSL2} yields the CS for the corresponding spin-$3$ extensions of the NR Maxwell algebra \eqref{hsmiandii}
\be\label{SMSEcontraction2}
\mathcal L_{\hsali} \xrightarrow[\ell\rightarrow\infty]\, \mathcal L_{\hsmi},\qquad
\mathcal L_{\hsalii} \xrightarrow[\ell\rightarrow\infty]\, \mathcal L_{\hsmii}.
\ee
The Maxwellian action can be independently constructed by repeating the same computation with the semigroup $S^{(2)}_E$ as starting point. This requires to replace \eqref{pairs2} by
\be\label{pairs22}
\bal
&i\diamond j=0\quad\rightarrow &\quad& (i,j)=\{(0,0)\},\\
&i\diamond j=1\quad\rightarrow &\quad& (i,j)=\{(0,1),(1,0)\}  ,\\
&i\diamond j=2\quad\rightarrow &\quad& (i,j)=\{(0,2),(2,0),(1,1),(2,2)\}. 
\eal
\ee
One could easily extend \eqref{SMSEcontraction2} to include underlying relativistic gravity theories in analogy with the diagrams \eqref{fulldiagram1adslor} and \eqref{fulldiagram2adslor} at the level of the gauge symmetries. The relativistic CS forms are obtained by extending the AdS-Lorentz higher-spin gravity action invariant under \eqref{hsadsL} \cite{Caroca:2017izc} using the CS forms associated to the extensions \eqref{otherexpsadslor}. As before, the CS theories defined by \eqref{hsCSNRAdSL1} and \eqref{hsCSNRAdSL2} could be alternatively obtained by evaluating the connections \eqref{expAs} and the curvatures \eqref{expFs} for $S=S^{(2)}_M$ in \eqref{CSform}, and using the explicit form of the invariant bilinear forms \eqref{itnwexp}-\eqref{itexp1}.


\providecommand{\href}[2]{#2}\begingroup\raggedright\endgroup

\end{document}